\newif\ifshowcomments
\showcommentstrue

\documentclass{egpubl}
\usepackage{egsr2022}
 
\SpecialIssuePaper         

\usepackage[T1]{fontenc}
\usepackage{dfadobe}  

\usepackage{cite}  
\BibtexOrBiblatex
\electronicVersion
\PrintedOrElectronic
\ifpdf \usepackage[pdftex]{graphicx} \pdfcompresslevel=9
\else \usepackage[dvips]{graphicx} \fi

\usepackage{egweblnk} 
\usepackage{amsfonts}

\usepackage{graphicx}
\usepackage{amsmath}
\usepackage{overpic}
\usepackage{enumitem} 
\usepackage{overpic} 
\usepackage{color}
\usepackage[export]{adjustbox}
\usepackage{nicefrac}
\usepackage{multirow}

\definecolor{turquoise}{cmyk}{0.65,0,0.1,0.3}
\definecolor{purple}{rgb}{0.65,0,0.65}
\definecolor{dark_green}{rgb}{0, 0.5, 0}
\definecolor{orange}{rgb}{0.8, 0.6, 0.2}
\definecolor{red}{rgb}{0.8, 0.2, 0.2}
\definecolor{darkred}{rgb}{0.6, 0.1, 0.05}
\definecolor{blueish}{rgb}{0.0, 0.3, .6}
\definecolor{light_gray}{rgb}{0.7, 0.7, .7}
\definecolor{pink}{rgb}{1, 0, 1}
\definecolor{greyblue}{rgb}{0.25, 0.25, 1}

\DeclareMathOperator*{\argmin}{arg\,min}

\newcommand{\loss}[1]{\mathcal{L}_\mathrm{#1}}
\newcommand{\weight}[1]{\lambda_\mathrm{#1}}

\newcommand{\Real}{\mathbb{R}}
\newcommand{\calM}{\mathcal{M}}
\newcommand{\bx}{\boldsymbol{x}}
\newcommand{\bom}{\boldsymbol{\omega}}
\newcommand{\bomi}{\bom_\mathrm{i}}
\newcommand{\bomo}{\bom_\mathrm{o}}
\newcommand{\btheta}{\boldsymbol{\theta}}
\newcommand{\param}{\theta}
\newcommand{\D}{\mathrm{d}}

\newcommand{\sph}{\mathbb{S}^2}
\newcommand{\Li}{L_\mathrm{i}}
\newcommand{\brdf}{f_\mathrm{r}}
\newcommand{\vel}{v_{\bot}}

\newcommand{\thetaG}{\btheta_\mathrm{g}}
\newcommand{\thetaA}{\btheta_\mathrm{a}}
\newcommand{\thetaS}{\btheta_\mathrm{s}}

\newcommand{\sdf}{\phi}
\newcommand{\calMi}{\calM}
\newcommand{\calMe}{\tilde{\calM}}

\newcommand{\kd}{k_\mathrm{d}}
\newcommand{\ks}{k_\mathrm{s}}
\newcommand{\rough}{\alpha}

\newcommand{\nrf}{\mathcal{B}}
\newcommand{\render}{\mathcal{R}}
\newcommand{\img}{\mathcal{I}}
\newcommand{\mask}{\mathcal{S}}

\renewcommand{\paragraph}[1]{\vspace{1em}\noindent\textbf{#1}.}

\newlength{\resLen}
\newcommand{\clrbar}[2]{
	\textbf{#1: 0} \includegraphics[height=0.1in]{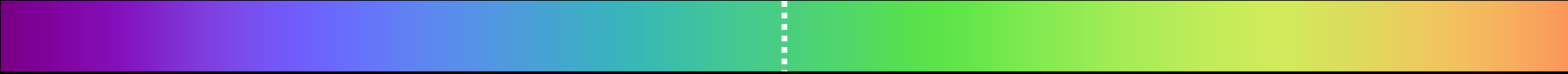} \textbf{#2}
}

\ifshowcomments
	\newcommand{\sz}[1]{\textcolor{dark_green}{[SZ: #1]}}
	\newcommand{\zd}[1]{\textcolor{red}{[ZD: #1]}}
	\newcommand{\gc}[1]{\textcolor{purple}{[GC: #1]}}
\else
	\newcommand{\sz}[1]{}
	\newcommand{\zd}[1]{}
	\newcommand{\gc}[1]{}
\fi

\iffalse
	\newcommand{\rev}[1]{{\color{darkred}{#1}}}
\else
	\newcommand{\rev}[1]{#1}
\fi

\begin{document}

\title[%
	Physics-Based Inverse Rendering using Combined Implicit and Explicit Geometries
]{%
	Physics-Based Inverse Rendering using\\Combined Implicit and Explicit Geometries
}


\author[Cai et al.]{%
	\parbox{\textwidth}{\centering 
		G. Cai$^{1,2}$,
		K. Yan$^{1,2}$,
		Z. Dong$^2$,
		I. Gkioulekas$^3$,
		S. Zhao$^1$
	}
	\\
	\parbox{\textwidth}{\centering
		$^1$University of California, Irvine
		\hspace{1cm}
		$^2$Meta Reality Labs Research
		\hspace{1cm}
		$^3$Carnegie Mellon University
	}
}

\teaser{%
	\includegraphics[width=\textwidth]{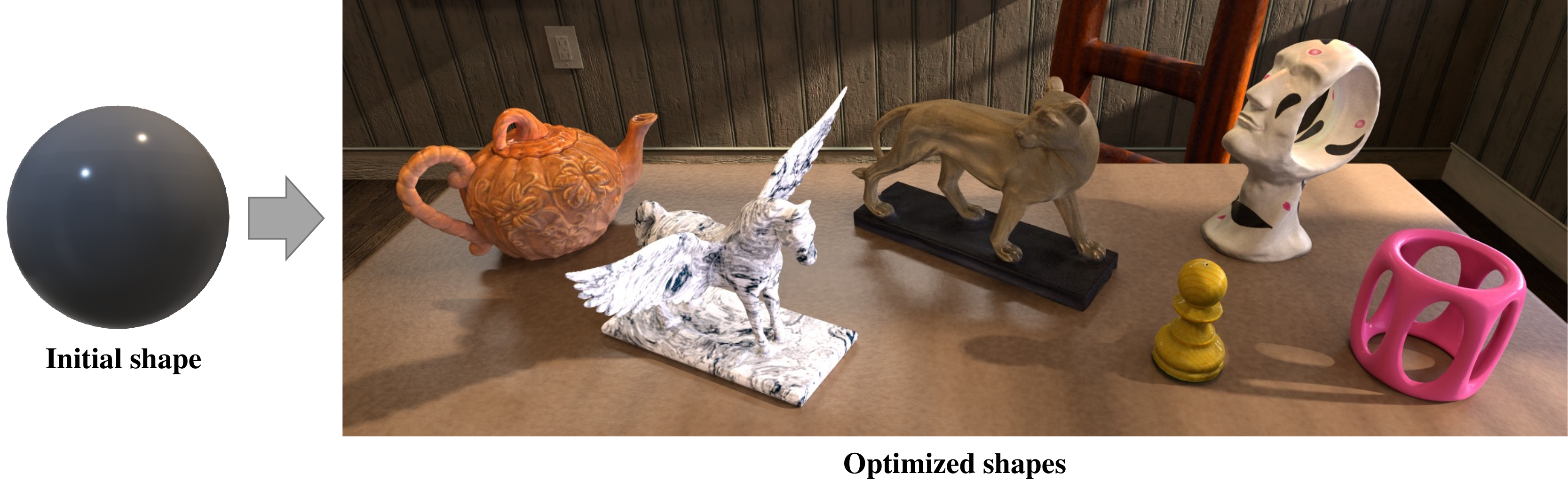}
	\caption{\label{fig:teaser}
		We introduce a new technique that combines implicit and explicit geometric representations to solve physics-based inverse-rendering problems.
		When optimizing object geometry, our technique is capable of deforming simple spheres into detailed shapes with complex topologies.
		In this figure, we show re-renderings of 3D models (i.e., objects on the table and the chair behind them) reconstructed individually using our technique.
	}
}

\maketitle

\begin{abstract}
	Mathematically representing the shape of an object is a key ingredient for solving inverse rendering problems.
	Explicit representations like meshes are efficient to render in a differentiable fashion but have difficulties handling topology changes.
	Implicit representations like signed-distance functions, on the other hand, offer better support of topology changes but are much more difficult to use for physics-based differentiable rendering.
	We introduce a new physics-based inverse rendering pipeline that uses both implicit and explicit representations.
	Our technique enjoys the benefit of both representations by supporting both topology changes and differentiable rendering of complex effects such as environmental illumination, soft shadows, and interreflection.
	We demonstrate the effectiveness of our technique using several synthetic and real examples.

		

\end{abstract}

\section{Introduction}
\label{sec:intro}
Inverse rendering---the inference of object shape and appearance from 2D images---has been a long-standing problem in computer vision and graphics.

A key ingredient to solving inverse-rendering problems is the mathematical description of object surfaces. Previously, most vision and graphics applications utilize \emph{explicit} representations like polygonal meshes that enjoys high flexibility and easy editability. Additionally, state-of-the-art physics-based differentiable rendering techniques---which requires Monte Carlo sampling over not only object surfaces, but also discontinuity (e.g., visibility) boundaries---have largely focused on mesh-based representations.
However, when solving inverse-rendering problems using gradient-based methods, explicit representations suffer from one major limitation: the difficulty in handling topology changes.
This causes most existing mesh-based inverse rendering technique to require good initial geometry and/or frequent remeshing during optimization.

On the contrary, \emph{implicit} geometric representations define a surface as the zero level set of some function (such as signed-distance functions).
Compared to their explicit counterparts, these representations enjoy a major advantage of allowing easy topology changes and, therefore, have been gaining popularity for solving inverse-rendering problems.
However, implicit representations suffer from their own limitations.
First, computing the intersection between a ray and an implicit surface (using, for example, sphere tracing~\cite{hart1996sphere}) can be much more expensive than computing intersections with polygons (like triangles).
More importantly, since implicitly expressed surfaces are known to be difficult to parameterize, Monte Carlo sampling over these surfaces (or curves embedded in them) is also challenging.
This makes it difficult to adopt state-of-the-art physics-based differentiable rendering methods~\cite{li2018differentiable,zhang2019differential,Zhang:2020:PSDR} to handle implicit surfaces.

In this paper, we introduce a new physics-based inverse rendering technique that uses both implicit and explicit representations and, thus, enjoys the benefits of both worlds.
Taking as input a collection of (geometrically calibrated) input images, our technique is capable of generating high-quality 3D reconstructions of object shape and appearance without the need of good initial geometries.
The output of our technique is in ``standard'' format (i.e., a mesh with spatially varying BRDF maps) and enjoys easy editability and wide applicability.

Concretely, our contributions include:
\begin{itemize}
	\item A physics-based differentiable rendering method for implicit surfaces (\S\ref{ssec:implicit}).
	Leveraging differentiable iso-surface extraction~\cite{MeshSDF,guillard2021deepmesh}, we first convert an implicit surface to the corresponding explicit form (i.e., a mesh), and then apply physics-based differentiable rendering to the mesh.
	We demonstrate that, compared to existing differentiable-sphere-tracing based methods (e.g.,  \cite{niemeyer2020differentiable,yariv2020multiview}), our pipeline is not only more general---by supporting more complex light-transport effects such as environmental illumination, soft shadow, and interreflection---but also computationally more efficient.
	\item A two-stage inverse-rendering pipeline (\S\ref{ssec:implicit} and \S\ref{ssec:explicit}) that enables high-fidelity 3D reconstruction of objects with complex topologies.
\end{itemize}
We demonstrate the effectiveness of our technique via several synthetic and real examples.

\section{Related Works}
\label{sec:related}
\paragraph{Shape and reflectance reconstruction}
Reconstructing the shape and/or reflectance of an object has been a central problem in computer vision.
Many techniques like multi-view stereo (MVS)~\cite{seitz1999photorealistic,vogiatzis2005multi,schoenberger2016mvs}, shape from shading (SfS)~\cite{horn1970shape,ikeuchi1981numerical,queau2017variational}, and photometric stereo (PS)~\cite{woodham1980photometric,Holroyd2008,zhou2010ring,queau2016unbiased} have been developed to recover object geometry and reflectance.
Unfortunately, these techniques typically only recovers diffuse albedo for reflectance.
Further, they rely on assumptions about object appearance---such as diffuse-dominated reflectance or being sufficiently textured---and can produce low-quality reconstructions if these assumptions are violated.

Recently, Luan~et~al.~\shortcite{Luan2021} have demonstrated that, using physics-based differentiable rendering~\cite{li2018differentiable,zhang2019differential,Zhang:2020:PSDR}, the reconstruction of object shape and reflectance can be formulated as an \emph{inverse rendering} (aka. \emph{analysis by synthesis}) problem and solved using gradient-based methods.
This technique offers high generality and robustness as it relies on few assumptions about object shape or appearance.
On the other hand, since their method is fully mesh-based, it has difficulties refining object topology and, therefore, requires nontrivial initialization of object geometry (by, for example, using \texttt{COLMAP}~\cite{schoenberger2016mvs}).

In this paper, we address this problem by using both implicit and explicit representations for object geometry.

\paragraph{Differentiable rendering of meshes}
Specialized differentiable renderers have long existed in computer graphics and vision \cite{gkioulekas2013inverse,gkioulekas2016evaluation,azinovic2019inverse,tsai2019beyond,che2020towards}. Recently, several general-purpose ones like \texttt{redner}~\cite{li2018differentiable}, \texttt{Mitsuba 2}~\cite{nimier2019mitsuba}, and \texttt{psdr-cuda} \linebreak \cite{Zhang:2020:PSDR} have been developed.
A key technical challenge in differentiable rendering is to estimate gradients with respect to object geometry (e.g., positions of mesh vertices). To this end, several approximated methods~\cite{liu2019soft,loubet2019reparameterizing,ravi2020accelerating} have been proposed. Unfortunately, inaccuracies introduced by these techniques can lead to degraded result quality, as demonstrated by Luan~et~al.~\shortcite{Luan2021}.
On the contrary, recent techniques specifically handling visibility boundaries~\cite{li2018differentiable,zhang2019differential,Zhang:2020:PSDR} provides unbiased gradient estimates capable of producing higher-quality reconstructions.

\paragraph{Differentiable rendering of implicit surfaces}
Forward rendering of implicit surfaces has been studied long ago, and techniques like sphere tracing~\cite{hart1996sphere} have allowed efficient rendering of implicit geometries.
Differentiable rendering implicit surfaces, on the other hand, used to be largely under-explored.
Several recent works (e.g, \texttt{DVR}~\cite{niemeyer2020differentiable} and \texttt{IDR}~\cite{yariv2020multiview}) have derived differentiable sphere tracing---which differentiates the ray-implicit-surface intersection computations---and introduced inverse-rendering solutions based on this technique.
Unfortunately, physics-based rendering is beyond ray-surface intersection: To support effects like full-pixel anti-aliasing, area and environmental illumination, soft shadows and interreflection, contributions of rays need to be integrated (over a pixel or the hemisphere).
Differentiating these integrals for implicit surfaces 
\rev{is challenging.
	In a concurrent work, Vicini~et~al.~\cite{Vicini2022sdf} have introduced a new technique for differentiable rendering SDFs.
	We consider comparisons between this approach and our technique---and potential combinations of the two---interesting future topics.
}

We overcome this challenge by using meshes as an intermediate representation, which can then be rendered using state-of-the-art mesh-based methods~\cite{li2018differentiable,zhang2019differential,Zhang:2020:PSDR}.
\rev{%
	To our knowledge, although the high-level idea of using explicit and implicit geometries has been experimented before in different contexts (e.g., generative 3D modeling~\cite{poursaeed2020coupling}), our technique is one of the first (among  concurrent works like \cite{munkberg2022extracting}) to realize this idea for physics-based inverse rendering.
}

\paragraph{Differentiable mesh generation}
Recently, several techniques have been developed to generate meshes in a differentiable fashion.
Remelli~et~al.~\shortcite{MeshSDF}, for instance, have introduced \texttt{MeshSDF}---a differentiable marching cube algorithm that converts signed-distance functions (SDFs) into meshes.
This technique was later extended to handle general implicit representations~\cite{guillard2021deepmesh}.
Additionally, Peng~et~al.~\shortcite{Peng2021SAP} have presented a technique that creates meshes from point clouds.
We utilize \texttt{MeshSDF} in our differentiable rendering pipeline.

\begin{figure*}[t!]
	\centering
	\small
	\includegraphics[height=1.2in]{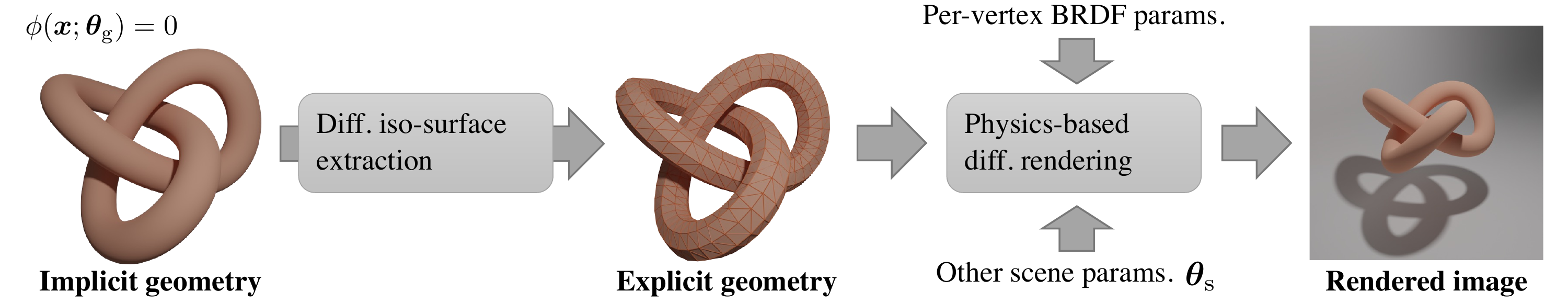}
	\caption{\label{fig:pipeline_implicit}
		Our pipeline for \textbf{physics-based differentiable rendering of implicit surfaces}:
		Instead of directly rendering implicit geometry using techniques like differentiable sphere tracing, we leverage differentiable iso-surface extraction to covert implicit geometries into explicit ones (i.e., meshes) and render the latter using state-of-the-art differentiable rendering techniques capable of handling complex light-transport effects like environmental/area lighting, soft shadows, and interreflection.
	}
\end{figure*}

\section{Preliminaries on Differentiable Rendering}
\label{sec:prelim}
In what follows, we briefly revisit some key aspects in physics-based differentiable rendering.
For more comprehensive introductions, please refer to recent surveys and course materials (e.g, \cite{kato2020differentiable,zhao2020physics}).

\paragraph{Physics-based differentiable rendering}
At the core of physics-based (forward) rendering is the \emph{rendering equation}~\cite{kajiya1986rendering}.
This integral equation states that, on a non-emissive surface $\calM$, the radiance $L$ leaving some point $\bx \in \calM$ with direction $\bomo \in \sph$ is given by:
\begin{equation}
	\label{eqn:RE}
	L(\bx, \bomo) = \int_{\sph} \Li(\bx, \bomi) \,\brdf(\bx, \bomi, \bomo) \,\D\sigma(\bomi),
\end{equation}
where $\Li$ denotes incident radiance; $\brdf$ is the cosine-weighted BRDF (given by the original BRDF multiplied with the dot product of the incident direction $\bomi$ and the surface normal at $\bx$); and $\sigma$ is the solid-angle measure.

Recent works in physics-based differentiable rendering~\cite{li2018differentiable,zhang2019differential,Zhang:2020:PSDR} have shown that, in general, differentiating the rendering equation~\eqref{eqn:RE} with respect to some scene parameter $\param$ yields the \emph{differential rendering equation}:
\begin{multline}
	\label{eqn:dRE}
	\frac{\D}{\D\param} L(\bx, \bomo) = \int_{\sph} \frac{\D}{\D\param} \left[\Li(\bx, \bomi) \,\brdf(\bx, \bomi, \bomo) \right] \D\sigma(\bomi) \\
	+ \int_{\partial\sph} \vel(\bomi) \,\Delta\Li(\bx, \bomi) \,\brdf(\bx, \bomi, \bomo) \,\D\ell(\bomi),
\end{multline}
where the right-hand side consists of two terms.
The first \emph{interior} integral is simply provided by differentiating the integrand of the rendering equation~\eqref{eqn:RE}.
The second \emph{boundary} integral, on the other hand, is unique to differentiable rendering:
The domain $\partial\sph$ is comprised of all jump discontinuity points of $\Li$ with respect to $\bomi$ (assuming the BRDF $\brdf$ to be continuous); $\vel(\bomi)$ captures the change rate (with respect to $\param$) of the discontinuity boundary at $\bomi$; and $\Delta\Li$ denotes the difference in $\Li$ across the boundary.
We note that the discontinuity of the incident radiance $\Li$ can arise from visibility boundaries---which always exist regardless of how smooth the surface geometries are.

In practice, when the parameter $\param$ controls the scene geometry (such as the pose of an object or the position of a mesh vertex), the boundary integral in Eq.~\eqref{eqn:dRE} should not be neglected as it can be the major contributor to the resulting gradient $\nicefrac{\D L}{\D\param}$.

\paragraph{Geometric representations}
Physics-based forward rendering methods have largely been agnostic to geometric representations: Most algorithms (such as path tracing) only requires ray-surface intersection to be efficiently computed---which can be achieved using acceleration data structures (e.g., Kd-trees) for meshes and sphere tracing~\cite{hart1996sphere} for implicit surfaces.

Physics-based differentiable rendering techniques, on the contrary, are more dependent on geometric representations due to their need for estimating the boundary integral in Eq.~\eqref{eqn:dRE}.
When the scene geometries are expressed using meshes, the discontinuities curves $\partial\sph$ are simply (projected) face edges.
When the scene geometries are implicit, on the other hand, so are the discontinuities curves.
Although there are prior works capable of generating (approximately) uniform point samples on implicit curves/surfaces (e.g., \cite{Witkin1994}), they are generally unsuitable for estimating integrals since the probability distributions remain unknown.
To overcome this obstacle, we introduce a new pipeline in \S\ref{ssec:implicit} that combines implicit and explicit representations.

\section{Our Method}
\label{sec:ours}
We now present our technique that combines implicit and explicit geometries for physics-based inverse rendering.

Our optimization pipeline involves the following two main stages.
The first \emph{implicit} stage utilizes implicit representations like signed-distance functions (SDFs) for the object geometry.
The main objective of this stage is to obtain a coarse reconstruction of the object's 3D shape with the right \emph{topology}.
Since implicit geometries are used in this stage, the optimization enjoys the flexibility of allowing topology changes.

Our second \emph{explicit} stage, on the contrary, is purely mesh-based.
Specifically, this stage starts with taking the implicit geometry from the previous stage and applying (non-differentiable) iso-surface extraction to obtain the corresponding explicit expression.
Then, we jointly optimize the object shape (by changing mesh vertex positions) and appearance (by changing pixel values in SVBRDF maps).

In what follows, we detail each of the two stages.

\subsection{Implicit Stage}
\label{ssec:implicit}
Our implicit stage computes coarse reconstructions of an object's shape and appearance.
In this stage, the object surface $\calMi \subset \Real^3$ is represented \emph{implicitly} as the zero level set of some function $\sdf$ that is, in turn, controlled by some abstract set of \emph{geometric parameters}~$\thetaG$.
Namely,
\begin{equation}
    \calMi(\thetaG) := \left\{ \bx \in \Real^3 \;:\; \sdf(\bx;\,\thetaG) = 0 \right\}.
\end{equation}

Additionally, to describe the appearance of the object, we utilize the microfacet BRDF model.
For any surface point $\bx \in \calMi(\thetaG)$, we specify the diffuse albedo $\kd$, specular albedo $\ks$, and surface roughness $\rough$ at this point using a \emph{reflectance field}~$\nrf$ controlled by an abstract set of \emph{optical parameters}~$\thetaA$:
\begin{equation}
    \nrf(\bx; \thetaA) = \begin{pmatrix}
        \kd & \ks & \rough
    \end{pmatrix}.
\end{equation}
%

\paragraph{Differentiable rendering}
To render the surface represented with $\calM$ and $\nrf$, as opposed to prior works that directly compute differentiable ray intersection against the SDF $\sdf$, we instead (i)~apply differentiable iso-surface extraction (e.g., MeshSDF~\cite{MeshSDF}) to obtain a triangle mesh $\calMe \approx \calMi$, and (ii)~use physics-based differentiable rendering to render this mesh.
As illustrated in Figure~\ref{fig:pipeline_implicit}, this process effectively implements a \emph{render function}~$\render(\thetaG, \thetaA, \thetaS)$ that takes as input the geometric parameters $\thetaG$ and optical ones $\thetaA$---among other scene parameters~$\thetaS$ such as illumination and viewing conditions---and returns an image.

Note that, since the conversion from the implicit representation $\sdf$ to the mesh $\calMe$ is differentiable, the gradients of mesh vertex positions with respect to the geometric parameters $\thetaG$ are automatically available during differentiable rendering.
This makes the entire rendering function~$\render$ end-to-end differentiable (with respect to $\thetaG$).

\paragraph{Inverse-rendering optimization}
Similar to prior works, we take as input $n$ images $\img_j$ of an object with known illumination and viewing conditions $\thetaS^{j}$, for $j = 1, 2, \ldots, n$.
Then, we reconstruct the shape and appearance of the object by solving the following inverse-rendering optimization:
\begin{equation}
    (\thetaG^*, \thetaA^*) = \argmin_{\thetaG, \thetaA} \loss{}(\thetaG, \thetaA),
\end{equation}
where the loss~$\loss{}$ is further given by
\begin{equation}
    \label{eqn:loss}
    \loss{}(\thetaG, \thetaA) = \loss{img}(\thetaG, \thetaA) + \weight{reg}\,\loss{reg}(\thetaG, \thetaA).
\end{equation}
In Eq.~\eqref{eqn:loss}, $\loss{img}$ denotes the image loss given by
\begin{equation}
	\label{eqn:loss_img}
    \loss{img}(\thetaG, \thetaA) = \sum_{j = 1}^n \left\| \render\left(\thetaG, \thetaA, \thetaS^{(i)}\right) - \img_j \right\|_1,
\end{equation}
and $\loss{reg}$ is a regularization term with a hyper-parameter $\weight{reg} \in \Real_{\geq 0}$ controlling its weight.
Optionally, when each input image $\img_j$ is supplemented with an anti-aliased mask image $\mask_j$, our optimization can get mask supervision by including an extra loss term:
\begin{equation}
    \label{eqn:loss_mask}
    \weight{mask}\,\loss{mask}(\thetaG) = \sum_{j = 1}^n \left\| \render_\mathrm{mask}\left(\thetaG, \thetaS^{(i)}\right) - \mask_j \right\|_1,
\end{equation}
where $\render_\mathrm{mask}$ is similar to the rendering function $\render$ but returns only an anti-aliased mask.
In practice, $\render(\thetaG, \thetaA, \thetaS^{(i)})$ and $\render_\mathrm{mask}(\thetaG, \thetaS^{(i)})$ can be computed simultaneously from one rendering pass. 

\paragraph{Discussion}
By using both implicit and explicit geometries, our approach enjoys the benefit of both representations: It not only allows topological changes during the optimization, but also is capable of capturing light transport effects---such as environmental illumination and soft shadows---that cannot be easily handled by soft rasterizers or differentiable sphere tracers.
We will demonstrate the practical advantage of our technique in \S\ref{sec:results}.

\subsection{Explicit Stage}
\label{ssec:explicit}
Based on the coarse reconstruction provided by our implicit stage (\S\ref{ssec:implicit}), our explicit stage further refines the result using a purely mesh-based approach.
To start, we convert the implicit representation returned by the previous stage into its explicit counterpart.
Specifically, we apply 
iso-surface extraction \rev{(e.g., marching cube)} to the optimized function $\sdf(\cdot; \thetaG^*)$ returned by the previous stage and obtain a mesh $\calMe_0$.
\rev{We note that this extraction step does not need to be differentiable since it is essentially a preprocessing step for the explicit stage.}

\rev{Provided the initial mesh $\calMe_0$,} we then use the boundary first flatting (BFF) method~\cite{BFF} to UV-parameterize the mesh and, in turn, generate the spatially varying BRDF (SVBRDF) maps based on 
the reflectance field~$\nrf(\cdot; \thetaA^*)$ given by the implicit stage.

Using this mesh and the SVBRDF maps for the initial geometry and appearance, respectively, we then apply a second inverse-rendering optimization. 
Our pipeline is based on the mesh-based framework introduced by Luan~et~al.~\shortcite{Luan2021} and jointly optimizes the position of each mesh vertex and the value of each pixel in the SVBRDF maps.
Additionally, we utilize the recent work by Nicolet et al.~\cite{Nicolet2021Large} to update mesh vertex positions---as opposed to using the \texttt{elTopo} geometric processing library~\cite{brochu2009robust} like Luan~et~al. did.
This has led to significantly better performance in practice, since it does not require expensive (continuous) collision detection and handling.

\subsection{Implementation Details}
\label{ssec:impl}
For our implicit stage (\S\ref{ssec:implicit}), we adopt the neural networks from the \texttt{idr} work by Yariv et al.~\shortcite{yariv2020multiview}.
Specifically, we use their occupancy network for our function~$\phi$ and adopt their rendering network for our reflectance field~$\nrf$.
In this case, the geometric parameters $\thetaG$ and optical ones $\thetaA$ become weights of edges in the two neural networks, respectively.

Further, we set the regularization loss as
\begin{equation}
	\loss{reg}(\thetaG, \thetaA) = \sum_i (\| \nabla_{\bx}\,\sdf(\bx_i; \thetaG) \| - 1)^2,
\end{equation}
that encourages $\sdf$ to be a signed-distance function (i.e., with unit-length gradients).

We utilize the \texttt{MeshSDF} technique~\cite{MeshSDF} for differentiable iso-surface extraction and cache the BRDF parameter values (i.e., diffuse albedo~$\kd$, specular albedo~$\ks$, and surface roughness~$\rough$) at each mesh vertex (as opposed to evaluate the neural reflectance field~$\nrf$ per pixel on the fly) for better rendering performance.
Additionally, during optimization, we reuse one mesh produced by \texttt{MeshSDF} to render multiple camera views (from the same minibatch), further reducing the amortized computational overhead.

Lastly, we build our Monte Carlo differentiable renderer---which is used in both stages---based on Zhang~et~al.'s path-space technique~\cite{Zhang:2020:PSDR} and use the Adam method~\cite{kingma2014adam} implemented in \texttt{pyTorch} for our inverse-rendering optimizations in both stages---except for updating mesh vertex positions in the explicit stage where we use Nicolet~et~al.'s approach~\shortcite{Nicolet2021Large}.

\section{Results}
\label{sec:results}
In what follows, we validate in \S\ref{ssec:res_eval} our differentiable rendering of implicit surfaces.
Then, we demonstrate the effectiveness of our inverse-rendering technique in \S\ref{ssec:res_inverse_render}.
Please refer to the supplement for more results.

\setlength{\resLen}{1.05in}
\begin{figure}[t]
	\centering
	\addtolength{\tabcolsep}{-4pt}
	\small
	\begin{tabular}{ccc}
		\multicolumn{3}{c}{\scriptsize Negative \includegraphics[height=0.08in]{images/cbar.jpg} Positive}\\
		\includegraphics[height=\resLen]{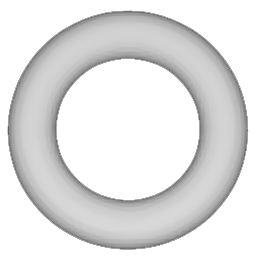} &
		\includegraphics[height=\resLen]{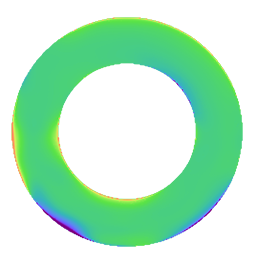} &
		\includegraphics[height=\resLen]{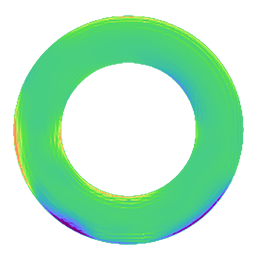}\\[-10pt]
		\includegraphics[height=\resLen]{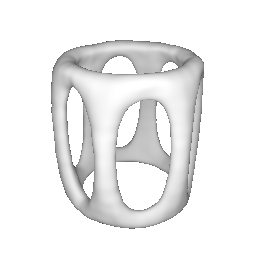} &
		\includegraphics[height=\resLen]{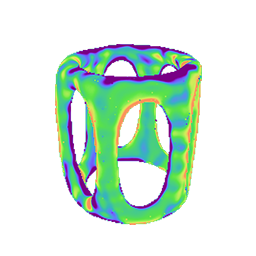} &
		\includegraphics[height=\resLen]{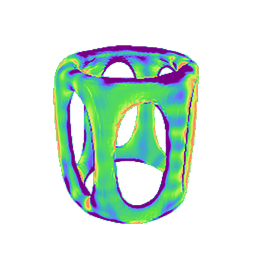}\\[-5pt]
		(a) Ordinary & (b) Finite differences & (c) Ours
	\end{tabular}
	\caption{
		\label{fig:res_validate}
		We \textbf{validate} the correctness of gradients computed by our method (\S\ref{ssec:implicit}) by comparing to finite-difference references.
		Each derivative image is computed with respect to the weight of one edge in the occupancy network (which specifies the implicit surface).
	}
\end{figure}

\subsection{Validation}
\label{ssec:res_eval}
We validate our differentiable rendering of implicit surfaces (described in \S\ref{ssec:implicit}) by comparing gradient images (with respect to the weight of one edge in the occupancy network) generated with our method and finite differences (FD), as shown in Figure~\ref{fig:res_validate}.
In this figure, each derivative image is computed with respect to the weight of one edge in the occupancy neural network representing $\sdf$.
When generating the ordinary and finite-difference (FD) results, we use sphere tracing to compute ray-surface intersection directly (without converting the implicit representations to meshes).
The results demonstrate that gradients estimated by our method closely match the FD references.
The 
\rev{stair-case-like artifacts in our results are caused by the use of marching cube with relatively low grid resolutions.}
Fortunately, as we will demonstrate in the rest of this section, these approximations do not affect the performance of our inverse-rendering solution.

\setlength{\resLen}{1.2in}

\begin{figure*}[t]
	\centering
	\addtolength{\tabcolsep}{-7pt}
	\small
	\rev{\begin{tabular}{ccccccc}
		& (a) \textbf{Ground truth} & (b) \textbf{Initial} & (c) \textbf{IDR*} & (d) \textbf{Mesh-based} & (e1) \textbf{Ours (impl.)} & (e2) \textbf{Ours (full)}
		\\[-2pt]
		\raisebox{40pt}{\rotatebox[origin=c]{90}{\bfseries Bunny temple}} &
		\includegraphics[width=\resLen]{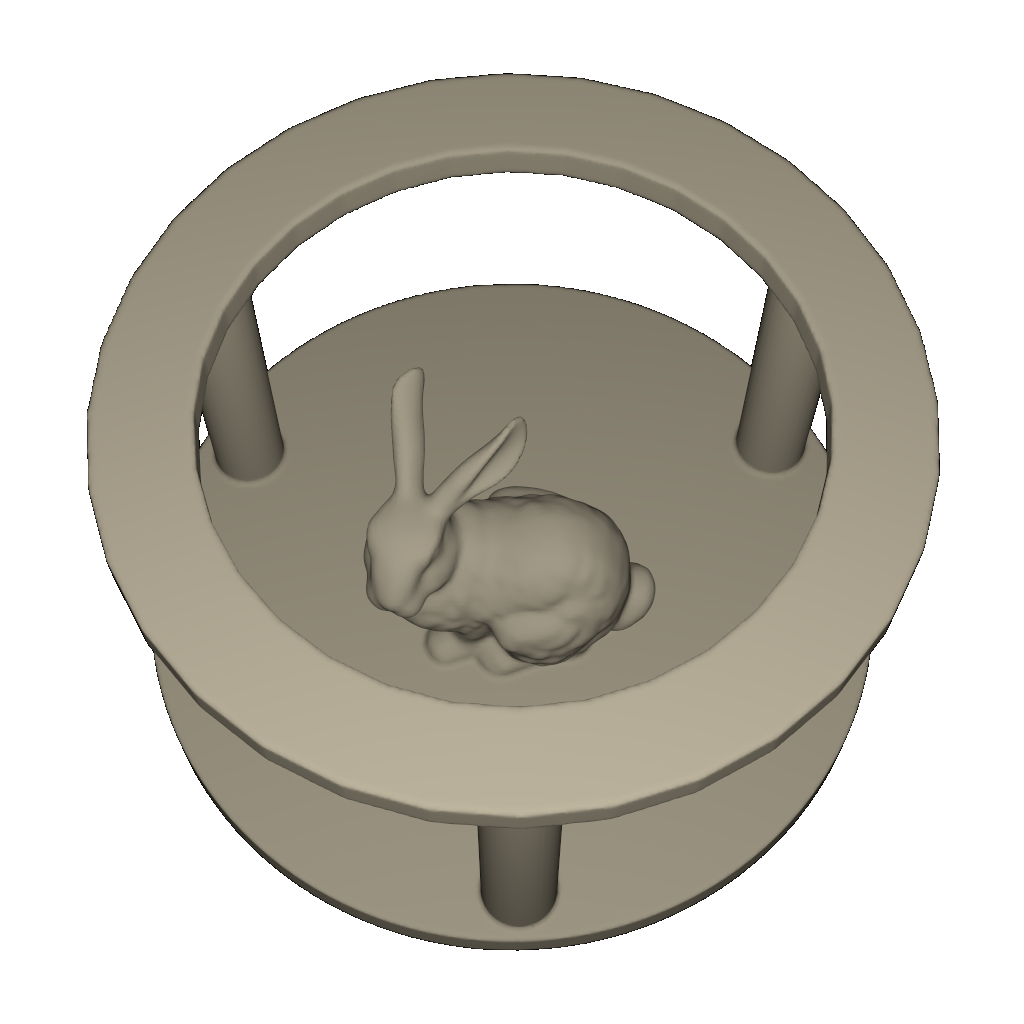} &
		\includegraphics[width=\resLen]{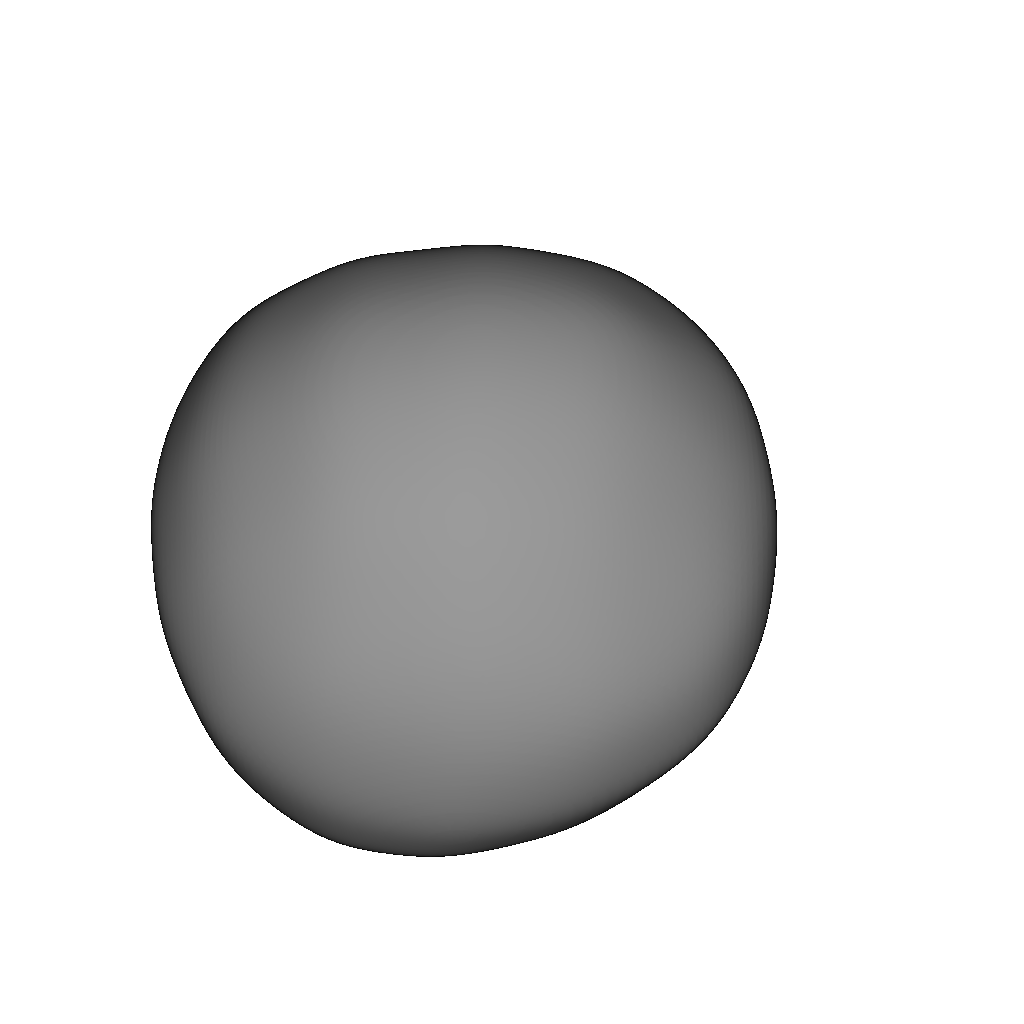} &
		\includegraphics[width=\resLen]{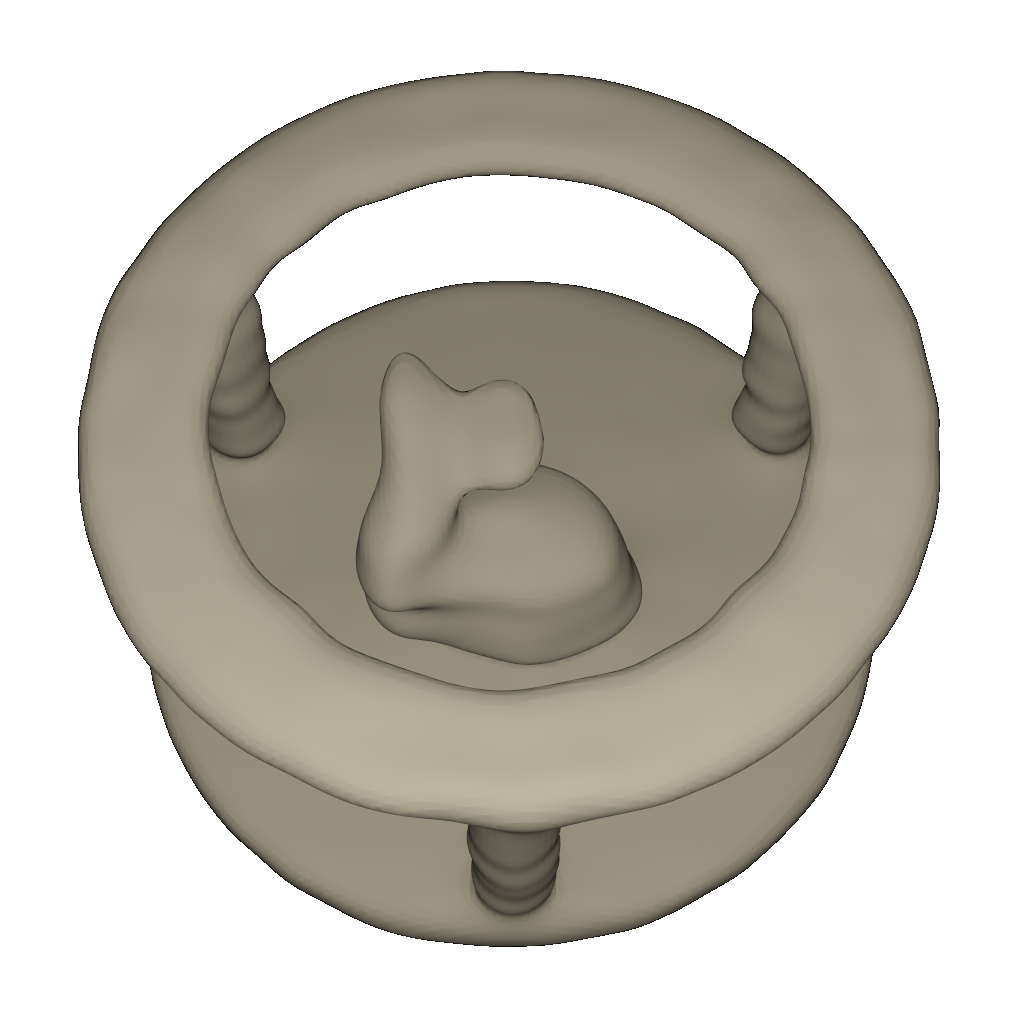} &
		\includegraphics[width=\resLen]{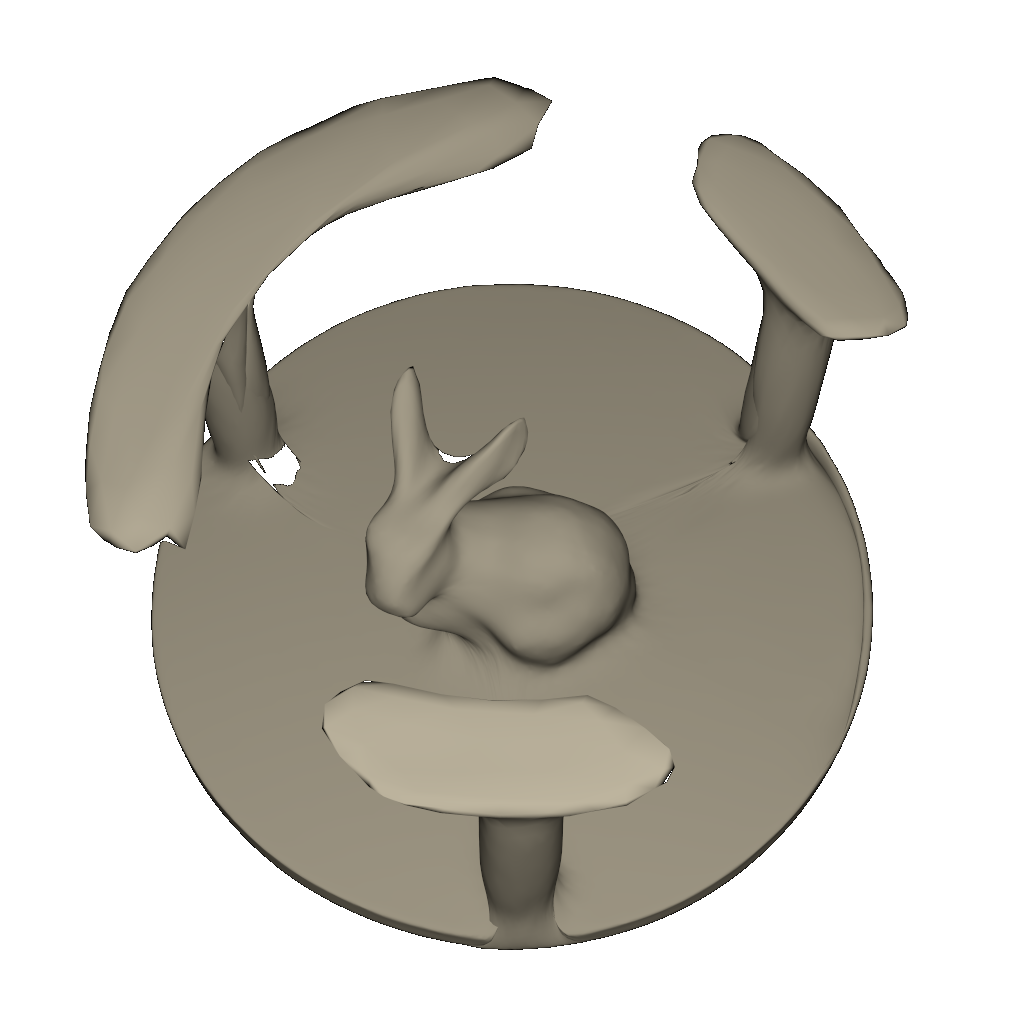} &
		\includegraphics[width=\resLen]{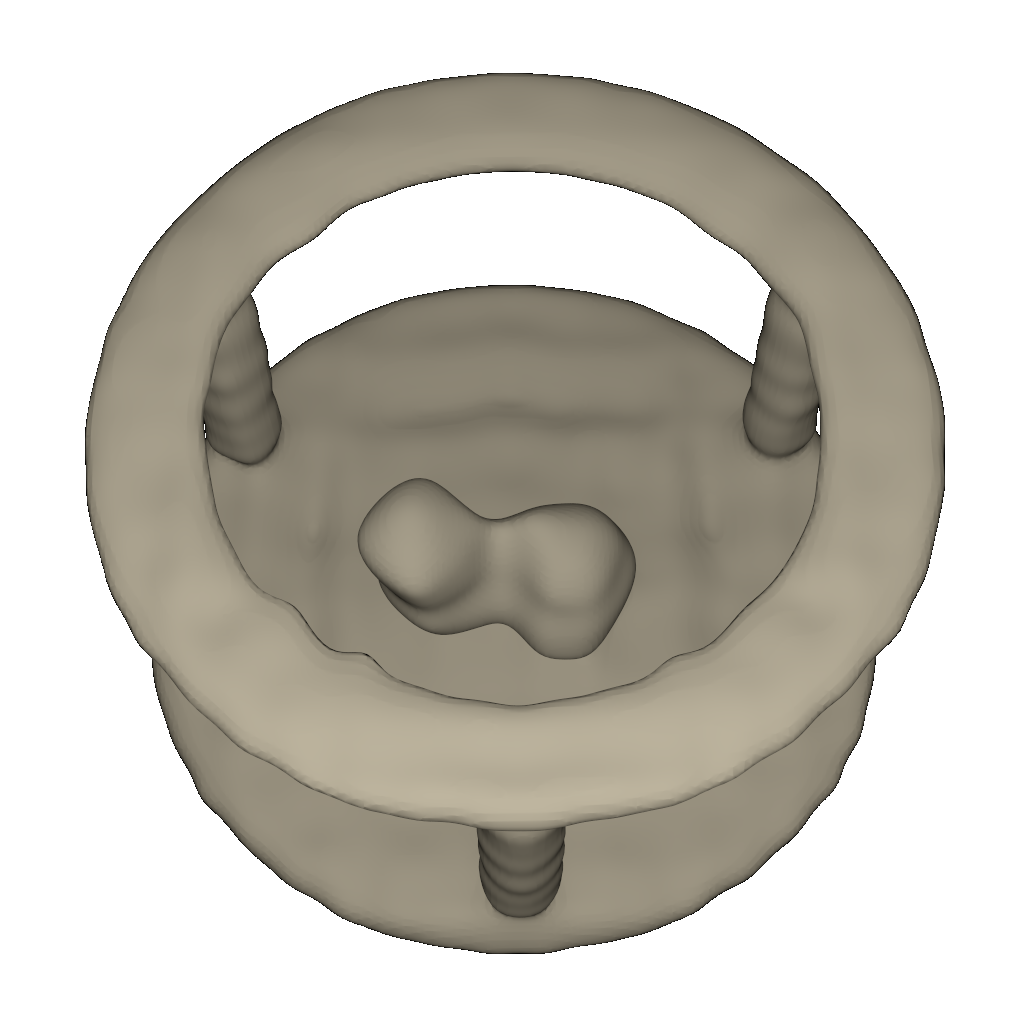} &
		\includegraphics[width=\resLen]{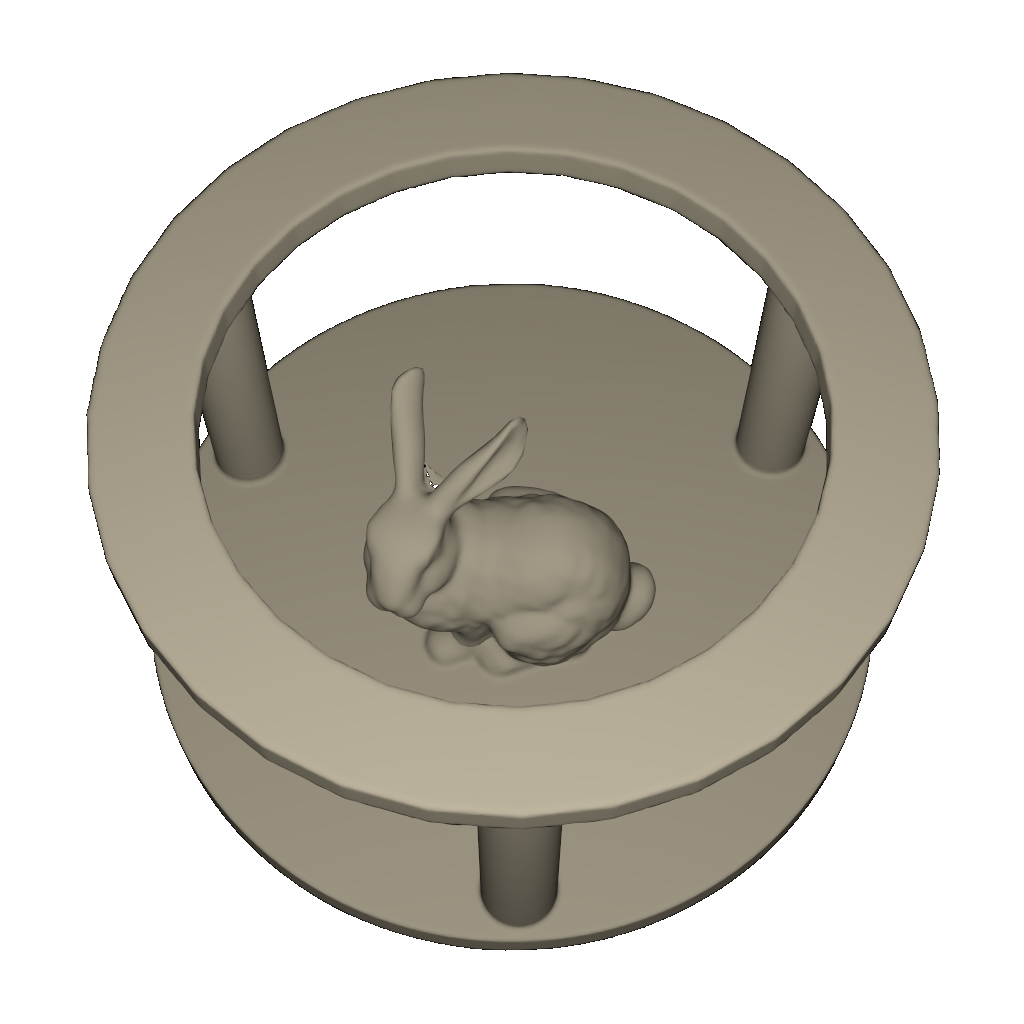}
		\\[-4pt]
		& \textbf{Chamfer dist.:} -- & 0.3586 & 0.0191 & 0.0210 & 0.0125 & \textbf{0.0002}\\
		& \textbf{Hausdorff dist.:} -- & 0.4394 & 0.1191 & 0.4812 & 0.2103 & \textbf{0.0531}\\
		& \textbf{Genus:} 3 & 0 & 3 & 0 & 3 & 3
		\\
		\raisebox{40pt}{\rotatebox[origin=c]{90}{\bfseries Chair}} &
		\adjincludegraphics[width=\resLen,trim={{.06\height} {.06\height} {.06\height} {.06\height}},clip]{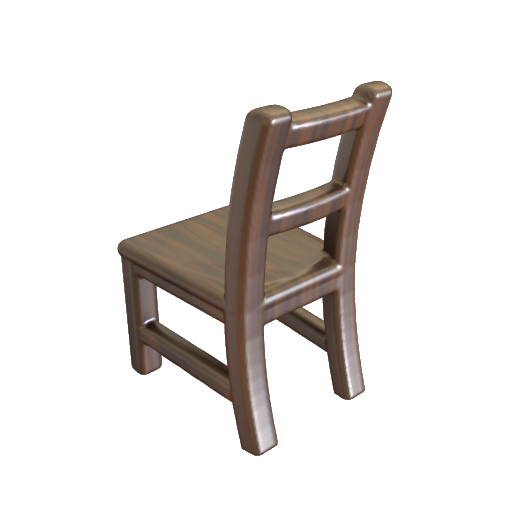} &
		\adjincludegraphics[width=\resLen,trim={{.06\height} {.06\height} {.06\height} {.06\height}},clip]{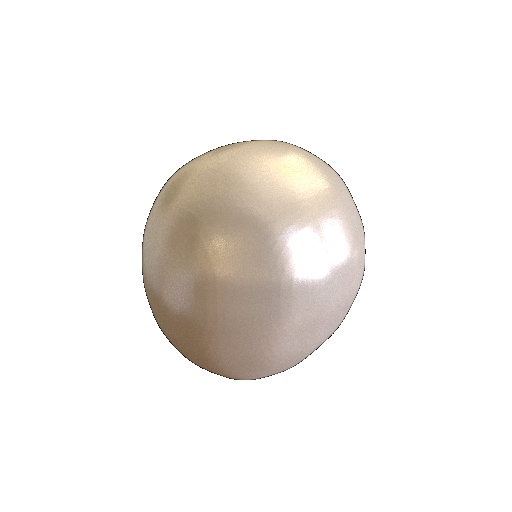} &
		\raisebox{.45\resLen}{--} &
		\adjincludegraphics[width=\resLen,trim={{.06\height} {.06\height} {.06\height} {.06\height}},clip]{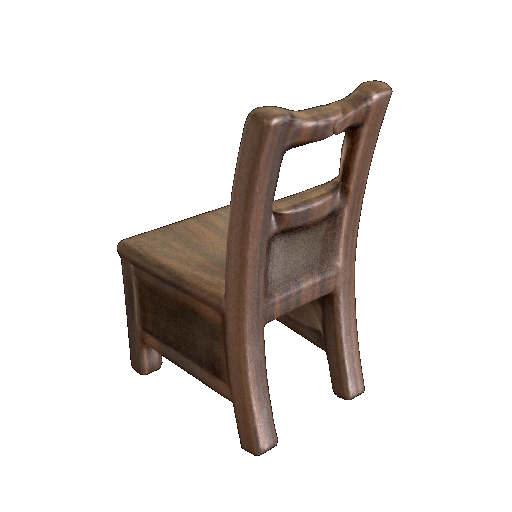} &
		\adjincludegraphics[width=\resLen,trim={{.06\height} {.06\height} {.06\height} {.06\height}},clip]{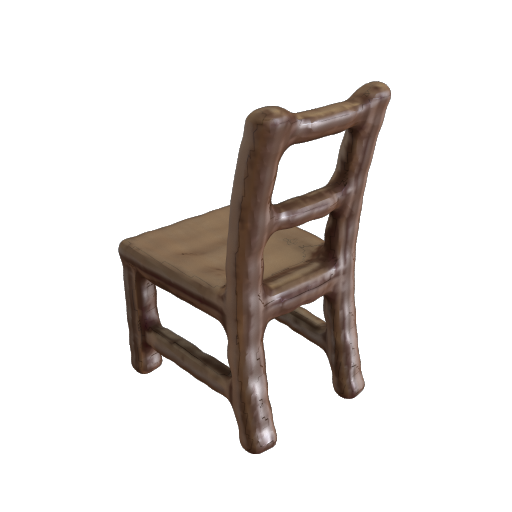} &
		\adjincludegraphics[width=\resLen,trim={{.06\height} {.06\height} {.06\height} {.06\height}},clip]{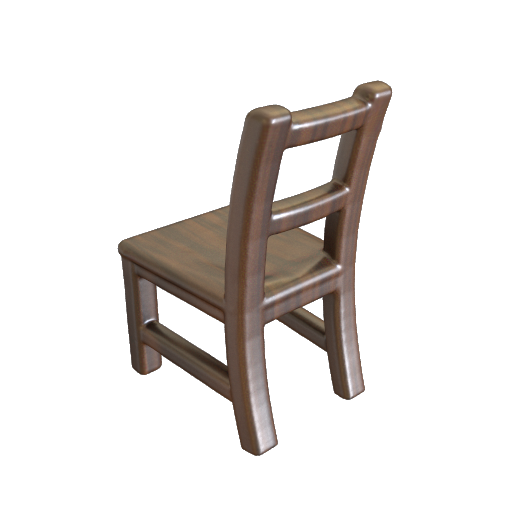}
		\\[-5pt]
		& \textbf{Chamfer dist.:} -- & 0.2561 & -- & 0.0173 & 0.0090 & \textbf{0.0054} \\
		& \textbf{Hausdorff dist.:} -- & 0.4491 & -- & 0.1330 & 0.0558 & \textbf{0.0551} \\
		& \textbf{Genus:} 4 & 0 & -- & 0 & 4 & 4 \\
	\end{tabular}}
	\caption{\label{fig:inv_comp1}
		\textbf{Inverse-rendering comparisons:}
		We show reconstruction results generated using IDR*---a modified version of IDR that uses physics-based shading---in (c), Luan~et~al.'s mesh-based method~\cite{Luan2021} in (d), our implicit stage in (e1), and our full pipeline in (e2).
		All methods shared identical initializations shown in (b).
		The \rev{numbers} below each reconstruction result \rev{indicate} the Chamfer~\cite{barrow1977parametric} \rev{and Hausdorff~\cite{aspert2002mesh} distances} between the reconstructed and groundtruth geometries (normalized so that the GT has a unit bounding box).
		\rev{Additionally, we show genus numbers of all results to validate their topologies.}
	}
\end{figure*}

\setlength{\resLen}{1.16in}

\newcommand{\invRenderSync}[9]{%
	\raisebox{#2}{\rotatebox{90}{\bfseries #1}} &
	\includegraphics[height=\resLen]{images/#3/scene1.png} &
	\includegraphics[height=\resLen]{images/#3/scene2.png} &
	\includegraphics[height=\resLen]{images/#3/implicit1.png} &
	\includegraphics[height=\resLen]{images/#3/implicit2.png} &
	\includegraphics[height=\resLen]{images/#3/explicit1.png} &
	\includegraphics[height=\resLen]{images/#3/explicit2.png}
	\\[#4]
	&
	\multirow{2}{*}{\includegraphics[height=.25\resLen]{images/#3/env1.png}} &
	\multirow{2}{*}{\includegraphics[height=.25\resLen]{images/#3/env2.png}} &
	\multicolumn{2}{c|}{#5} & \multicolumn{2}{c}{#6}
	\\
	& & &
	\multicolumn{2}{c|}{#7} & \multicolumn{2}{c}{#8}
	\\[#9]
}

\begin{figure*}[t]
	\centering
	\addtolength{\tabcolsep}{-6.5pt}
	\small
	\rev{\begin{tabular}{ccc|cc|cc}
		& \multicolumn{2}{c}{(a) \textbf{Ground truth}}
		& \multicolumn{2}{c}{(b) \textbf{Ours (implicit)}}
		& \multicolumn{2}{c}{(c) \textbf{Ours (full)}}
		\\[-9pt]
		\invRenderSync{Frame}{20pt}{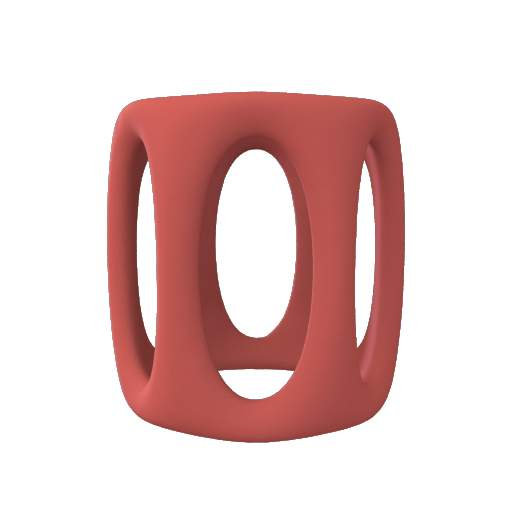}{-8pt}{Chamfer distance: 0.0068}{Chamfer distance: 0.0008}{Hausdorff distance: 0.0222}{Hausdorff distance: 0.0138}{5pt}
		\invRenderSync{Pegasus}{20pt}{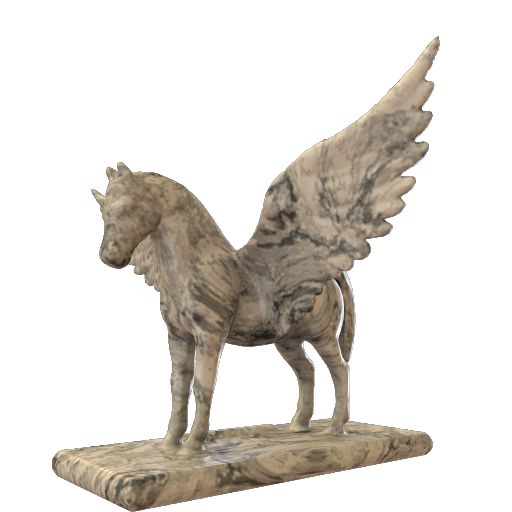}{2pt}{Chamfer distance: 0.0051}{Chamfer distance: 0.0009}{Hausdorff distance: 0.0572}{Hausdorff distance: 0.0403}{0pt}
	\end{tabular}}
	\caption{\label{fig:inv_results1}
		\textbf{Inverse-rendering results (synthetic):}
		Renderings of groundtruth~(a) and reconstructed models generated by our implicit stage~(b) and full pipeline~(c).
		The left image in each column is rendered using the training illumination and a \emph{novel} viewing angel;
		The right image is rendered using \emph{novel} illumination and viewing conditions.
		All optimizations use the same initializations for the occupancy network (that yield sphere-like shapes similar to Figure~\protect\ref{fig:inv_comp1}-b) and neural reflectance fields (that produce near-constant BRDF parameters).
	}
\end{figure*}

\begin{table}[t]
	\small
	\caption{\label{tab:perf}
		Performance statistics for our inverse-rendering results in Figures~\protect\ref{fig:inv_comp1}, \protect\ref{fig:inv_results1}, and \protect\ref{fig:inv_results2}.
		The ``MC'' column shows the grid resolution used for differentiable iso-surface extraction.
		The ``Time'' column contains per-iteration optimization time (in seconds) measured on a workstation with an NVIDIA Titan RTX graphics card.
	}
	\centering
	\addtolength{\tabcolsep}{-2pt}
	\begin{tabular}{l|r|rrr|rr}
		\multirow{2}{*}{\textbf{Example}} &
		\textbf{\# Input} &
		\multicolumn{3}{c|}{\bfseries Implicit stage} &
		\multicolumn{2}{c}{\bfseries Explicit stage}
		\\
		& \textbf{images} & MC & \# Iter. & Time & \# Iter. & Time
		\\
		\hline
		Bunny temple &  96 &  $96^3$ & 1000 &  1.19  & 1000 & 0.73\\
		Chair        & 196 & $128^3$ &  300 &  9.24  &  500 & 2.71\\
		Frame        &  96 & $128^3$ & 3500 & 11.24  &  700 & 1.35\\
		Pegasus      & 396 &  $96^3$ & 2000 &  5.69  &  200 & 3.14\\
		Chess        &  74 & $128^3$ &  400 &  2.49  &  200 & 3.34\\
		Teapot       & 135 & $128^3$ & 1000 &  4.66  &  200 & 3.51\\
		Leopard      & 132 &  $96^3$ & 1500 &  6.32  &  500 & 3.47\\
		Head         & 126 &  $96^3$ &  750 &  5.97  &  350 & 3.38
	\end{tabular}
\end{table}

\subsection{Inverse-Rendering Results}
\label{ssec:res_inverse_render}
We now demonstrate the effectiveness of our technique using a few inverse-rendering results.
Please refer to Table~\ref{tab:perf} for performance numbers \rev{and the supplement for additional error visualizations and recovered SVBRDF maps}.

\paragraph{Synthetic results}
We compare our inverse-rendering pipeline with a few baselines using several synthetic examples in Figure~\ref{fig:inv_comp1}.
The first baseline---which we denote as \texttt{IDR*}---is a modified version of the \texttt{IDR} technique~\cite{yariv2020multiview} where the neural rendering network is replaced with a neural reflectance field (NeRF) that outputs per-pixel BRDF parameters.
Using these parameters with ray-surface intersection results provided by differentiable sphere tracing, we implemented a (differentiable) physics-based shading step that produces the final rendered image.
Our second baseline is a purely mesh-based method that directly applies our explicit stage (\S\ref{ssec:explicit}).
We set the initial geometry for \texttt{IDR*} and our implicit stage using network weights from \texttt{IDR} that lead to sphere-like shapes (see Figure~\ref{fig:inv_comp1}-b).
For the mesh-based baseline, we obtain the initial mesh by applying (non-differentiable) iso-surface extraction to the initial implicit geometry.
As the \texttt{IDR*} baseline supports only one-bounce light transport, we configure all methods to render direct illumination only.

\setlength{\resLen}{1.18in}

\begin{figure*}[t]
	\centering
	\addtolength{\tabcolsep}{-7.5pt}
	\small
	\begin{tabular}{ccc|cc|cc}
		& \multicolumn{2}{c|}{(a) \textbf{Ground truth}} & \multicolumn{2}{c|}{(b) \textbf{Ours (implicit)}} & \multicolumn{2}{c}{(c) \textbf{Ours (full)}}
		\\
		\raisebox{50pt}{\rotatebox[origin=c]{90}{\bfseries Chess}} &
		\adjincludegraphics[width=.9\resLen,trim={{.1\width} 0 {.1\width} {.35\height}},clip]{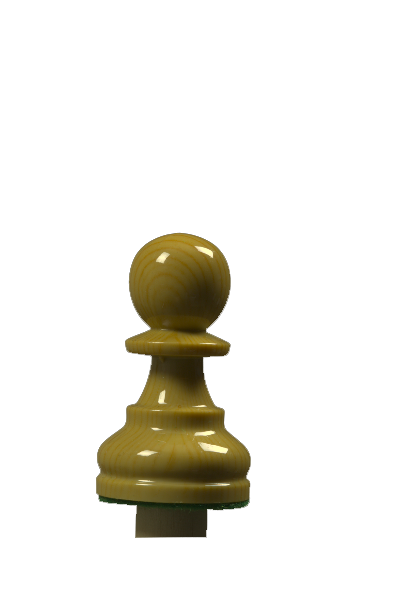} &
		\adjincludegraphics[width=.9\resLen,trim={{.05\width} 0 {.05\width} {.25\height}},clip]{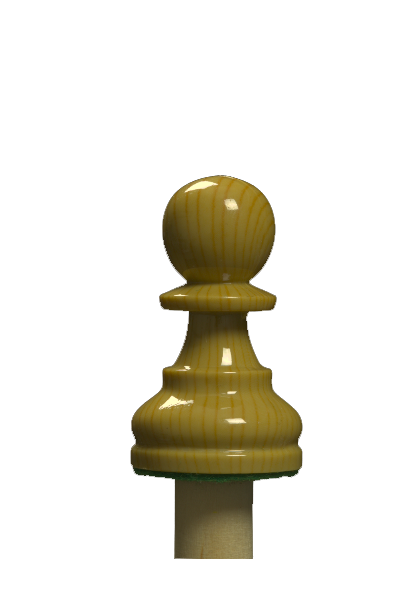} &
		\adjincludegraphics[width=.9\resLen,trim={{.1\width} 0 {.1\width} {.35\height}},clip]{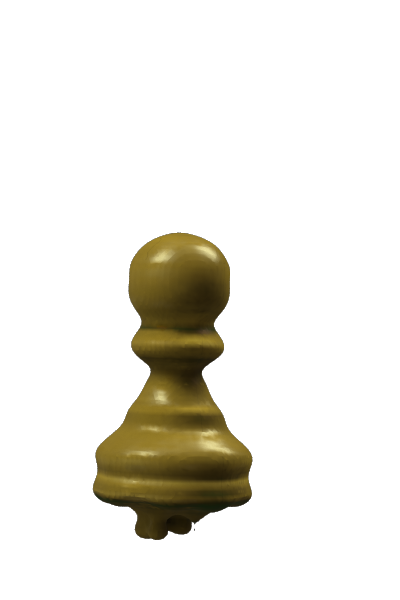} &
		\adjincludegraphics[width=.9\resLen,trim={{.05\width} 0 {.05\width} {.25\height}},clip]{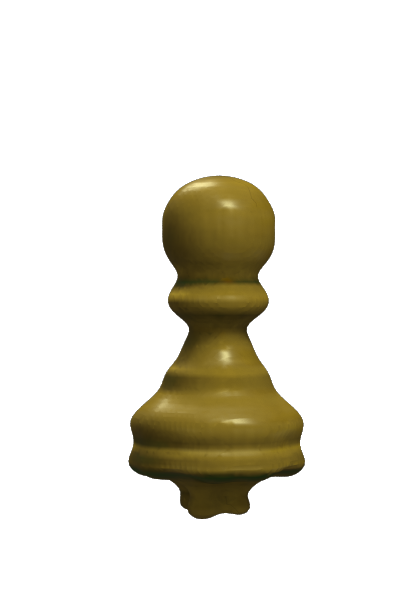} &
		\adjincludegraphics[width=.9\resLen,trim={{.1\width} 0 {.1\width} {.35\height}},clip]{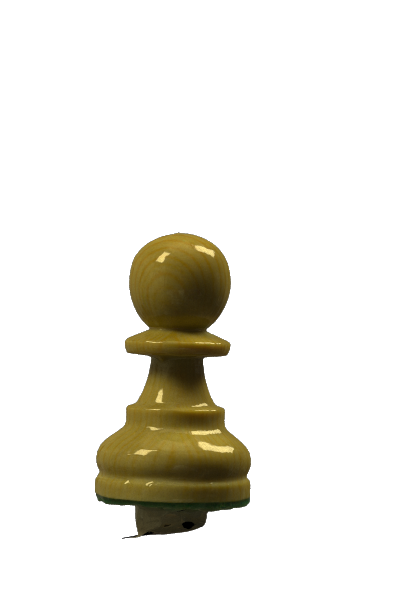} &
		\adjincludegraphics[width=.9\resLen,trim={{.05\width} 0 {.05\width} {.25\height}},clip]{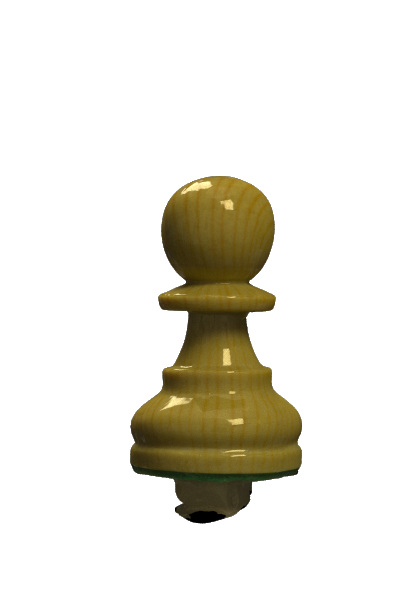}
		\\[-10pt]
		\raisebox{20pt}{\rotatebox[origin=c]{90}{\bfseries Teapot}} &
		\adjincludegraphics[width=\resLen,trim={{0.1\width} {.1\height} {.1\width} {.2\height}},clip]{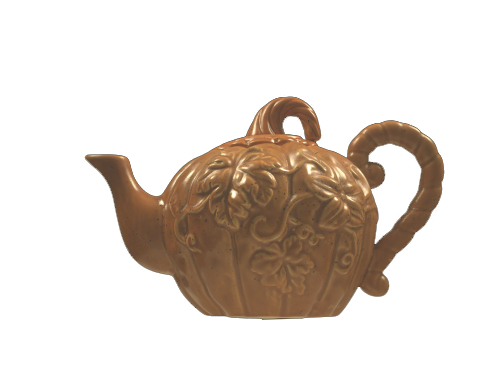} &
		\adjincludegraphics[width=\resLen,trim={0 0 0 {.2\height}},clip]{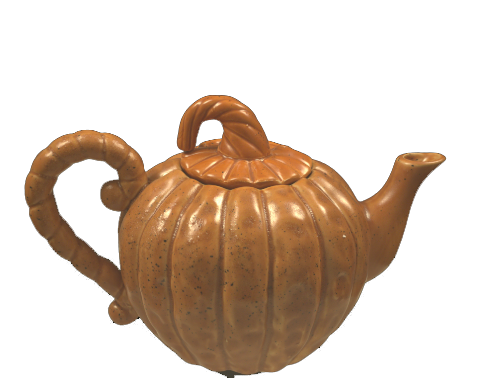} &
		\adjincludegraphics[width=\resLen,trim={{0.1\width} {.1\height} {.1\width} {.2\height}},clip]{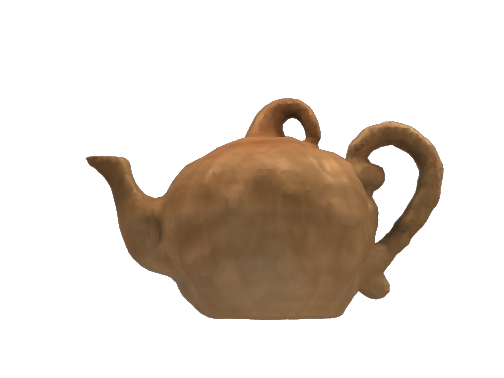} &
		\adjincludegraphics[width=\resLen,trim={0 0 0 {.2\height}},clip]{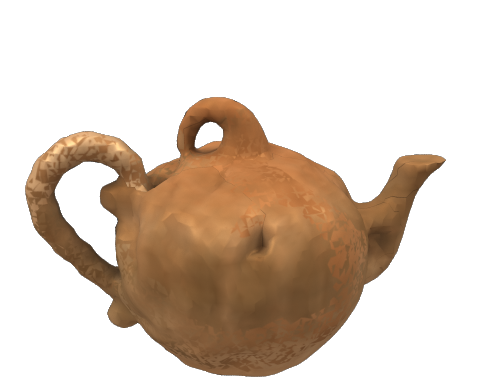} &
		\adjincludegraphics[width=\resLen,trim={{0.1\width} {.1\height} {.1\width} {.2\height}},clip]{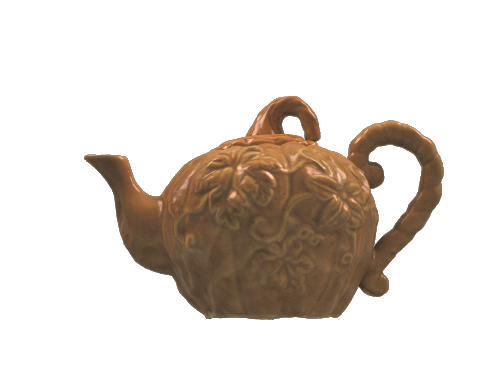} &
		\adjincludegraphics[width=\resLen,trim={0 0 0 {.2\height}},clip]{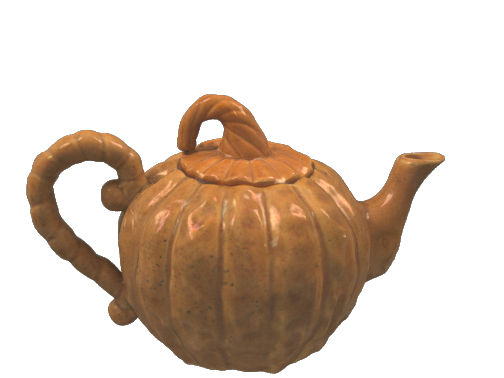}
		\\
		\raisebox{25pt}{\rotatebox[origin=c]{90}{\bfseries Leopard}} &
		\includegraphics[width=\resLen]{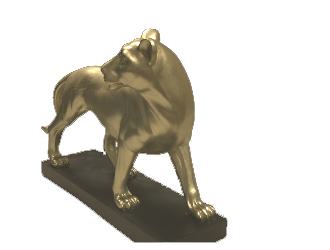} &
		\includegraphics[width=\resLen]{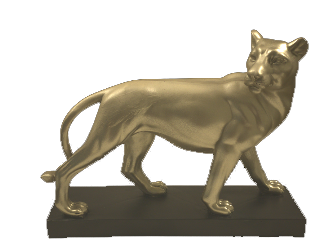} &
		\includegraphics[width=\resLen]{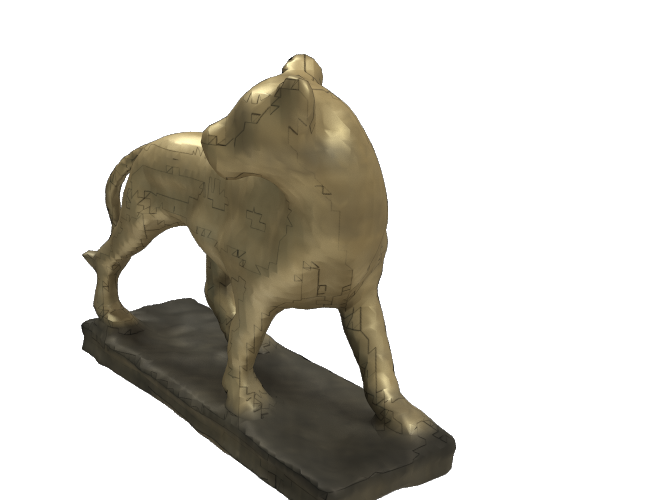} &
		\includegraphics[width=\resLen]{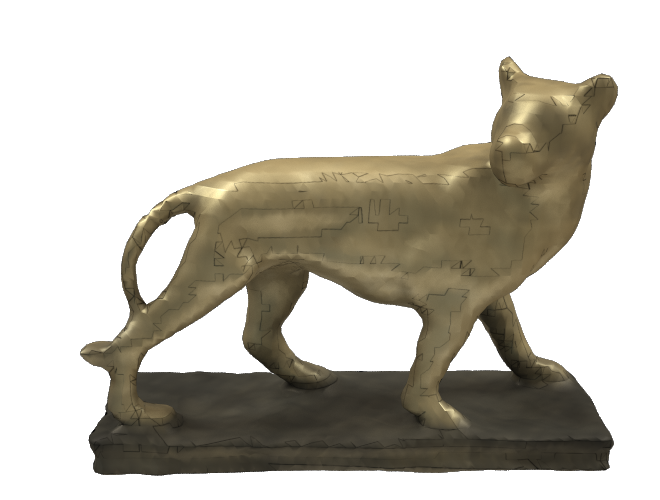} &
		\includegraphics[width=\resLen]{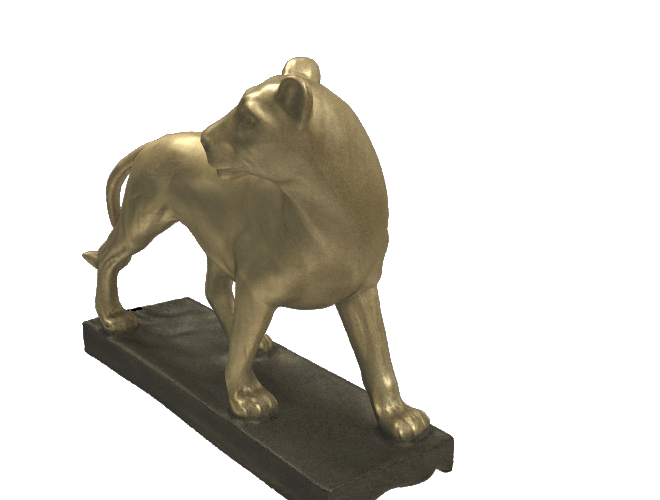} &
		\includegraphics[width=\resLen]{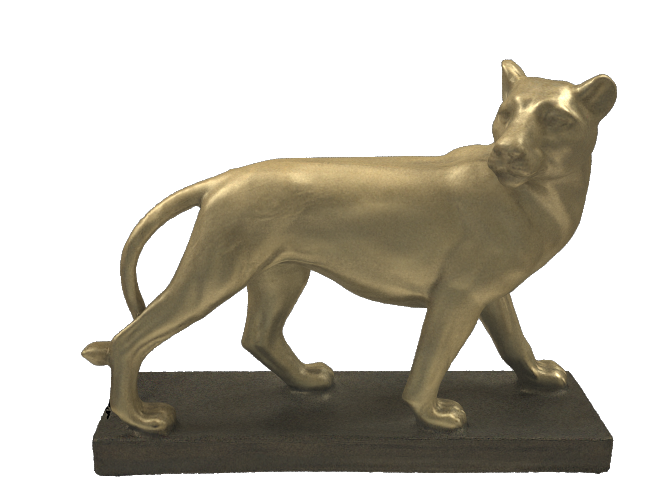}
		\\[-14pt]
		\raisebox{45pt}{\rotatebox[origin=c]{90}{\bfseries Head}} &
		\adjincludegraphics[width=\resLen,trim={{0.1\width} 0 {.1\width} {.2\height}},clip]{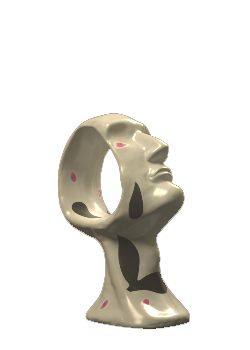} &
		\includegraphics[width=\resLen]{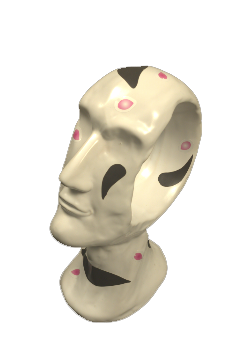} &
		\adjincludegraphics[width=\resLen,trim={{0.1\width} 0 {.1\width} {.2\height}},clip]{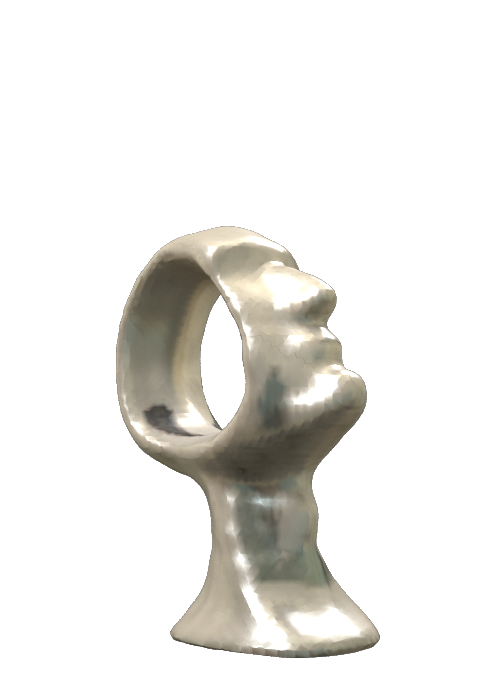} &
		\includegraphics[width=\resLen]{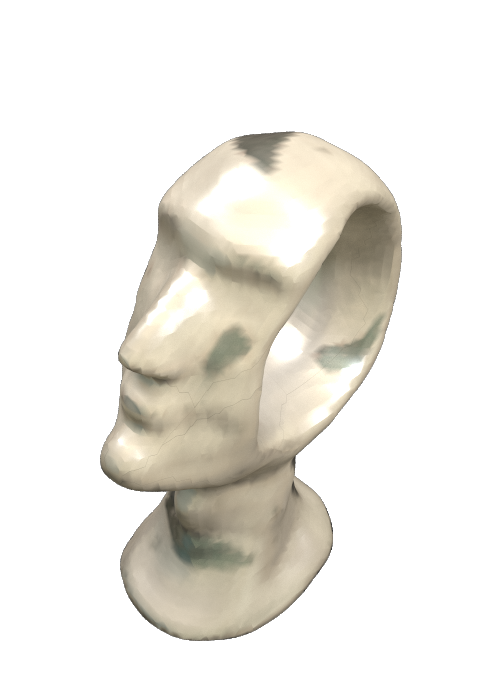} &
		\adjincludegraphics[width=\resLen,trim={{0.1\width} 0 {.1\width} {.2\height}},clip]{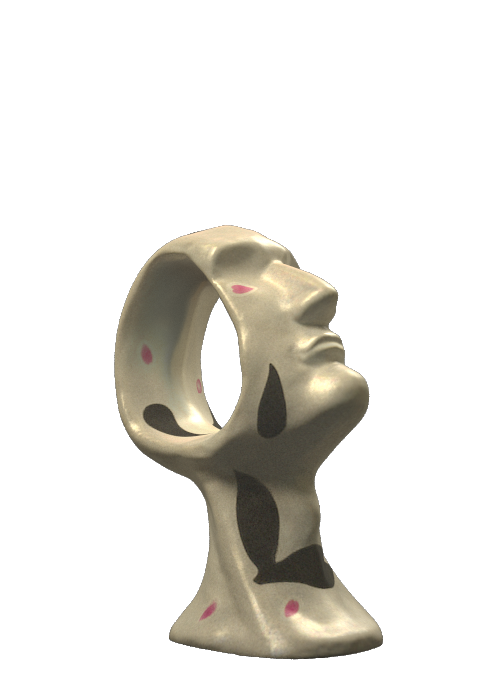} &
		\includegraphics[width=\resLen]{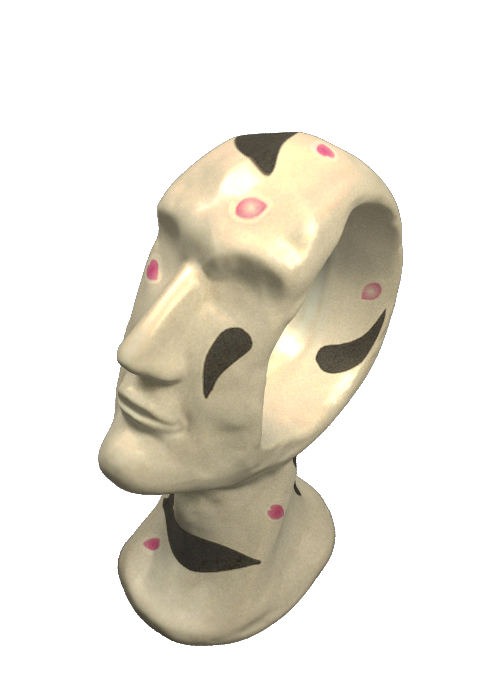}
	\end{tabular}
	\caption{\label{fig:inv_results2}
		\textbf{Inverse-rendering results (real):}
		Photographs of real-world objects (a) and renderings of reconstructed models generated by our implicit stage (b) and full pipeline (c) under two \emph{novel} views.
		Similar to the synthetic results, all optimizations use the same sphere-like initial geometry and near-constant BRDF parameter values.
	}
\end{figure*}

The \textbf{bunny temple} example in Figure~\ref{fig:inv_comp1} has a bunny sitting within a temple-like structure, leading to a fairly complex overall topology.
In this example, we use 96 target images under collocated configurations (that is, with a point light source collocated with the camera) for all methods and show renderings from one novel viewing direction in the figure.
Initialized with a sphere-like geometry (Figure~\ref{fig:inv_comp1}-b), \texttt{IDR*} successfully recovers the overall shape of the object but fails to recover some geometric details of the bunny (Figure~\ref{fig:inv_comp1}-c).
The mesh-based method, on the contrary, successfully recovers some of the details but fails to obtain correct overall topology (Figure~\ref{fig:inv_comp1}-d).

\setlength{\resLen}{0.82in}

\begin{figure*}[t]
	\centering
	\small
	\addtolength{\tabcolsep}{-4pt}
	\begin{tabular}{cc|cc|cc|cc}
		\multicolumn{2}{c|}{(a) \textbf{Target}} &
		\multicolumn{2}{c|}{(b) \textbf{Initial}} &
		\multicolumn{2}{c|}{(c) \textbf{Intermediate}} &
		\multicolumn{2}{c}{(d) \textbf{Optimized}}
		\\[2pt]
		\begin{overpic}[width=\resLen]{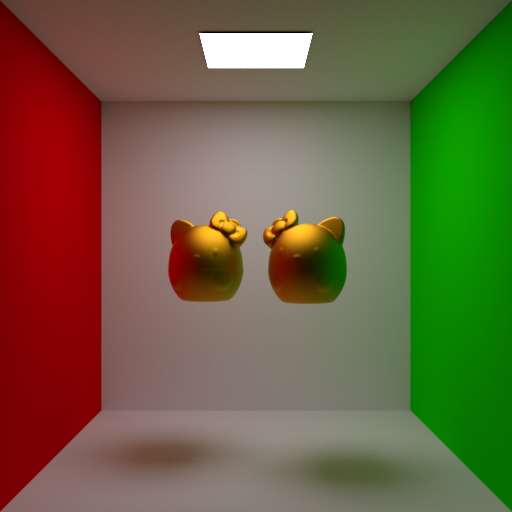}
			\put(2,3){\color{white} \bfseries \small Pose 1}
		\end{overpic}
		&
		\begin{overpic}[width=\resLen]{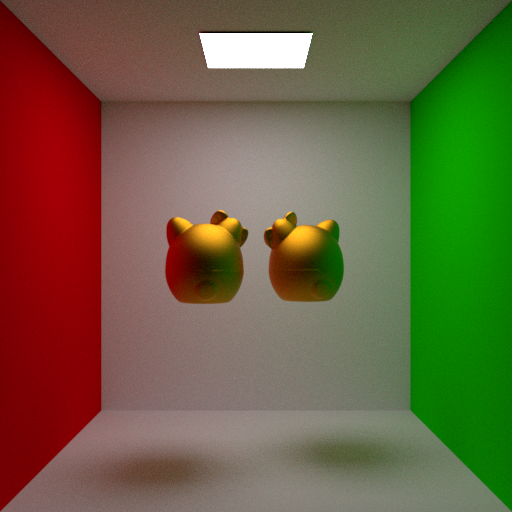}
			\put(2,3){\color{white} \bfseries \small Pose 2}
		\end{overpic}
		&
		\begin{overpic}[width=\resLen]{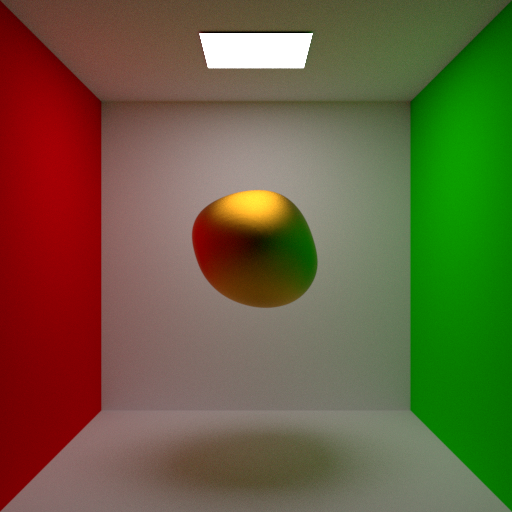}
			\put(2,3){\color{white} \bfseries \small Pose 1}
		\end{overpic}
		&
		\begin{overpic}[width=\resLen]{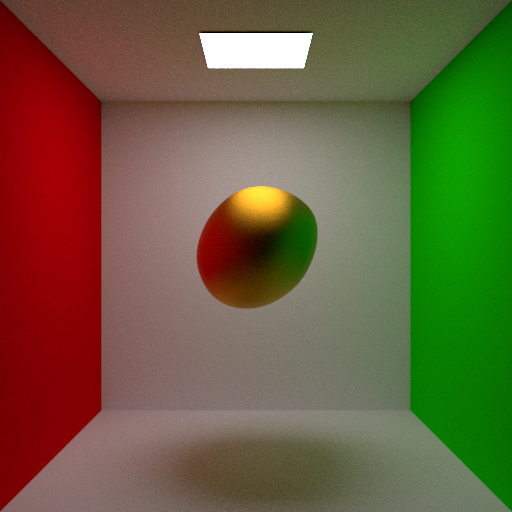}
			\put(2,3){\color{white} \bfseries \small Pose 2}
		\end{overpic}
		&
		\begin{overpic}[width=\resLen]{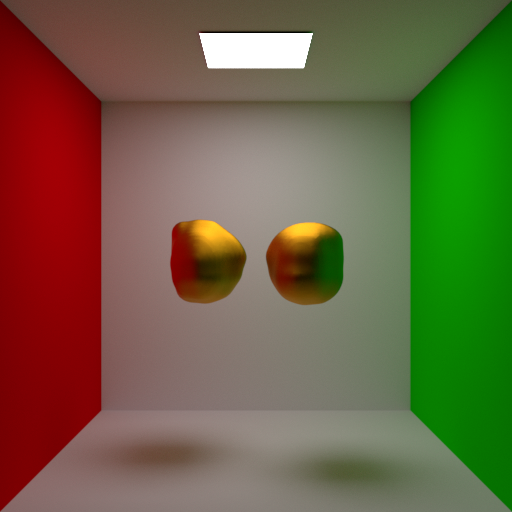}
			\put(2,3){\color{white} \bfseries \small Pose 1}
		\end{overpic}
		&
		\begin{overpic}[width=\resLen]{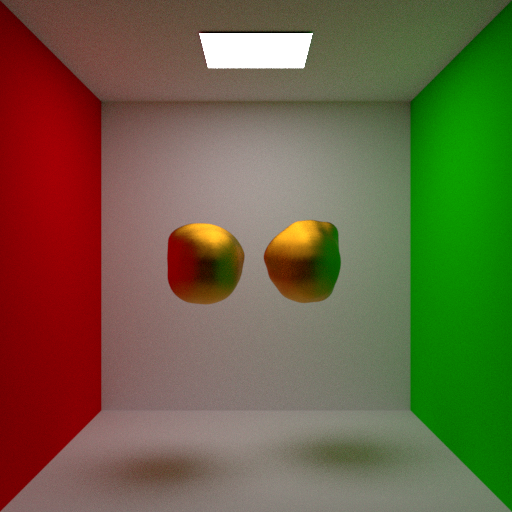}
			\put(2,3){\color{white} \bfseries \small Pose 2}
		\end{overpic}
		&
		\begin{overpic}[width=\resLen]{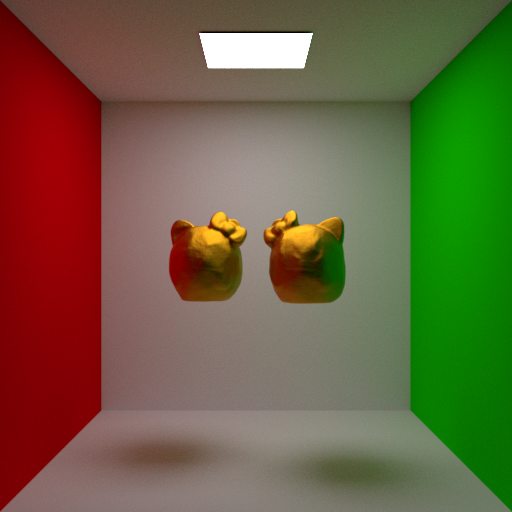}
			\put(2,3){\color{white} \bfseries \small Pose 1}
		\end{overpic}
		&
		\begin{overpic}[width=\resLen]{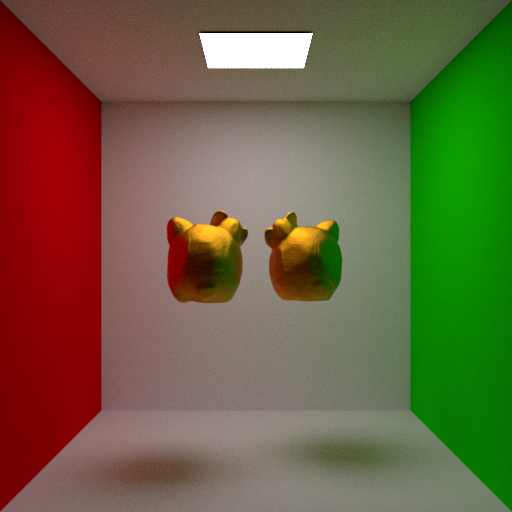}
			\put(2,3){\color{white} \bfseries \small Pose 2}
		\end{overpic}
		\\[2pt]
		\begin{overpic}[width=\resLen]{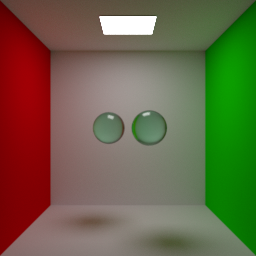}
			\put(2,3){\color{white} \bfseries \small Pose 1}
		\end{overpic}
		&
		\begin{overpic}[width=\resLen]{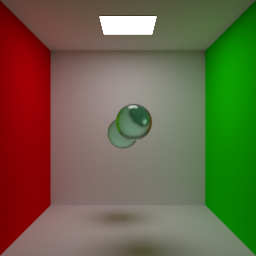}
			\put(2,3){\color{white} \bfseries \small Pose 2}
		\end{overpic}
		&
		\begin{overpic}[width=\resLen]{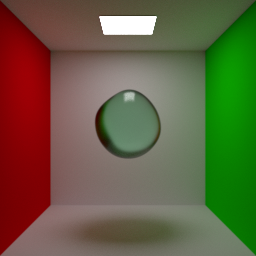}
			\put(2,3){\color{white} \bfseries \small Pose 1}
		\end{overpic}
		&
		\begin{overpic}[width=\resLen]{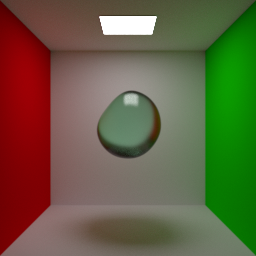}
			\put(2,3){\color{white} \bfseries \small Pose 2}
		\end{overpic}
		&
		\begin{overpic}[width=\resLen]{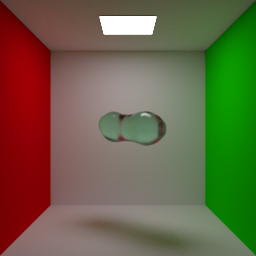}
			\put(2,3){\color{white} \bfseries \small Pose 1}
		\end{overpic}
		&
		\begin{overpic}[width=\resLen]{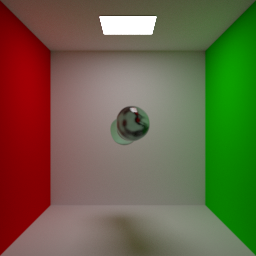}
			\put(2,3){\color{white} \bfseries \small Pose 2}
		\end{overpic}
		&
		\begin{overpic}[width=\resLen]{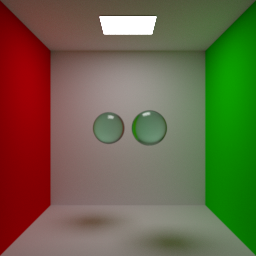}
			\put(2,3){\color{white} \bfseries \small Pose 1}
		\end{overpic}
		&
		\begin{overpic}[width=\resLen]{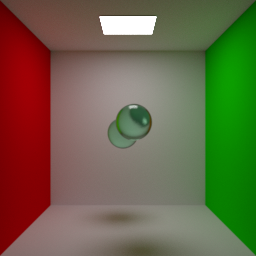}
			\put(2,3){\color{white} \bfseries \small Pose 2}
		\end{overpic}
		\\[2pt]
		\begin{overpic}[width=\resLen]{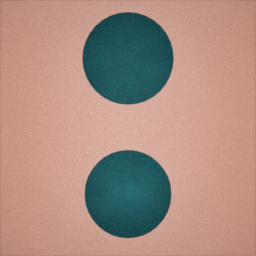}
			\put(2,3){\color{white} \bfseries \small Pose 1}
		\end{overpic}
		&
		\begin{overpic}[width=\resLen]{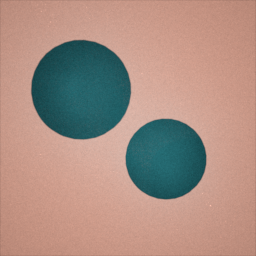}
			\put(2,3){\color{white} \bfseries \small Pose 2}
		\end{overpic}
		&
		\begin{overpic}[width=\resLen]{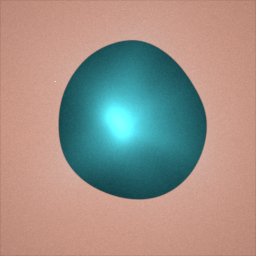}
			\put(2,3){\color{white} \bfseries \small Pose 1}
		\end{overpic}
		&
		\begin{overpic}[width=\resLen]{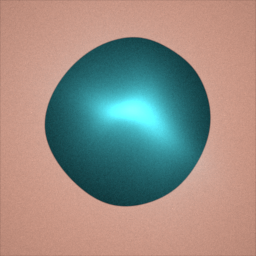}
			\put(2,3){\color{white} \bfseries \small Pose 2}
		\end{overpic}
		&
		\begin{overpic}[width=\resLen]{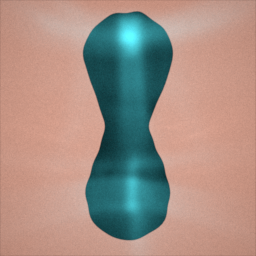}
			\put(2,3){\color{white} \bfseries \small Pose 1}
		\end{overpic}
		&
		\begin{overpic}[width=\resLen]{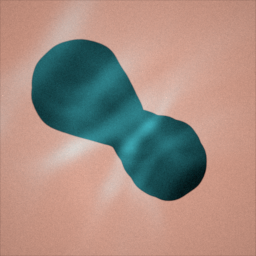}
			\put(2,3){\color{white} \bfseries \small Pose 2}
		\end{overpic}
		&
		\begin{overpic}[width=\resLen]{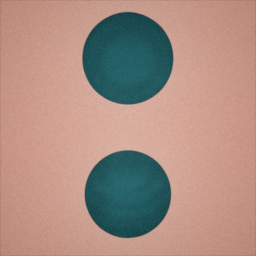}
			\put(2,3){\color{white} \bfseries \small Pose 1}
		\end{overpic}
		&
		\begin{overpic}[width=\resLen]{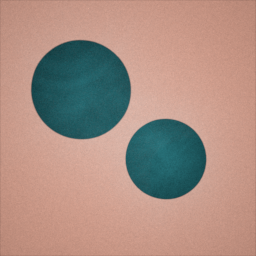}
			\put(2,3){\color{white} \bfseries \small Pose 2}
		\end{overpic}
	\end{tabular}
	\caption{\label{fig:spheres_caustic}
		\textbf{Proof-of-concept examples:}
		The top and middle examples consist of Cornell boxes containing two glossy kitty models and two rough glass spheres, respectively.
		In the bottom example, we use cast shadow (and caustics) images of two rough glass spheres lit by a small area light.
		For all examples, we infer the shape of the object using multiple input images (with two shown) and corresponding object orientations. Using one sphere as the initial shape, our inverse-rendering pipeline successfully recovers the target shapes with different topologies as the initial.
	}
\end{figure*}

Our implicit stage behave similarly to \texttt{IDR*}: It recovers slightly fewer geometric details (Figure~\ref{fig:inv_comp1}-e1) but runs approximately 3.2$\times$ faster: under identical configurations, our method takes 1.19 seconds per iteration while \texttt{IDR*} takes 3.90 seconds.
This is because (i)~we use relatively coarse (i.e., $96^3$) grids for iso-surface extractions, introducing low computational overhead; and (ii)~our underlying differentiable renderer is much faster than \texttt{IDR}'s sphere-tracing-based implementation.
Initialized with results from our implicit stage, our explicit stage successfully recovers the geometric details and produces a reconstruction with the best quality (Figure~\ref{fig:inv_comp1}-e2).

The \textbf{chair} example in Figure~\ref{fig:inv_comp1} involves a chair with spatially varying reflectance under environmental lighting.
We use 196 target images and show renderings under one novel view (top) and one novel illumination (bottom).
The \texttt{IDR*} baseline does not apply here due to the need to estimate the boundary integrals in Eq.~\eqref{eqn:dRE}---which is challenging since the domain of integration $\partial\sph$ is implicit.
The mesh-based method is still applicable.
But similar to the previous example, it has difficulties recovering the overall topology and introduces self intersections (Figure~\ref{fig:inv_comp1}-d).
Our implicit stage---thanks to its use of meshes as an intermediate representation---is capable of perform differentiable rendering under environmental illumination and successfully returns a coarse reconstruction of the chair (Figure~\ref{fig:inv_comp1}-e1).
This result is further refined by our explicit stage, leading to a high-fidelity reconstruction (Figure~\ref{fig:inv_comp1}-e2).

We show two additional inverse-rendering results in Figure~\ref{fig:inv_results1}.
For both examples, our implicit stage successfully produces coarse reconstructions with correct overall topology.
These reconstructions are then refined by our explicit stage to recover the geometric and reflectance details.

\paragraph{Real results}
We further apply our technique to reconstruct the shape and reflectance of real objects, as shown in Figure~\ref{fig:inv_results2}.
We take photographs of these objects under indoor illumination with two bright area lights to better show specular highlights.
Our technique accurately recovers the fine textures for the \textbf{chess}, the detailed surface geometries for the \textbf{teapot}, the glossiness for the \textbf{chess} and the \textbf{leopard}, and the large hole for the \textbf{head} model---all based on sphere-like initial shapes.

\paragraph{Proof-of-concept example}
To further motivate the advantage of our physics-based differentiable rendering of implicit surface, we show three proof-of-concept synthetic examples with prominent global-illumination effects in Figure~\ref{fig:spheres_caustic}.
\rev{The top and middle examples show Cornell boxes containing two glossy kitty models and two rough glass spheres, respectively.}
The bottom example involves the same rough-glass object lit by a small area light.
For all examples, we take as input multiple images (100 for the top, and 26 for the middle and bottom examples) of the object (with varying known orientations) and solve for the shape of the object.
We note that differentiating these renderings cannot be handled by existing differentiable renderers: Soft rasterizers and simple ray tracers have difficulties handling light-transport effects like soft shadows and interreflection; existing physics-based differentiable renderers, on the other hand, cannot be easily adopted to render implicit geometries.

Thanks to our new differentiable rendering pipeline described in \S\ref{ssec:implicit}, for \rev{all} 
examples, our inverse-rendering technique manages to 
\rev{recover the target shapes using initializations with varying topologies.}
We believe that the new level of generality enabled by our technique will enable interesting future applications in computer graphics, vision, and computational imaging.

\section{Discussion and Conclusion}
\label{sec:conclusion}
\paragraph{Limitations and future work}
%
%
\rev{%
	When jointly optimizing shape and spatially varying reflectance, the inverse-rendering problem can be under-constrained, causing illumination or geometric details to be ``baked'' into textures.
	Our technique, likely many (if not most) inverse-rendering techniques, can suffer from this problem---specifically when provided relatively few input images.
	Overcoming this challenge requires devising discriminative losses and effective regularizations, which we believe is an important topic for future research.
}

Additionally, the \texttt{MeshSDF} library~\cite{MeshSDF} relies on marching cube with regular grids.
Improving this step to more efficiently capture geometric details will allow our implicit stage to produce finer reconstructions, reducing the amount of refinement needed for the explicit stage.
Also, extending our pipeline to use other forms of differentiable mesh generation (e.g., \cite{Peng2021SAP}) is worth exploring.

\paragraph{Conclusion}
We introduced a new physics-based inverse rendering technique that uses both implicit and explicit representations of object geometry.
A key component of our technique is a new physics-based differentiable renderer for implicit surfaces: Instead of using differentiable sphere tracing, our method leverages differentiable iso-surface extraction to produce an intermediate mesh, and render this mesh using physics-based differentiable rendering.
Our technique enjoys the benefits from both implicit and explicit representations by allowing easy topology changes and supporting complex light-transport effects like environmental illumination, soft shadows, and interreflection.

\paragraph{Acknowledgments}
We thank the anonymous reviewers for their constructive feedback.
This work was supported in part by NSF grants 1900783, 1900849, and 1900927. Ioannis Gkioulekas was supported by a Sloan Research Fellowship.

\bibliographystyle{eg-alpha-doi}
\bibliography{main}

\newcommand{\etalchar}[1]{$^{#1}$}
\begin{thebibliography}{\uppercase{NDVZJ19}}

\bibitem[ALKN19]{azinovic2019inverse}
\textsc{Azinovic D., Li T.-M., Kaplanyan A., Nie{\ss}ner M.}:
\newblock Inverse path tracing for joint material and lighting estimation.
\newblock In \emph{Proc. IEEE/CVF CVPR} (2019), pp.~2447--2456.

\bibitem[ASCE02]{aspert2002mesh}
\textsc{Aspert N., Santa-Cruz D., Ebrahimi T.}:
\newblock Mesh: Measuring errors between surfaces using the hausdorff distance.
\newblock In \emph{Proceedings. IEEE international conference on multimedia and
  expo} (2002), vol.~1, IEEE, pp.~705--708.

\bibitem[BB09]{brochu2009robust}
\textsc{Brochu T., Bridson R.}:
\newblock Robust topological operations for dynamic explicit surfaces.
\newblock \emph{SIAM Journal on Scientific Computing 31}, 4 (2009), 2472--2493.

\bibitem[BTBW77]{barrow1977parametric}
\textsc{Barrow H.~G., Tenenbaum J.~M., Bolles R.~C., Wolf H.~C.}:
\newblock \emph{Parametric correspondence and chamfer matching: Two new
  techniques for image matching}.
\newblock Tech. rep., SRI INTERNATIONAL MENLO PARK CA ARTIFICIAL INTELLIGENCE
  CENTER, 1977.

\bibitem[CLZ{\etalchar{*}}20]{che2020towards}
\textsc{Che C., Luan F., Zhao S., Bala K., Gkioulekas I.}:
\newblock Towards learning-based inverse subsurface scattering.
\newblock \emph{ICCP} (2020), 1--12.

\bibitem[GLZ16]{gkioulekas2016evaluation}
\textsc{Gkioulekas I., Levin A., Zickler T.}:
\newblock An evaluation of computational imaging techniques for heterogeneous
  inverse scattering.
\newblock In \emph{ECCV} (2016), Springer, pp.~685--701.

\bibitem[GRL{\etalchar{*}}21]{guillard2021deepmesh}
\textsc{Guillard B., Remelli E., Lukoianov A., Richter S., Bagautdinov T.,
  Baque P., Fua P.}:
\newblock Deepmesh: Differentiable iso-surface extraction.
\newblock \emph{arXiv preprint arXiv:2106.11795} (2021).

\bibitem[GZB{\etalchar{*}}13]{gkioulekas2013inverse}
\textsc{Gkioulekas I., Zhao S., Bala K., Zickler T., Levin A.}:
\newblock Inverse volume rendering with material dictionaries.
\newblock \emph{ACM Trans. Graph. 32}, 6 (2013), 1--13.

\bibitem[Har96]{hart1996sphere}
\textsc{Hart J.~C.}:
\newblock Sphere tracing: A geometric method for the antialiased ray tracing of
  implicit surfaces.
\newblock \emph{The Visual Computer 12}, 10 (1996), 527--545.

\bibitem[HLHZ08]{Holroyd2008}
\textsc{Holroyd M., Lawrence J., Humphreys G., Zickler T.}:
\newblock A photometric approach for estimating normals and tangents.
\newblock \emph{ACM Trans. Graph. 27}, 5 (2008), 133:1--133:9.

\bibitem[Hor70]{horn1970shape}
\textsc{Horn B.~K.}:
\newblock Shape from shading: A method for obtaining the shape of a smooth
  opaque object from one view.

\bibitem[IH81]{ikeuchi1981numerical}
\textsc{Ikeuchi K., Horn B.~K.}:
\newblock Numerical shape from shading and occluding boundaries.
\newblock \emph{Artificial intelligence 17}, 1-3 (1981), 141--184.

\bibitem[Kaj86]{kajiya1986rendering}
\textsc{Kajiya J.~T.}:
\newblock The rendering equation.
\newblock In \emph{Proceedings of the 13th annual conference on Computer
  graphics and interactive techniques} (1986), pp.~143--150.

\bibitem[KB14]{kingma2014adam}
\textsc{Kingma D.~P., Ba J.}:
\newblock Adam: A method for stochastic optimization.
\newblock \emph{arXiv preprint arXiv:1412.6980} (2014).

\bibitem[KBM{\etalchar{*}}20]{kato2020differentiable}
\textsc{Kato H., Beker D., Morariu M., Ando T., Matsuoka T., Kehl W., Gaidon
  A.}:
\newblock Differentiable rendering: A survey.
\newblock \emph{arXiv preprint arXiv:2006.12057} (2020).

\bibitem[LADL18]{li2018differentiable}
\textsc{Li T.-M., Aittala M., Durand F., Lehtinen J.}:
\newblock Differentiable {Monte Carlo} ray tracing through edge sampling.
\newblock \emph{ACM Trans. Graph. 37}, 6 (2018), 1--11.

\bibitem[LHJ19]{loubet2019reparameterizing}
\textsc{Loubet G., Holzschuch N., Jakob W.}:
\newblock Reparameterizing discontinuous integrands for differentiable
  rendering.
\newblock \emph{ACM Trans. Graph. 38}, 6 (2019), 1--14.

\bibitem[LLCL19]{liu2019soft}
\textsc{Liu S., Li T., Chen W., Li H.}:
\newblock Soft rasterizer: A differentiable renderer for image-based {3D}
  reasoning.
\newblock In \emph{ICCV} (2019), pp.~7708--7717.

\bibitem[LZBD21]{Luan2021}
\textsc{Luan F., Zhao S., Bala K., Dong Z.}:
\newblock Unified shape and svbrdf recovery using differentiable monte carlo
  rendering.
\newblock \emph{Computer Graphics Forum 40}, 4 (2021), 101--113.

\bibitem[MHS{\etalchar{*}}22]{munkberg2022extracting}
\textsc{Munkberg J., Hasselgren J., Shen T., Gao J., Chen W., Evans A.,
  M{\"u}ller T., Fidler S.}:
\newblock Extracting triangular 3d models, materials, and lighting from images.
\newblock In \emph{Proceedings of the IEEE/CVF Conference on Computer Vision
  and Pattern Recognition} (2022), pp.~8280--8290.

\bibitem[NDVZJ19]{nimier2019mitsuba}
\textsc{Nimier-David M., Vicini D., Zeltner T., Jakob W.}:
\newblock Mitsuba 2: A retargetable forward and inverse renderer.
\newblock \emph{ACM Trans. Graph. 38}, 6 (2019), 1--17.

\bibitem[NJJ21]{Nicolet2021Large}
\textsc{Nicolet B., Jacobson A., Jakob W.}:
\newblock Large steps in inverse rendering of geometry.
\newblock \emph{ACM Transactions on Graphics 40}, 6 (2021), 248:1--248:13.

\bibitem[NMOG20]{niemeyer2020differentiable}
\textsc{Niemeyer M., Mescheder L., Oechsle M., Geiger A.}:
\newblock Differentiable volumetric rendering: Learning implicit {3D}
  representations without {3D} supervision.
\newblock In \emph{Proceedings of the IEEE/CVF Conference on Computer Vision
  and Pattern Recognition} (2020), pp.~3504--3515.

\bibitem[PFAK20]{poursaeed2020coupling}
\textsc{Poursaeed O., Fisher M., Aigerman N., Kim V.~G.}:
\newblock Coupling explicit and implicit surface representations for generative
  {3D} modeling.
\newblock In \emph{European Conference on Computer Vision} (2020), Springer,
  pp.~667--683.

\bibitem[PJL{\etalchar{*}}21]{Peng2021SAP}
\textsc{Peng S., Jiang C.~M., Liao Y., Niemeyer M., Pollefeys M., Geiger A.}:
\newblock Shape as points: A differentiable poisson solver.
\newblock In \emph{Advances in Neural Information Processing Systems (NeurIPS)}
  (2021).

\bibitem[QMC{\etalchar{*}}17]{queau2017variational}
\textsc{Qu{\'e}au Y., M{\'e}lou J., Castan F., Cremers D., Durou J.-D.}:
\newblock A variational approach to shape-from-shading under natural
  illumination.
\newblock In \emph{International Workshop on Energy Minimization Methods in
  Computer Vision and Pattern Recognition} (2017), pp.~342--357.

\bibitem[QMD16]{queau2016unbiased}
\textsc{Qu{\'e}au Y., Mecca R., Durou J.-D.}:
\newblock Unbiased photometric stereo for colored surfaces: A variational
  approach.
\newblock In \emph{Proc. IEEE CVPR} (2016), pp.~4359--4368.

\bibitem[RLR{\etalchar{*}}20]{MeshSDF}
\textsc{Remelli E., Lukoianov A., Richter S., Guillard B., Bagautdinov T.,
  Baque P., Fua P.}:
\newblock {MeshSDF}: Differentiable iso-surface extraction.
\newblock In \emph{Advances in Neural Information Processing Systems} (2020),
  Larochelle H., Ranzato M., Hadsell R., Balcan M.~F., Lin H., (Eds.), vol.~33,
  pp.~22468--22478.

\bibitem[RRN{\etalchar{*}}20]{ravi2020accelerating}
\textsc{Ravi N., Reizenstein J., Novotny D., Gordon T., Lo W.-Y., Johnson J.,
  Gkioxari G.}:
\newblock Accelerating {3D} deep learning with {PyTorch3D}.
\newblock \emph{arXiv preprint arXiv:2007.08501} (2020).

\bibitem[SC17]{BFF}
\textsc{Sawhney R., Crane K.}:
\newblock Boundary first flattening.
\newblock \emph{ACM Trans. Graph. 37}, 1 (2017), 5:1--5:14.

\bibitem[SD99]{seitz1999photorealistic}
\textsc{Seitz S.~M., Dyer C.~R.}:
\newblock Photorealistic scene reconstruction by voxel coloring.
\newblock \emph{International Journal of Computer Vision 35}, 2 (1999),
  151--173.

\bibitem[SZPF16]{schoenberger2016mvs}
\textsc{Sch\"{o}nberger J.~L., Zheng E., Pollefeys M., Frahm J.-M.}:
\newblock Pixelwise view selection for unstructured multi-view stereo.
\newblock In \emph{ECCV} (2016).

\bibitem[TSG19]{tsai2019beyond}
\textsc{Tsai C.-Y., Sankaranarayanan A.~C., Gkioulekas I.}:
\newblock Beyond volumetric albedo--a surface optimization framework for
  non-line-of-sight imaging.
\newblock In \emph{Proc. IEEE/CVF CVPR} (2019), pp.~1545--1555.

\bibitem[VSJ22]{Vicini2022sdf}
\textsc{Vicini D., Speierer S., Jakob W.}:
\newblock Differentiable signed distance function rendering.
\newblock \emph{ACM Trans. Graph. 41}, 4 (2022), 125:1--125:18.

\bibitem[VTC05]{vogiatzis2005multi}
\textsc{Vogiatzis G., Torr P.~H., Cipolla R.}:
\newblock Multi-view stereo via volumetric graph-cuts.
\newblock In \emph{Proc. IEEE CVPR} (2005), vol.~2, pp.~391--398.

\bibitem[WH94]{Witkin1994}
\textsc{Witkin A.~P., Heckbert P.~S.}:
\newblock Using particles to sample and control implicit surfaces.
\newblock In \emph{Proceedings of the 21st Annual Conference on Computer
  Graphics and Interactive Techniques} (1994), SIGGRAPH '94, p.~269–277.

\bibitem[Woo80]{woodham1980photometric}
\textsc{Woodham R.~J.}:
\newblock Photometric method for determining surface orientation from multiple
  images.
\newblock \emph{Optical engineering 19}, 1 (1980), 191139.

\bibitem[YKM{\etalchar{*}}20]{yariv2020multiview}
\textsc{Yariv L., Kasten Y., Moran D., Galun M., Atzmon M., Ronen B., Lipman
  Y.}:
\newblock Multiview neural surface reconstruction by disentangling geometry and
  appearance.
\newblock \emph{Advances in Neural Information Processing Systems 33} (2020).

\bibitem[ZJL20]{zhao2020physics}
\textsc{Zhao S., Jakob W., Li T.-M.}:
\newblock Physics-based differentiable rendering: from theory to
  implementation.
\newblock In \emph{ACM SIGGRAPH 2020 Courses}. 2020, pp.~1--30.

\bibitem[ZMY{\etalchar{*}}20]{Zhang:2020:PSDR}
\textsc{Zhang C., Miller B., Yan K., Gkioulekas I., Zhao S.}:
\newblock Path-space differentiable rendering.
\newblock \emph{ACM Trans. Graph. 39}, 4 (2020), 143:1--143:19.

\bibitem[ZT10]{zhou2010ring}
\textsc{Zhou Z., Tan P.}:
\newblock Ring-light photometric stereo.
\newblock In \emph{European Conference on Computer Vision} (2010),
  pp.~265--279.

\bibitem[ZWZ{\etalchar{*}}19]{zhang2019differential}
\textsc{Zhang C., Wu L., Zheng C., Gkioulekas I., Ramamoorthi R., Zhao S.}:
\newblock A differential theory of radiative transfer.
\newblock \emph{ACM Trans. Graph. 38}, 6 (2019), 1--16.

\end{thebibliography}

\newpage

\title{%
	{\large Physics-Based Inverse Rendering using Combined Implicit and Explicit Geometries}\\[5pt]
	Supplemental Materials
}


\author[Cai et al.]{%
	\parbox{\textwidth}{\centering 
		G. Cai$^{1,2}$,
		K. Yan$^{1,2}$,
		Z. Dong$^2$,
		I. Gkioulekas$^3$,
		S. Zhao$^1$
	}
	\\
	\parbox{\textwidth}{\centering
		$^1$University of California, Irvine
		\hspace{1cm}
		$^2$Meta Reality Labs Research
		\hspace{1cm}
		$^3$Carnegie Mellon University
	}
}

\maketitle

\appendix
\section{Extra Experiments}
\label{X:sec:results}
We now show additional comparisons using both synthetic and real data.

Figures~\ref{fig:inv_comp1} and \ref{fig:inv_comp2} are extended versions of Figure~3 of the main paper.
In both figures, we show per-vertex distance visualizations by projecting each vertex on the reconstructed meshes to the ground-truth (marked as ``\emph{Recon. to GT}'') and vice versa (marked as ``\emph{GT to recon.}'').
The \texttt{IDR*} baseline%
\footnote{This baseline is obtained by adopting the \texttt{IDR} technique. To compute the distance visualizations, we convert the resulting implicit shapes into meshes using high-resolution marching cube.}
has difficulties recovering the geometric details of the GT.
The mesh-based baseline, on the other hand, manages to capture the geometric details locally but fails to obtain  correct global topologies, resulting in missing or redundant geometric features.
Our technique, by using both implicit and explicit geometries, enjoys the advantage of both representations.

In Figures~\ref{fig:inv_comp3}--\ref{fig:inv_comp8}, we compare reconstruction results with the mesh-based baseline.
We note that \texttt{IDR*} is not applicable for these examples as they use environmental lighting that cannot be easily handled by the \texttt{IDR} framework (in a physics-based fashion).
Thus, we mainly compare to the purely mesh-based method.
All renderings in these are from novel views (that are not used for inverse-rendering optimizations).
To quantitatively compare the qualities of these renderings, we compute their PSNR (shown under each rendering).

\section{Social Impact}
We do not foresee our technique to negatively impact our society in any significant fashion.
Our research does not involve human subjects or human-derived data.
Further, our main application---the digitization of 3D objects and scenes---is more about building virtual environments (or meta-verses) than manipulating the physical one, and has little to no environmental impact.

\setlength{\resLen}{1.18in}

\begin{figure*}[t]
	\centering
	\addtolength{\tabcolsep}{-7pt}
	\small
	\rev{\begin{tabular}{rccccc}
		\multicolumn{1}{c}{(a) \textbf{Ground truth}} & (b) \textbf{Initial} & (c) \textbf{IDR*} & (d) \textbf{Mesh-based} & (e1) \textbf{Ours (impl.)} & (e2) \textbf{Ours (full)}
		\\[-2pt]
		\adjincludegraphics[width=\resLen,trim={0 {.1\height} 0 {.15\height}},clip]{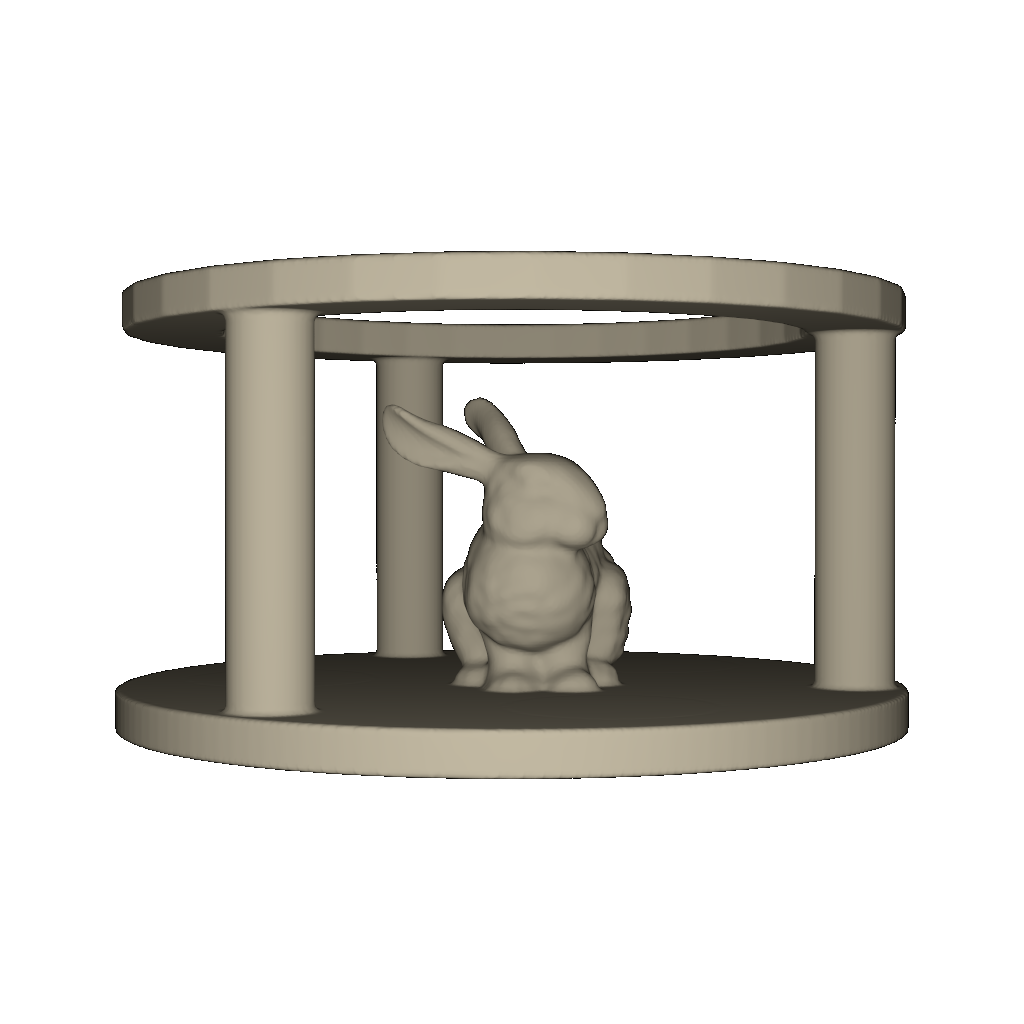} &
		\adjincludegraphics[width=\resLen,trim={0 {.1\height} 0 {.15\height}},clip]{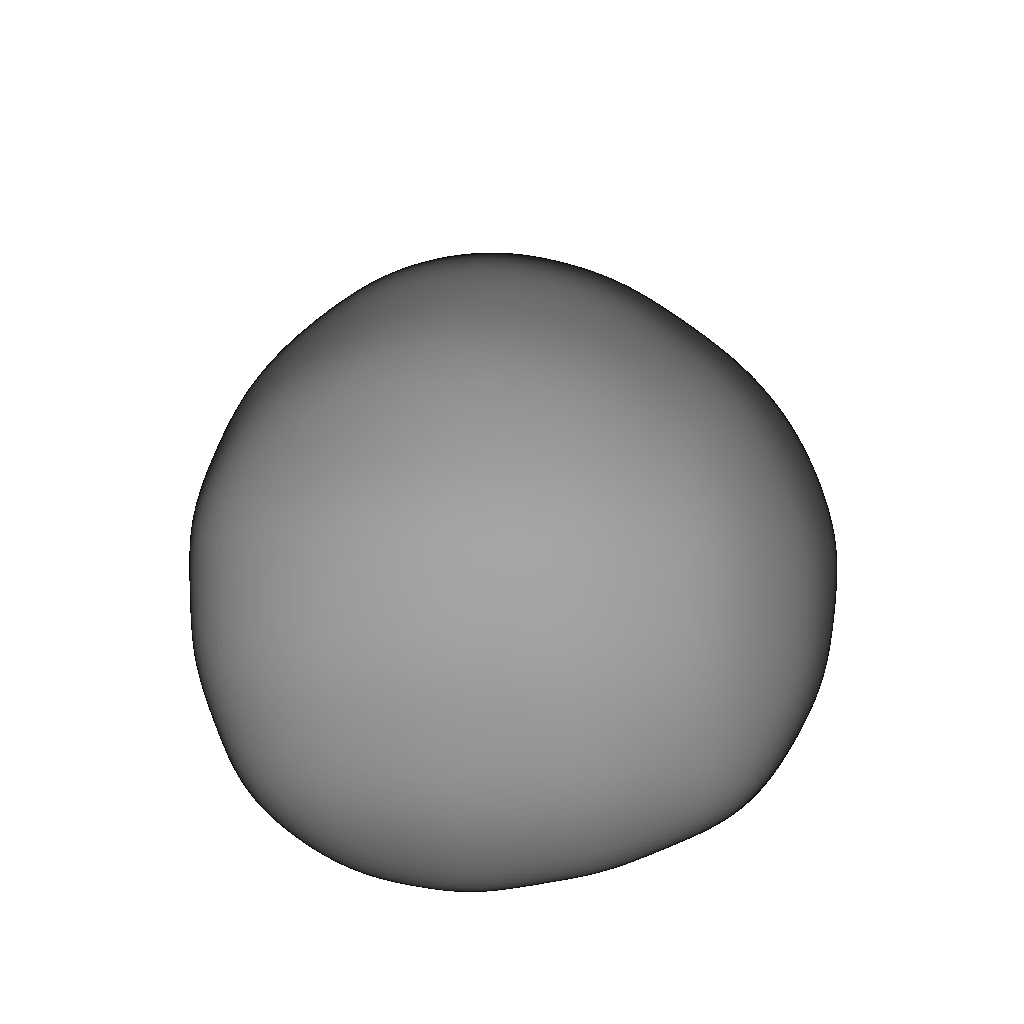} &
		\adjincludegraphics[width=\resLen,trim={0 {.1\height} 0 {.15\height}},clip]{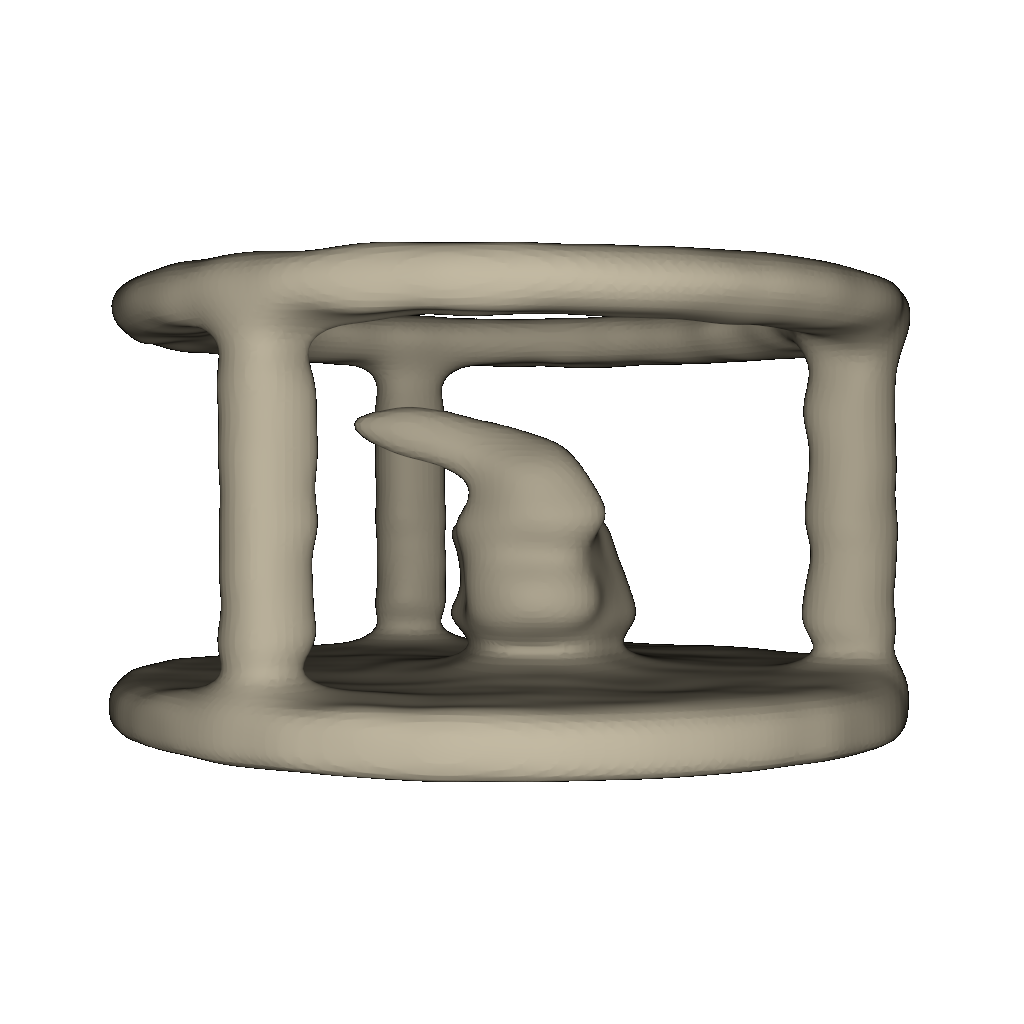} &
		\adjincludegraphics[width=\resLen,trim={0 {.1\height} 0 {.15\height}},clip]{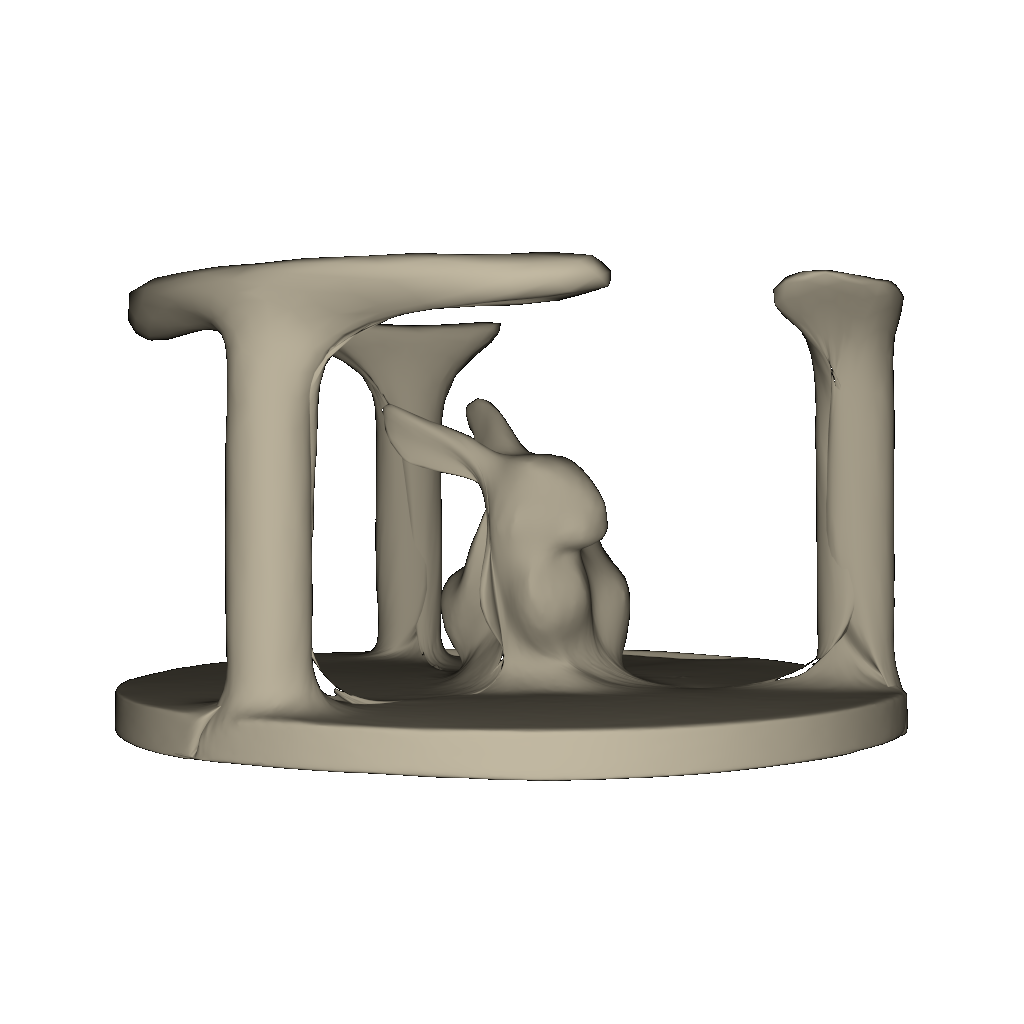} &
		\adjincludegraphics[width=\resLen,trim={0 {.1\height} 0 {.15\height}},clip]{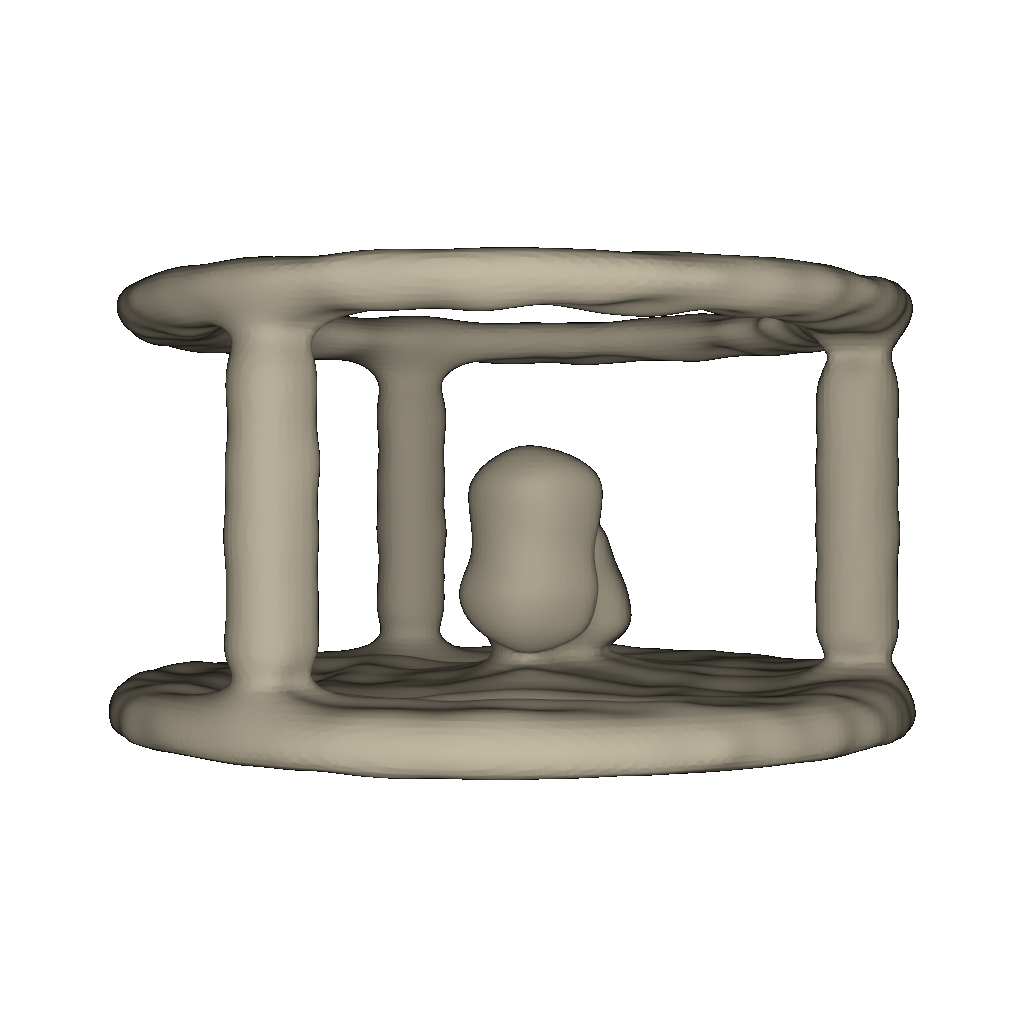} &
		\adjincludegraphics[width=\resLen,trim={0 {.1\height} 0 {.15\height}},clip]{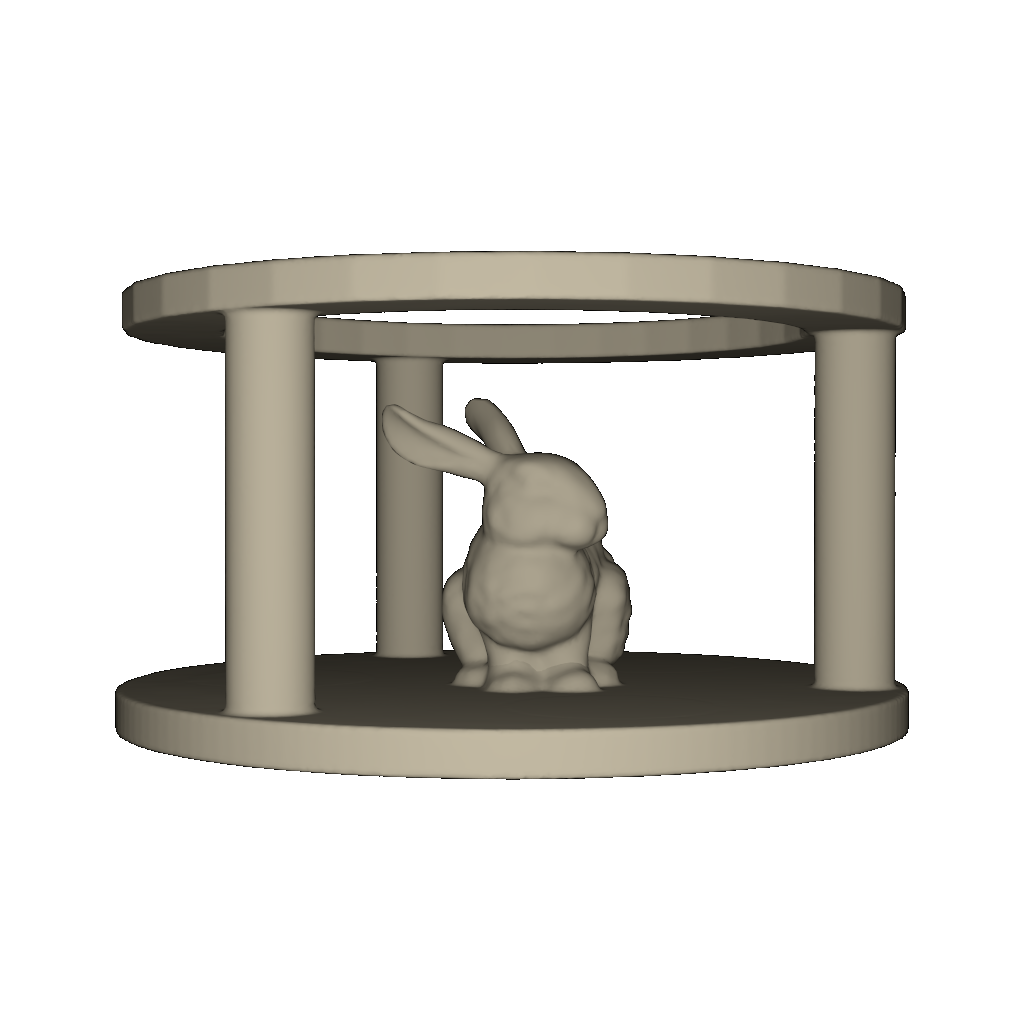}
		\\[-10pt]
		& \raisebox{30pt}{\bfseries Recon. to GT:} &
		\adjincludegraphics[width=\resLen,trim={0 {.1\height} 0 {.15\height}},clip]{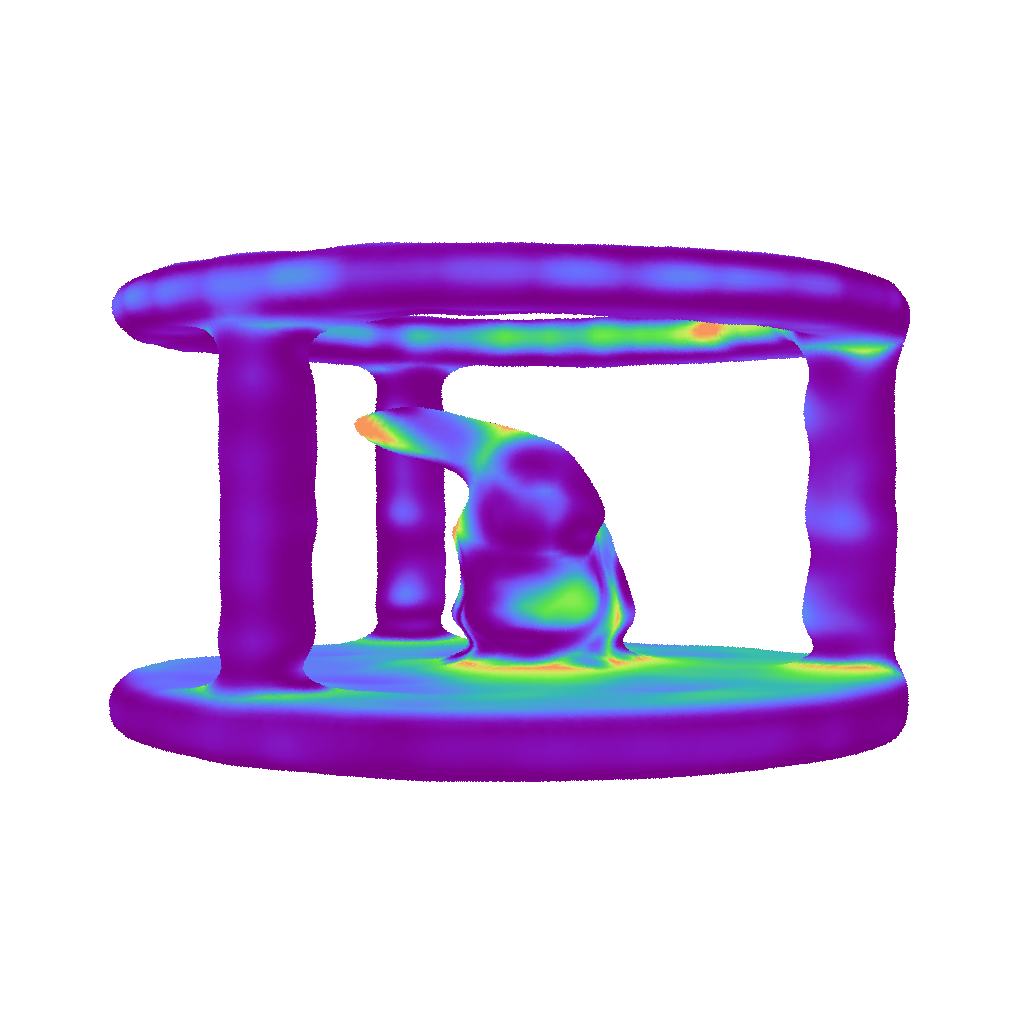} &
		\adjincludegraphics[width=\resLen,trim={0 {.1\height} 0 {.15\height}},clip]{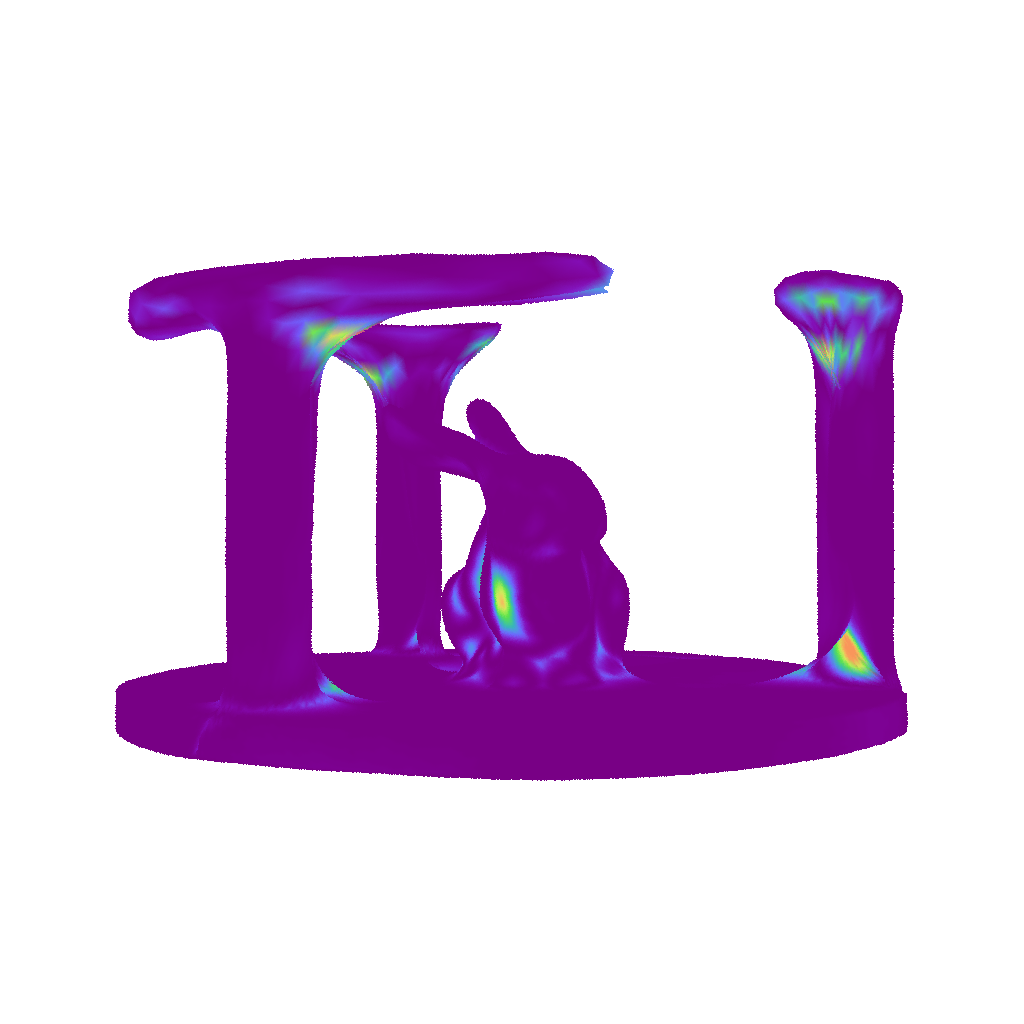} &
		\adjincludegraphics[width=\resLen,trim={0 {.1\height} 0 {.15\height}},clip]{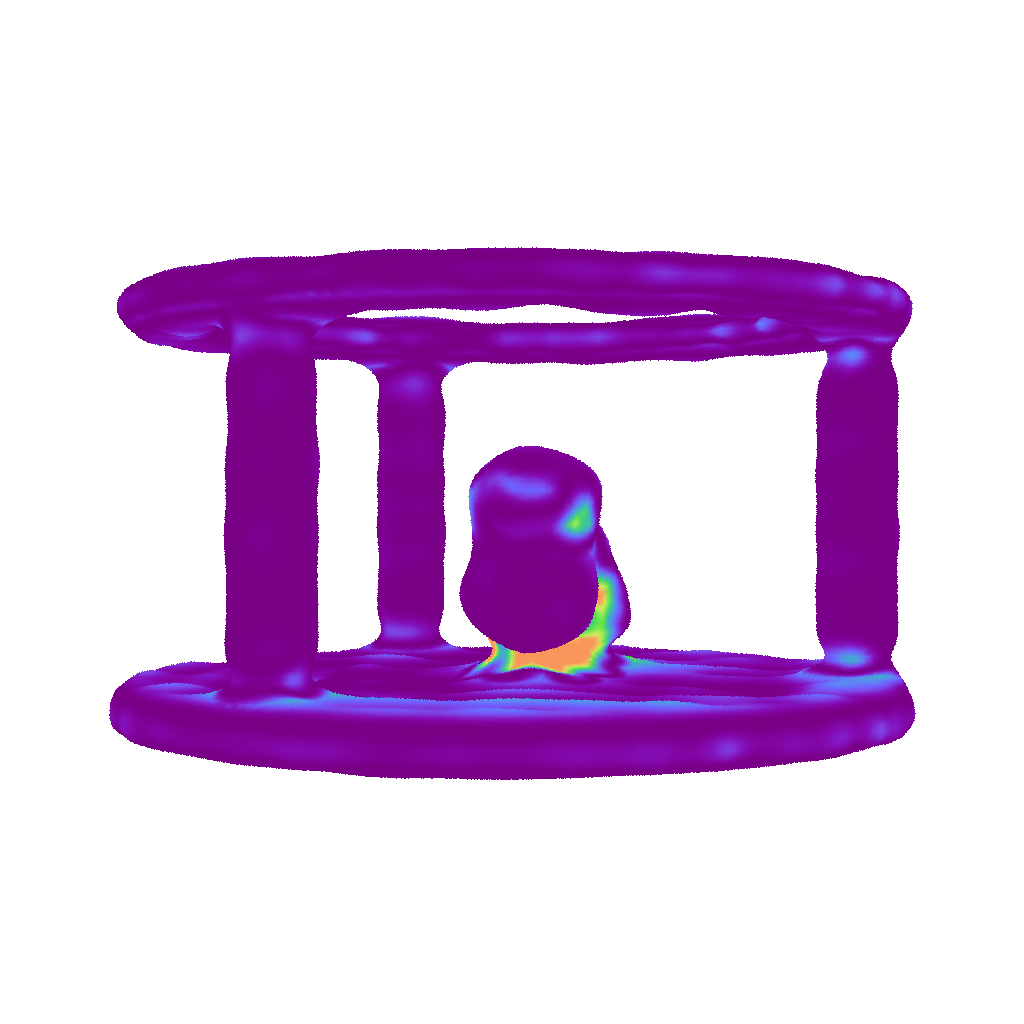} &
		\adjincludegraphics[width=\resLen,trim={0 {.1\height} 0 {.15\height}},clip]{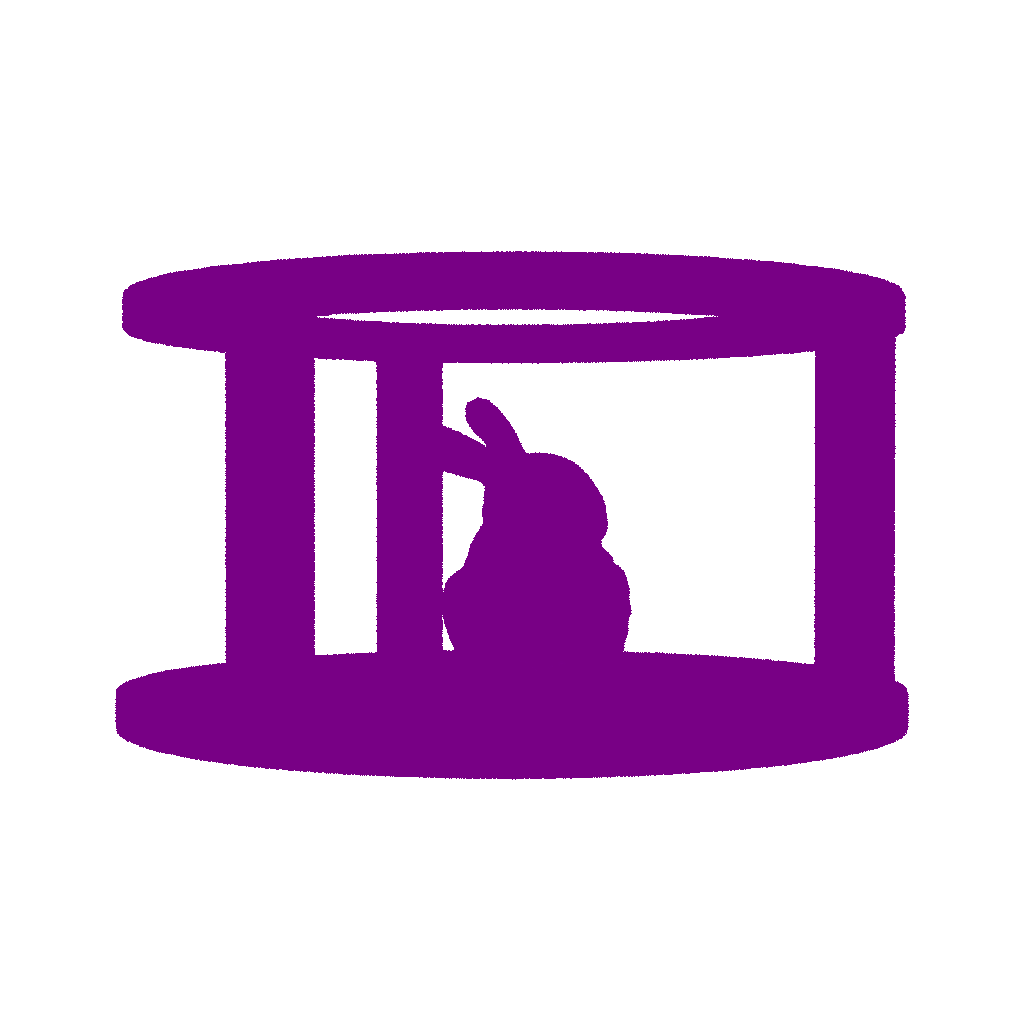}
		\\[-10pt]
		& \raisebox{30pt}{\bfseries GT to recon.:} &
		\adjincludegraphics[width=\resLen,trim={0 {.1\height} 0 {.15\height}},clip]{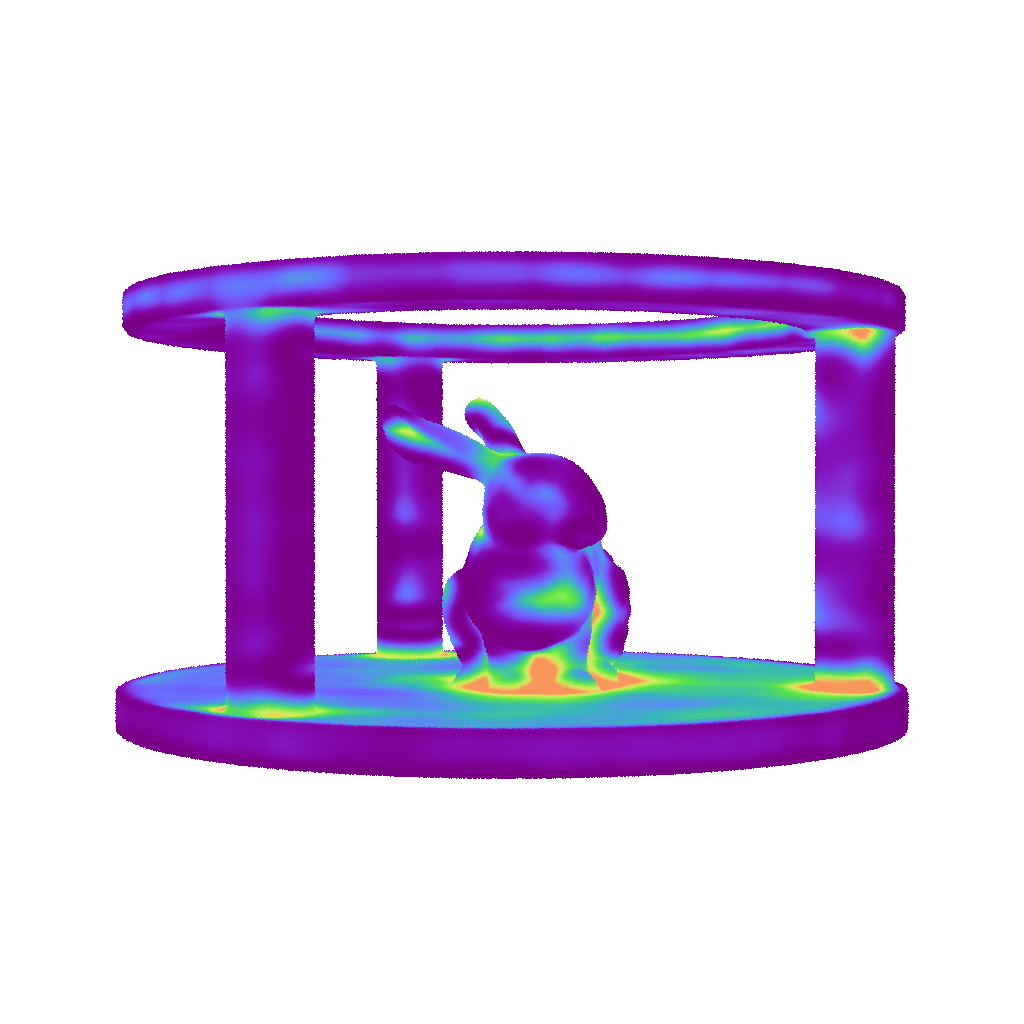} &
		\adjincludegraphics[width=\resLen,trim={0 {.1\height} 0 {.15\height}},clip]{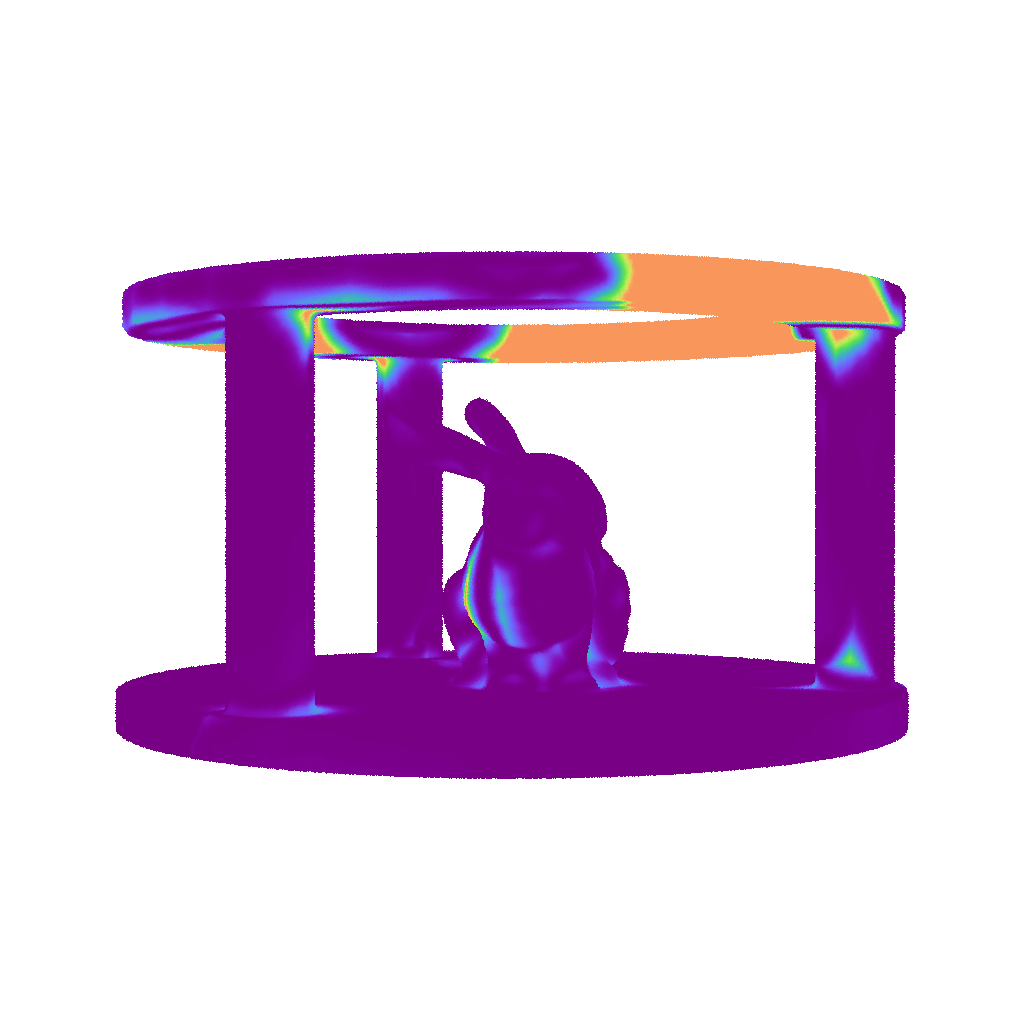} &
		\adjincludegraphics[width=\resLen,trim={0 {.1\height} 0 {.15\height}},clip]{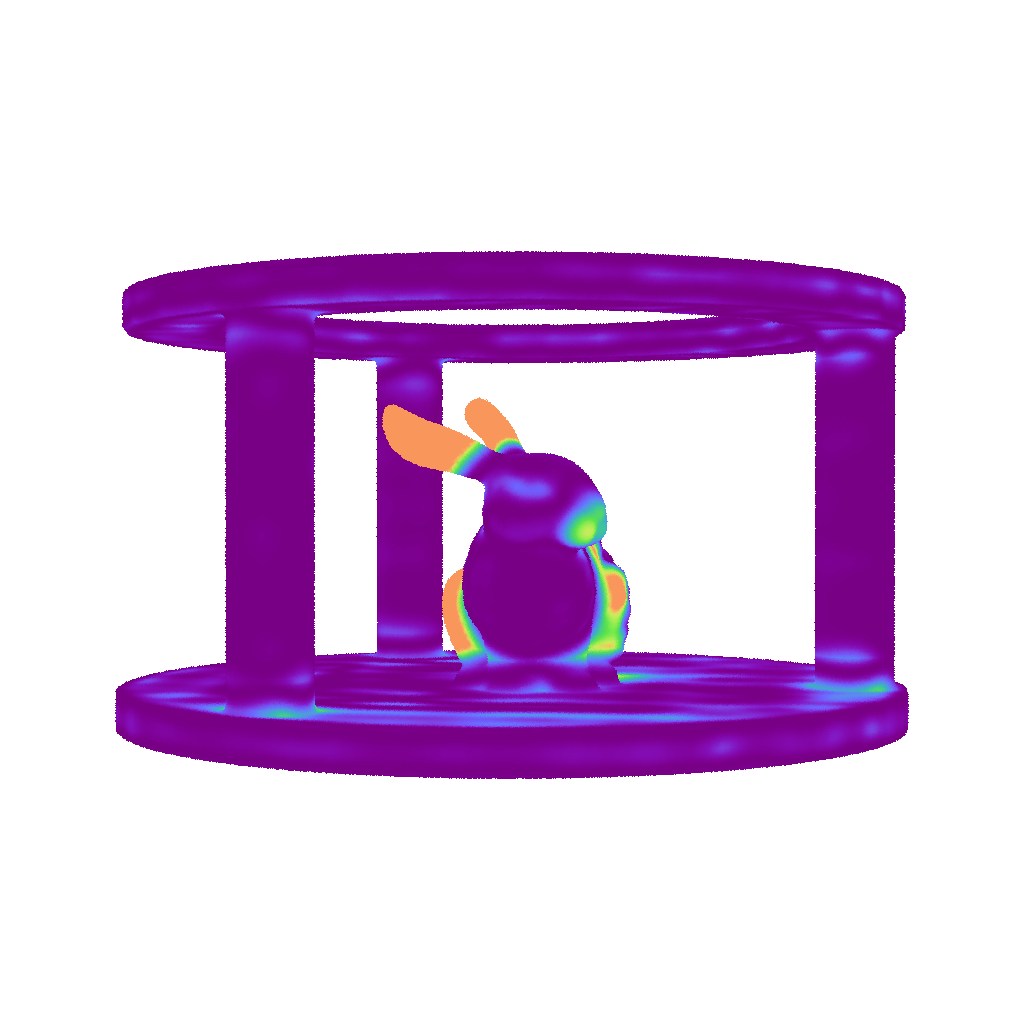} &
		\adjincludegraphics[width=\resLen,trim={0 {.1\height} 0 {.15\height}},clip]{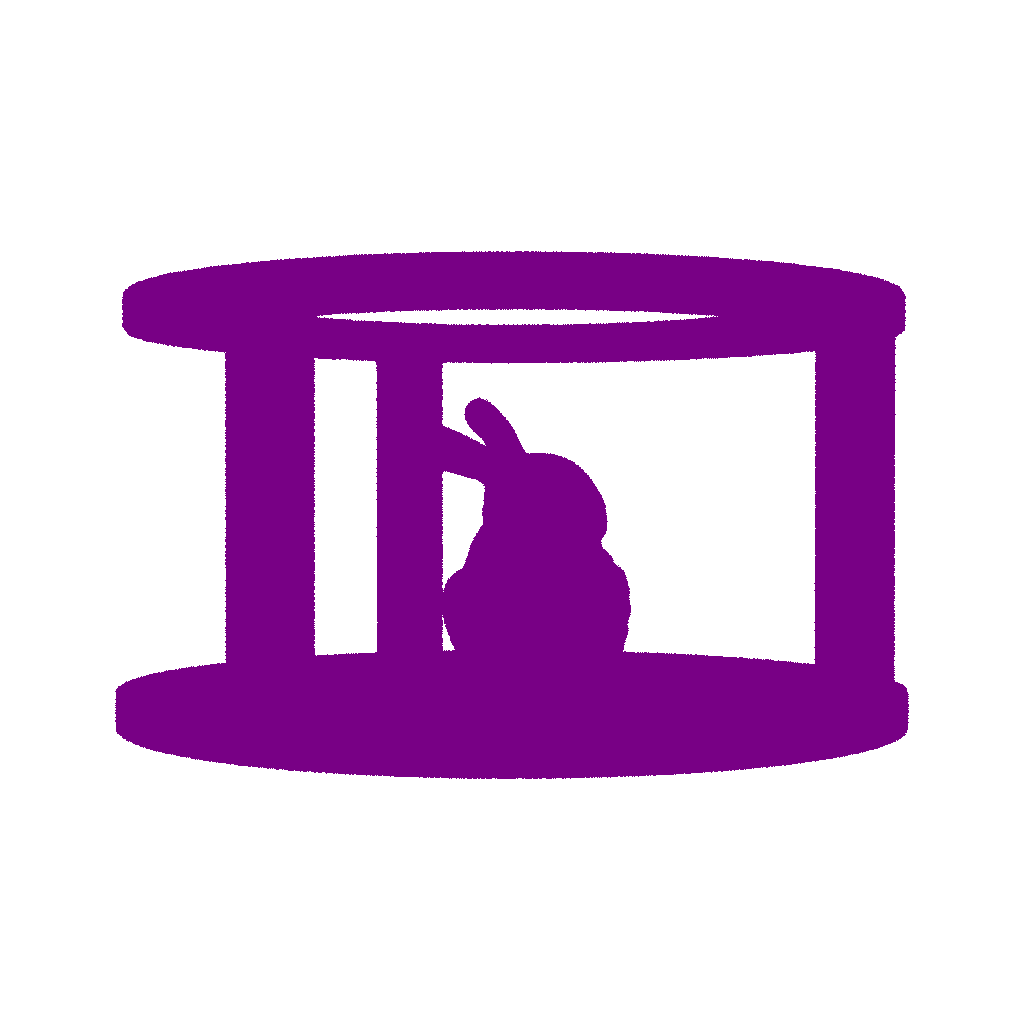}
		\\[-10pt]
		\includegraphics[width=\resLen]{images/comparison/target_1.png} &
		\includegraphics[width=\resLen]{images/comparison/init_1.png} &
		\includegraphics[width=\resLen]{images/comparison/idr_1.png} &
		\includegraphics[width=\resLen]{images/comparison/mesh_baseline_1.png} &
		\includegraphics[width=\resLen]{images/comparison/ours_implicit_1.png} &
		\includegraphics[width=\resLen]{images/comparison/ours_explicit_1.png}
		\\
		& \raisebox{35pt}{\bfseries Recon. to GT:} &
		\adjincludegraphics[width=\resLen,trim={0 {.07\height} 0 {.07\height}},clip]{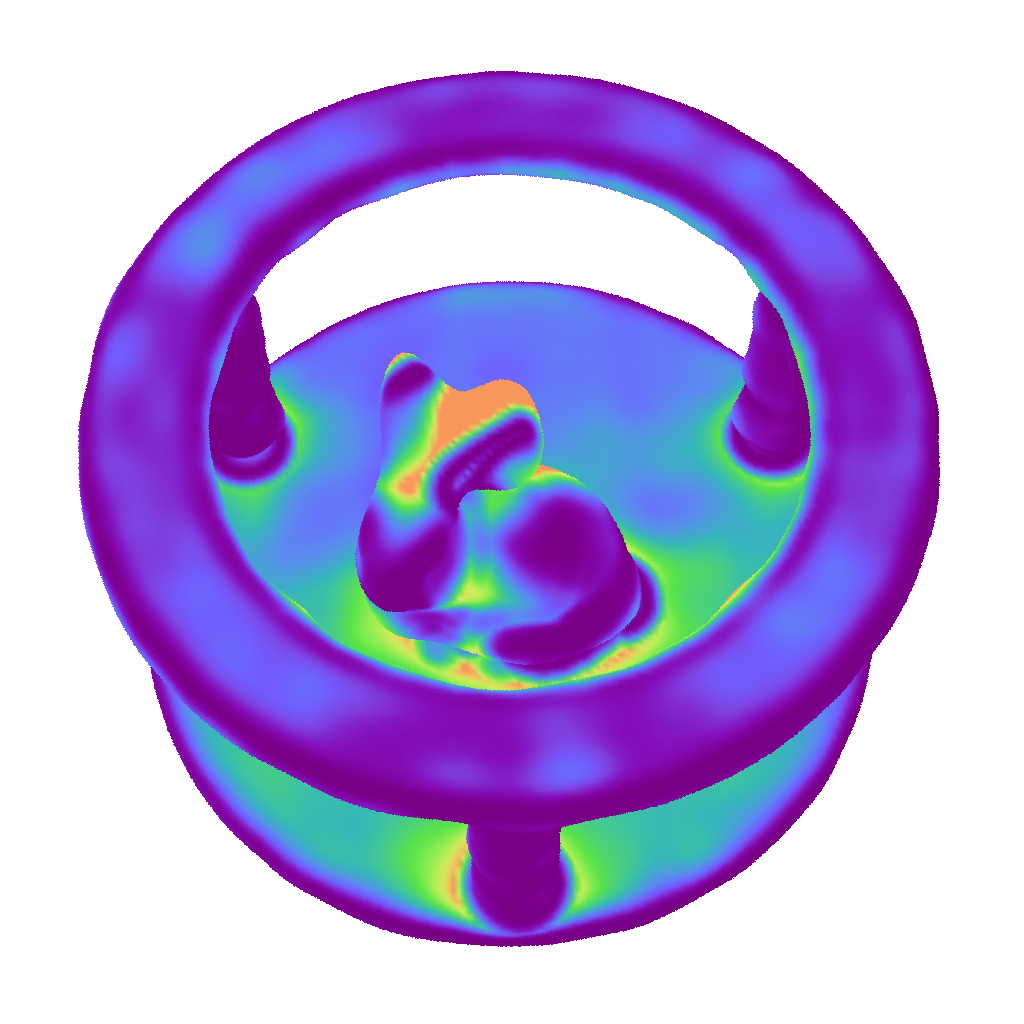} &
		\adjincludegraphics[width=\resLen,trim={0 {.07\height} 0 {.07\height}},clip]{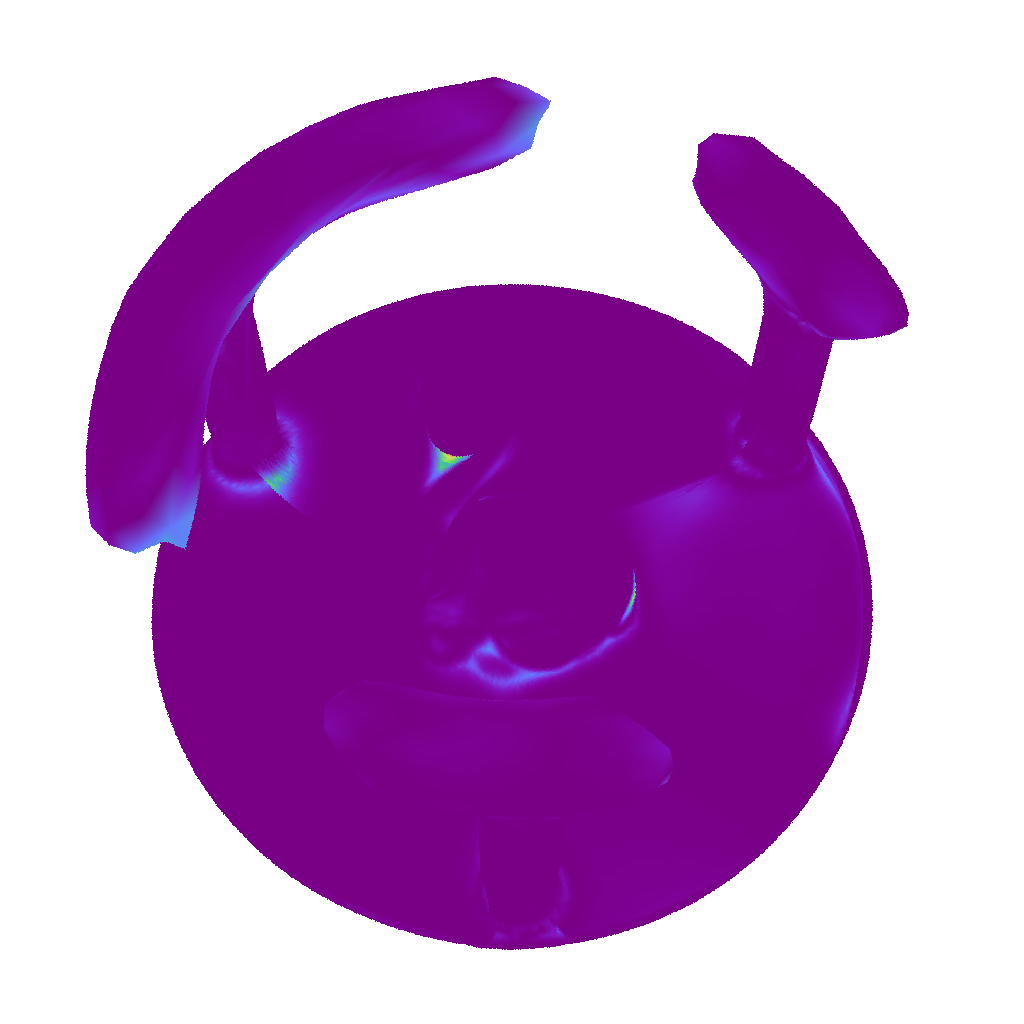} &
		\adjincludegraphics[width=\resLen,trim={0 {.07\height} 0 {.07\height}},clip]{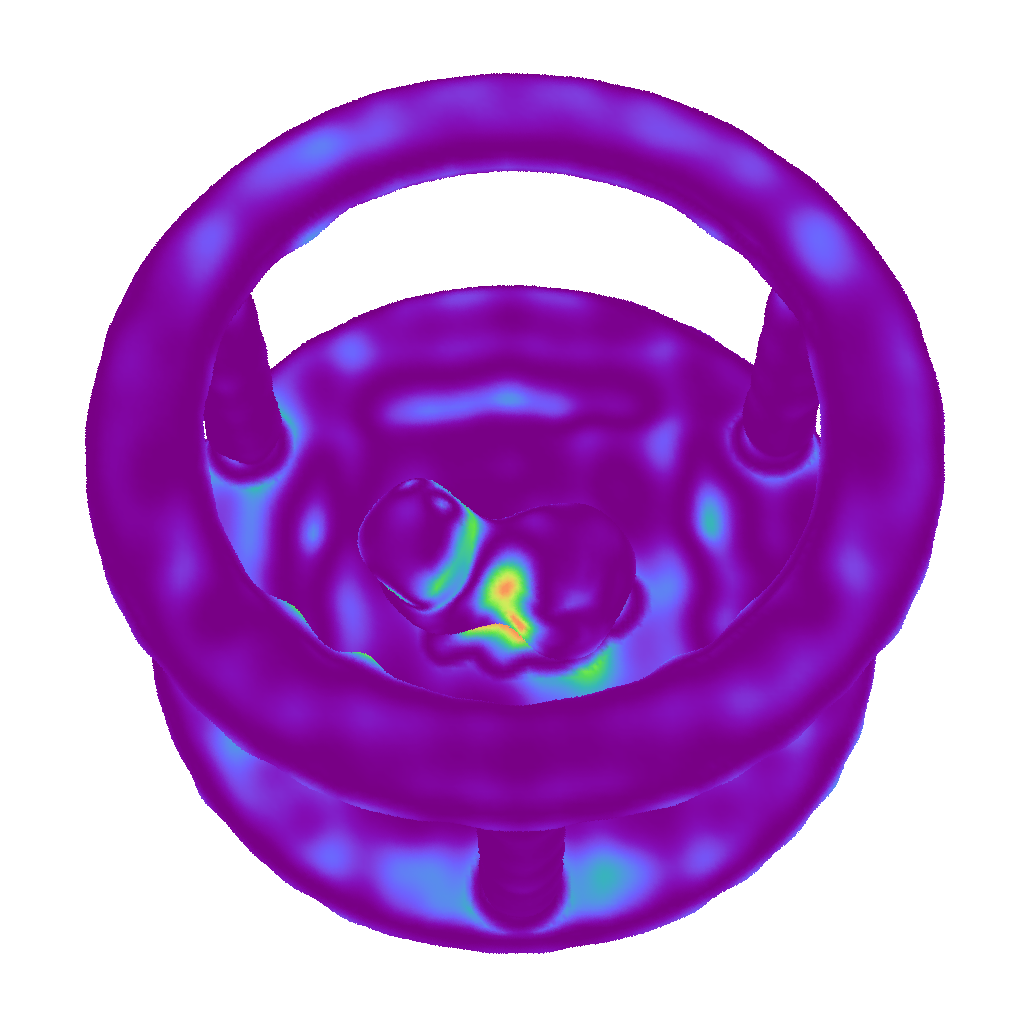} &
		\adjincludegraphics[width=\resLen,trim={0 {.07\height} 0 {.07\height}},clip]{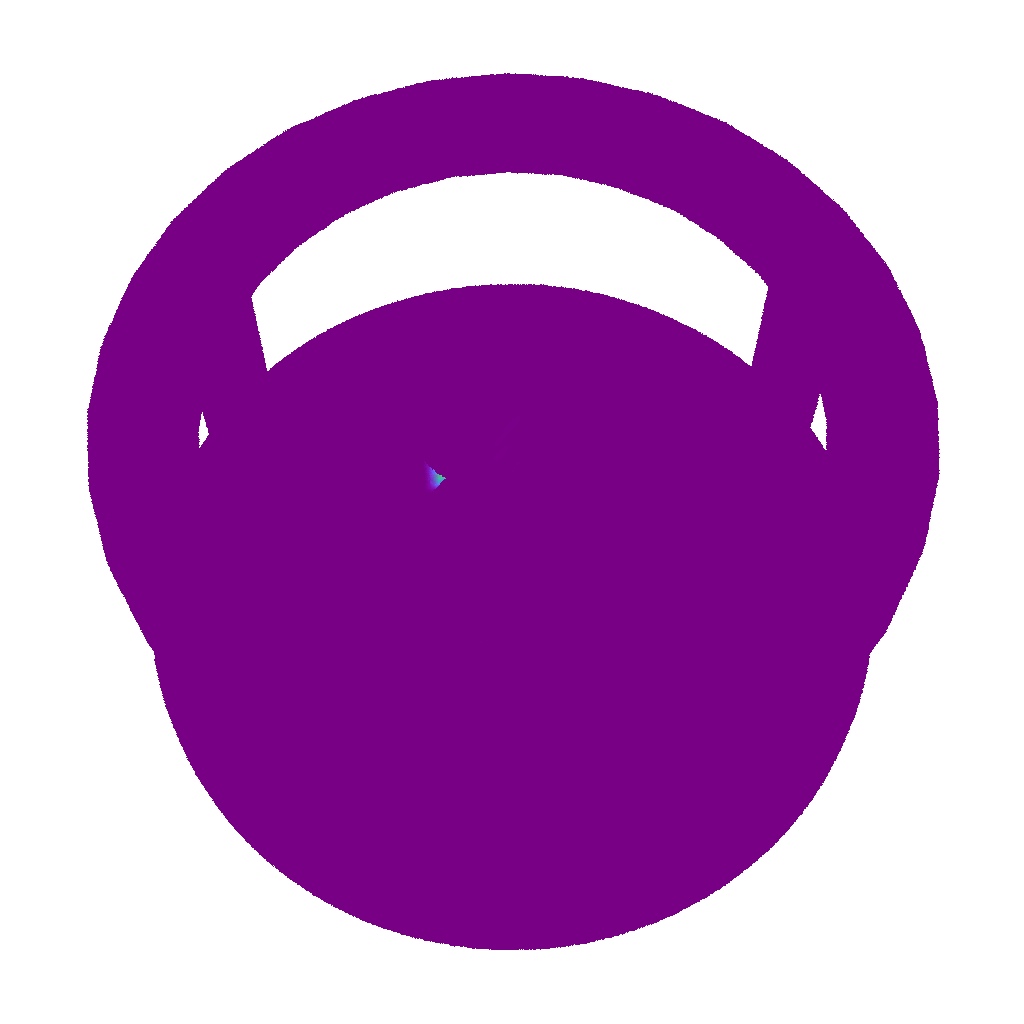}
		\\
		& \raisebox{35pt}{\bfseries GT to recon.:} &
		\adjincludegraphics[width=\resLen,trim={0 {.07\height} 0 {.07\height}},clip]{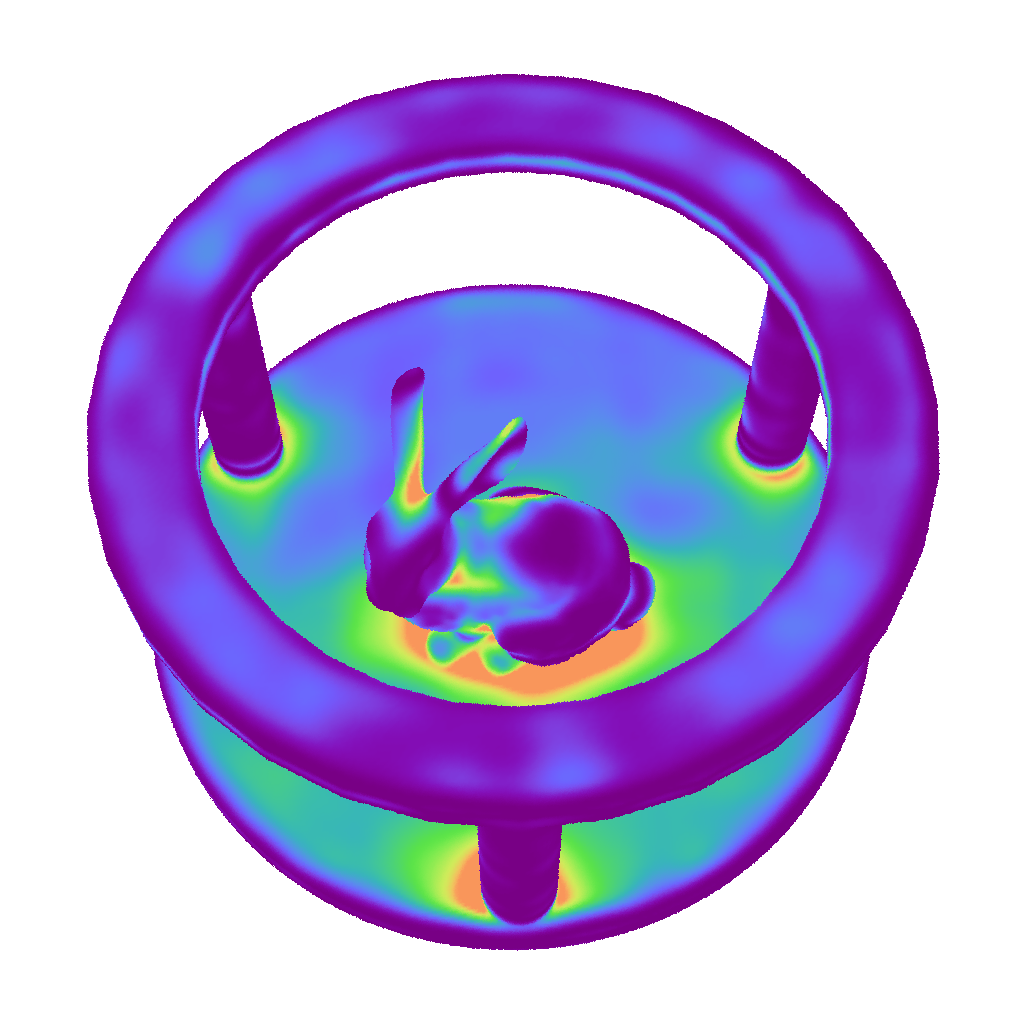} &
		\adjincludegraphics[width=\resLen,trim={0 {.07\height} 0 {.07\height}},clip]{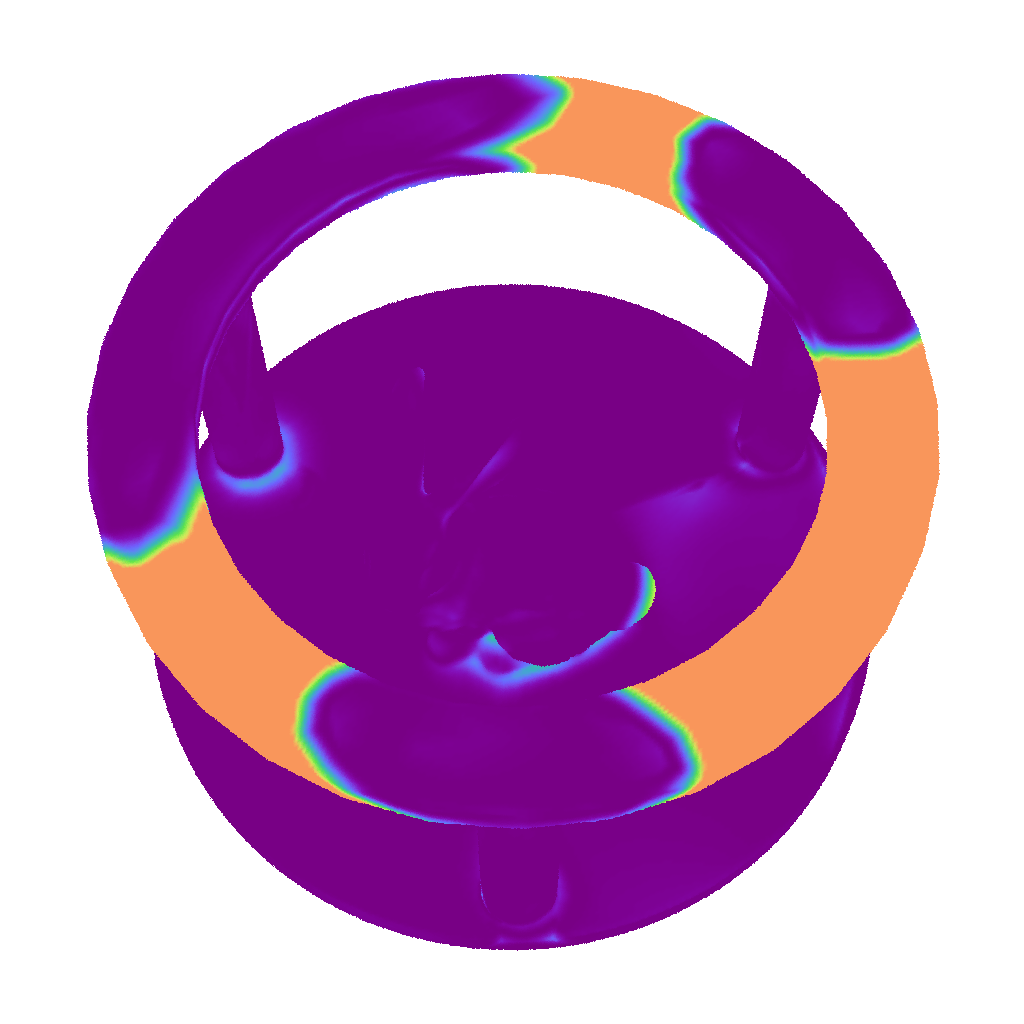} &
		\adjincludegraphics[width=\resLen,trim={0 {.07\height} 0 {.07\height}},clip]{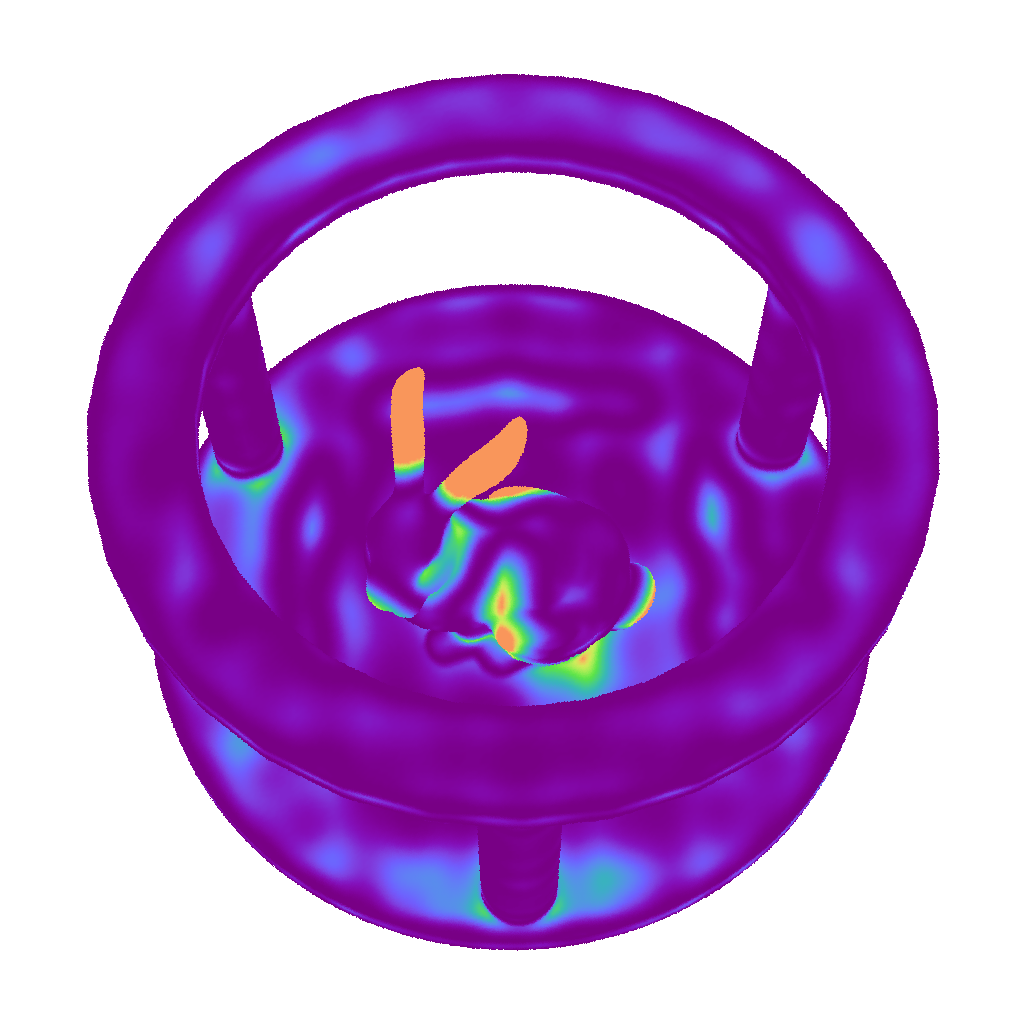} &
		\adjincludegraphics[width=\resLen,trim={0 {.07\height} 0 {.07\height}},clip]{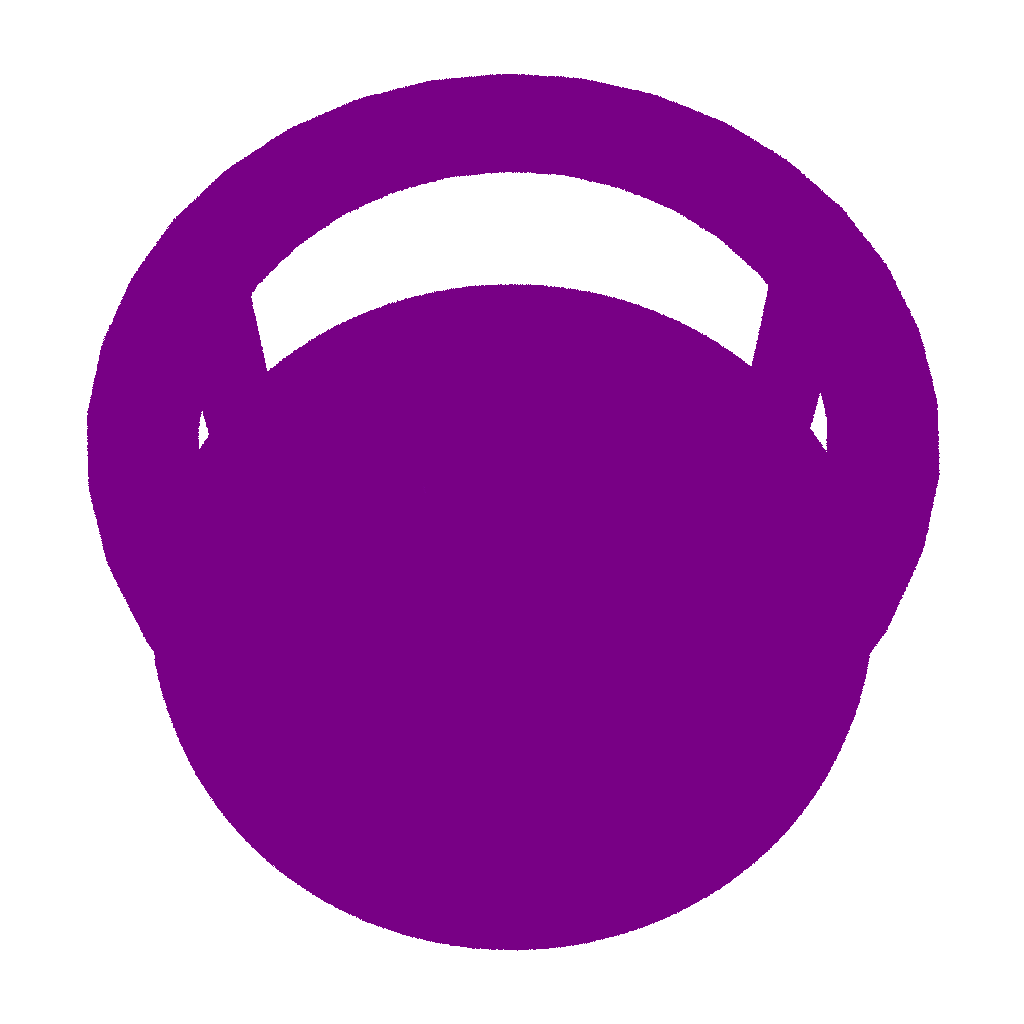}
		\\
		\textbf{Chamfer dist.:} -- & 0.3586 & 0.0191 & 0.0210 & 0.0125 & \textbf{0.0002} \\
		\textbf{Hausdorff dist.:} -- & 0.4394 & 0.1191 & 0.4812 & 0.2103 & \textbf{0.0531} \\
		\textbf{Genus:} 3 & 0 & 3 & 0 & 3 & 3 
		\\[3pt]
		& & \multicolumn{4}{c}{\clrbar{Abs. error}{0.001}}
	\end{tabular}}
	\caption{\label{fig:inv_comp1}
		\textbf{Inverse-rendering comparison (bunny temple):}
		We show reconstruction results generated using IDR*---a modified version of IDR that uses physics-based shading---in (c), mesh-based optimization in (d), our implicit stage in (e1), and our full pipeline in (e2).
		All methods shared identical initializations shown in (b).
		The number below each reconstruction result indicates the Chamfer distance between the reconstructed and groundtruth geometries (normalized so that the GT has a unit bounding box).
	}
\end{figure*}

\setlength{\resLen}{1.37in}

\begin{figure*}[t]
	\centering
	\addtolength{\tabcolsep}{-5pt}
	\small
	\rev{\begin{tabular}{rcccc}
		\multicolumn{1}{c}{(a) \textbf{Ground truth}} & (b) \textbf{Initial} & (c) \textbf{Mesh-based} & (d1) \textbf{Ours (impl.)} & (d2) \textbf{Ours (full)}
		\\[-5pt]
		\adjincludegraphics[width=\resLen,trim={{.06\height} {.06\height} {.06\height} {.06\height}},clip]{images/chair/scene1.png} &
		\adjincludegraphics[width=\resLen,trim={{.06\height} {.06\height} {.06\height} {.06\height}},clip]{images/chair/init1.png} &
		\adjincludegraphics[width=\resLen,trim={{.06\height} {.06\height} {.06\height} {.06\height}},clip]{images/chair/mesh1.png} &
		\adjincludegraphics[width=\resLen,trim={{.06\height} {.06\height} {.06\height} {.06\height}},clip]{images/chair/implicit1.png} &
		\adjincludegraphics[width=\resLen,trim={{.06\height} {.06\height} {.06\height} {.06\height}},clip]{images/chair/explicit1.png}
		\\[-15pt]
		& \raisebox{35pt}{\bfseries Recon. to GT:} &
		\adjincludegraphics[width=\resLen,trim={{.06\height} {.06\height} {.06\height} {.06\height}},clip]{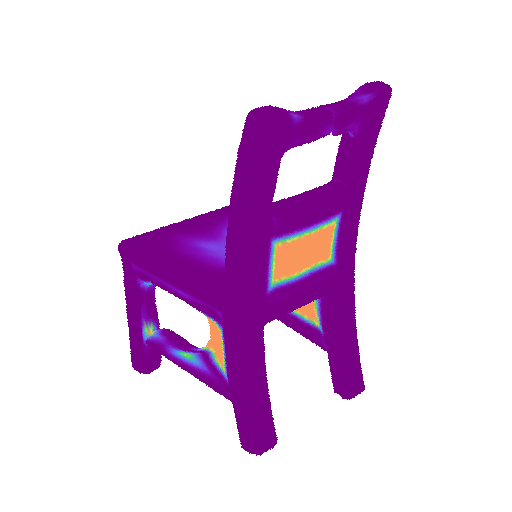} &
		\adjincludegraphics[width=\resLen,trim={{.06\height} {.06\height} {.06\height} {.06\height}},clip]{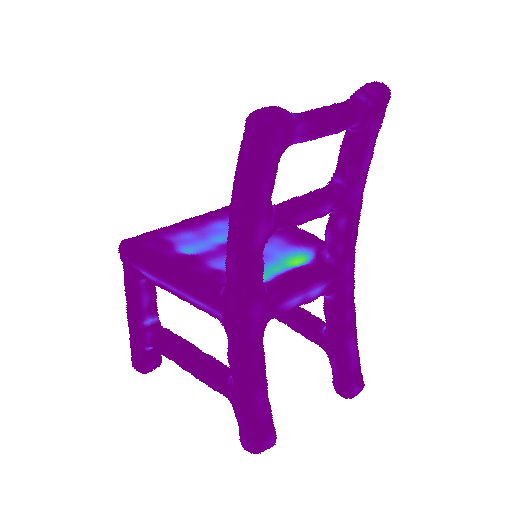} &
		\adjincludegraphics[width=\resLen,trim={{.06\height} {.06\height} {.06\height} {.06\height}},clip]{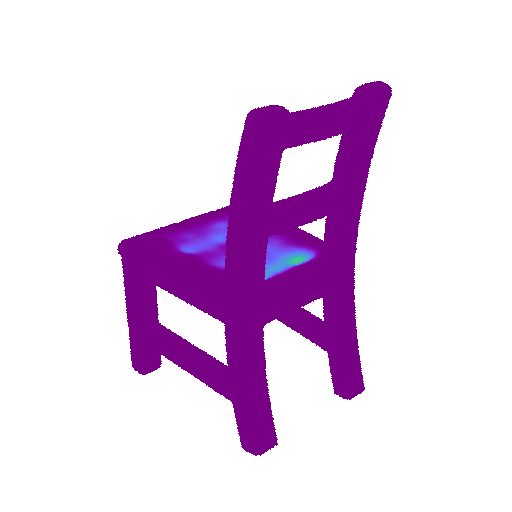}
		\\[-15pt]
		& \raisebox{35pt}{\bfseries GT. to recon.:} &
		\adjincludegraphics[width=\resLen,trim={{.06\height} {.06\height} {.06\height} {.06\height}},clip]{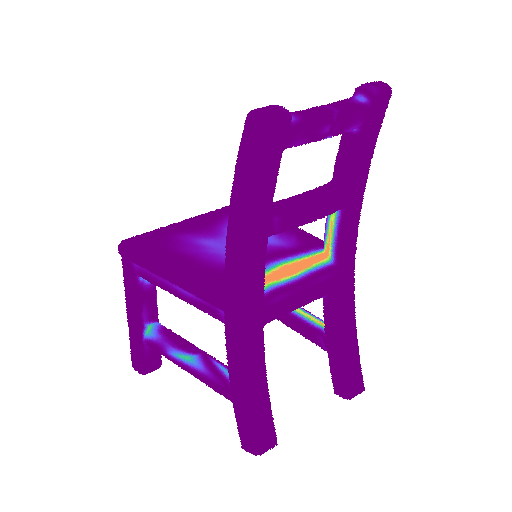} &
		\adjincludegraphics[width=\resLen,trim={{.06\height} {.06\height} {.06\height} {.06\height}},clip]{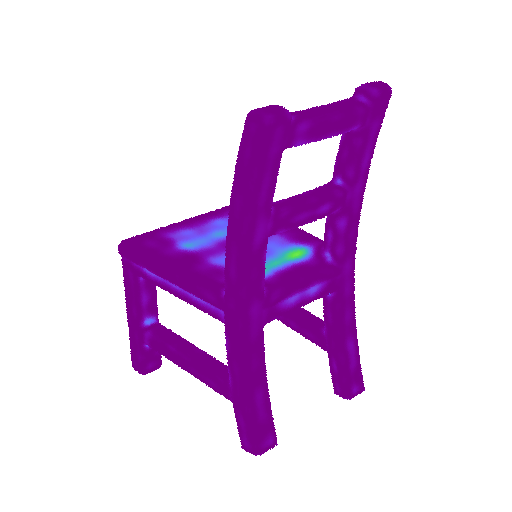} &
		\adjincludegraphics[width=\resLen,trim={{.06\height} {.06\height} {.06\height} {.06\height}},clip]{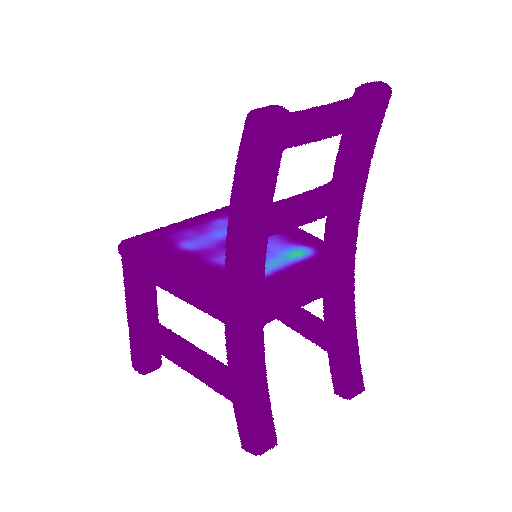}
		\\[-5pt]
		\adjincludegraphics[width=\resLen,trim={{.06\height} {.06\height} {.06\height} {.06\height}},clip]{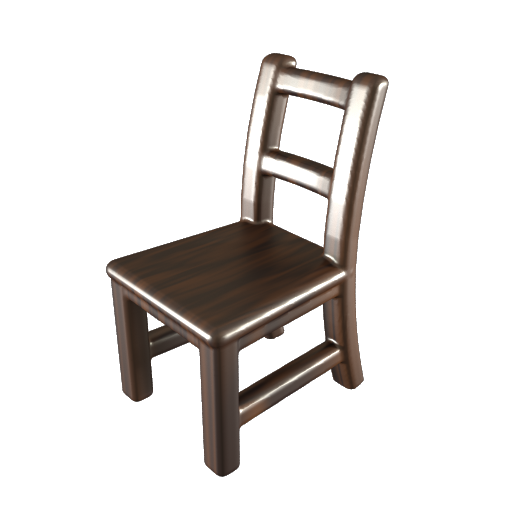} &
		\adjincludegraphics[width=\resLen,trim={{.06\height} {.06\height} {.06\height} {.06\height}},clip]{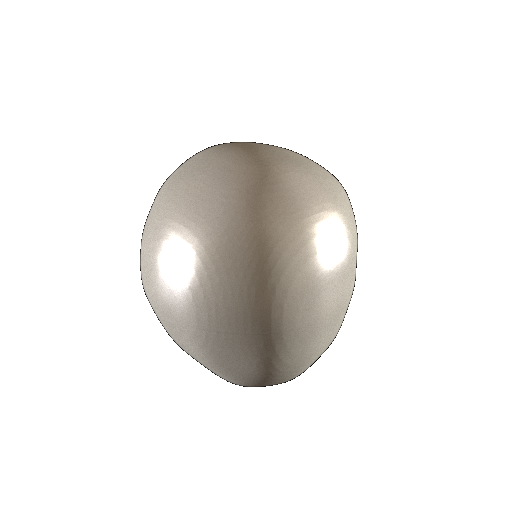} &
		\adjincludegraphics[width=\resLen,trim={{.06\height} {.06\height} {.06\height} {.06\height}},clip]{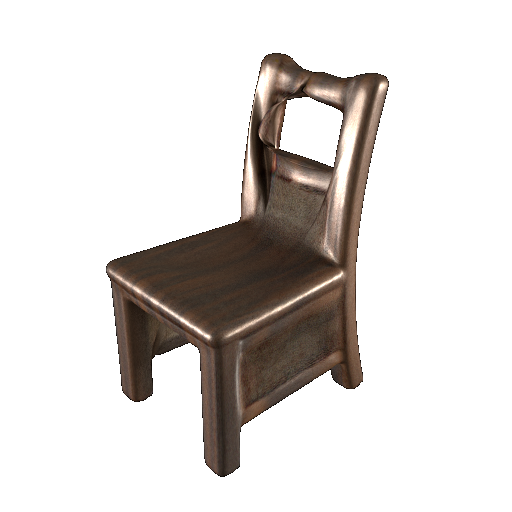} &
		\adjincludegraphics[width=\resLen,trim={{.06\height} {.06\height} {.06\height} {.06\height}},clip]{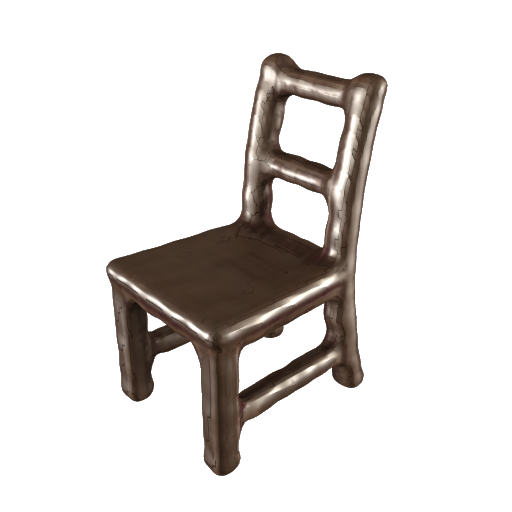} &
		\adjincludegraphics[width=\resLen,trim={{.06\height} {.06\height} {.06\height} {.06\height}},clip]{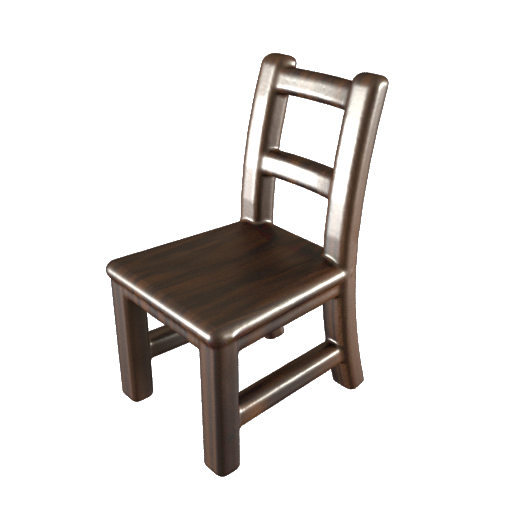}
		\\[-5pt]
		& \raisebox{40pt}{\bfseries Recon. to GT:} &
		\adjincludegraphics[width=\resLen,trim={{.06\height} {.06\height} {.06\height} {.06\height}},clip]{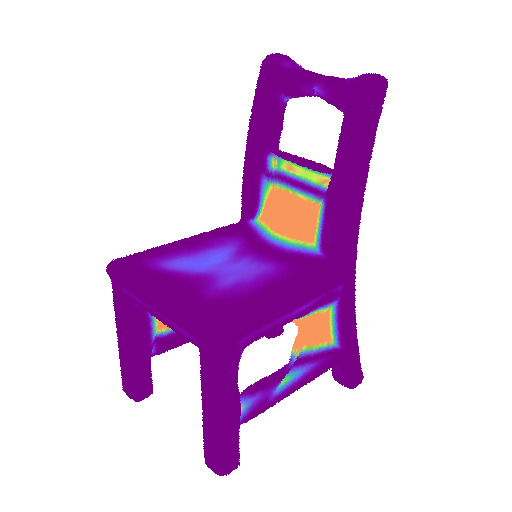} &
		\adjincludegraphics[width=\resLen,trim={{.06\height} {.06\height} {.06\height} {.06\height}},clip]{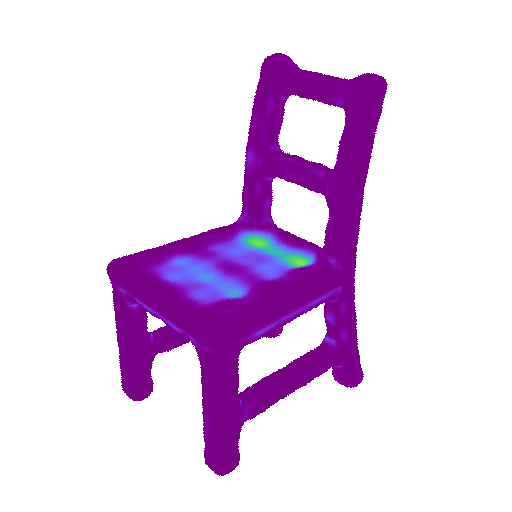} &
		\adjincludegraphics[width=\resLen,trim={{.06\height} {.06\height} {.06\height} {.06\height}},clip]{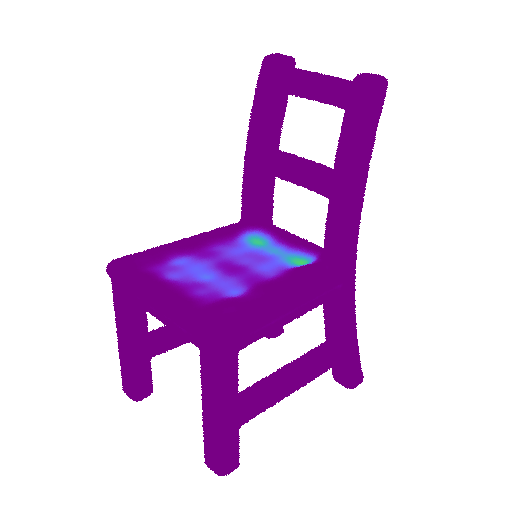}
		\\[-5pt]
		& \raisebox{40pt}{\bfseries GT. to recon.:} &
		\adjincludegraphics[width=\resLen,trim={{.06\height} {.06\height} {.06\height} {.06\height}},clip]{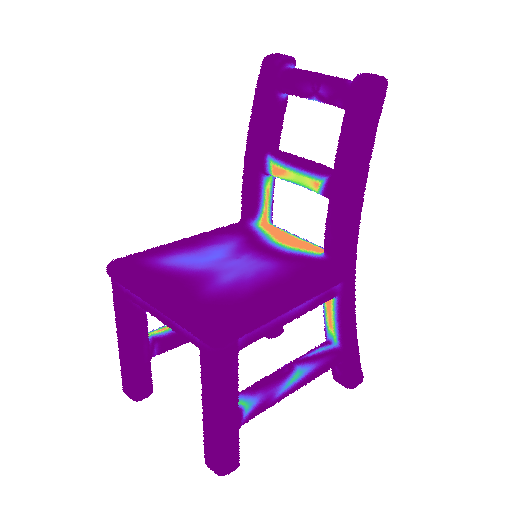} &
		\adjincludegraphics[width=\resLen,trim={{.06\height} {.06\height} {.06\height} {.06\height}},clip]{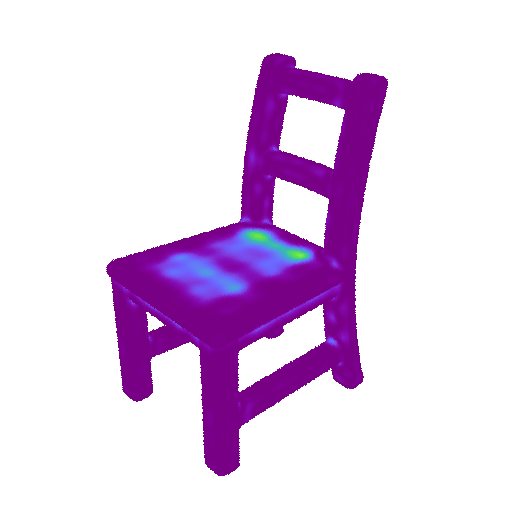} &
		\adjincludegraphics[width=\resLen,trim={{.06\height} {.06\height} {.06\height} {.06\height}},clip]{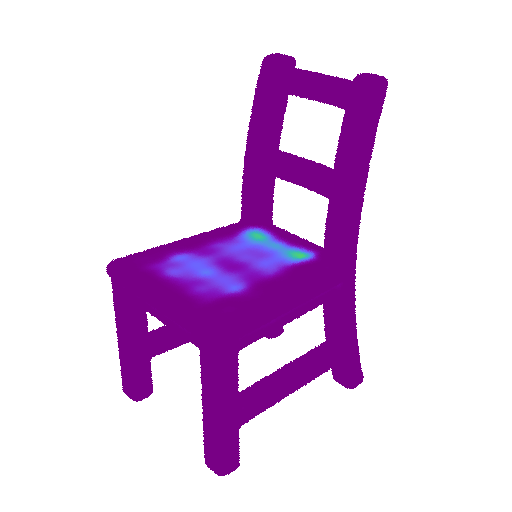}
		\\
		\textbf{Chamfer dist.:} -- & 0.2561 & 0.0173 & 0.0090 & \textbf{0.0054} \\
		\textbf{Hausdorff dist.:} -- & 0.4491 & 0.1330 & 0.0558 & \textbf{0.0551} \\
		\textbf{Genus:} 4 & 0 & 0 & 4 & 4
		\\[3pt]
		& & \multicolumn{3}{c}{\clrbar{Abs. error}{0.001}}
	\end{tabular}}
	\caption{\label{fig:inv_comp2}
		\textbf{Inverse-rendering comparison (chair):}
		We show reconstruction results generated using mesh-based optimization in (c), our implicit stage in (d1), and our full pipeline in (d2).
		All methods share identical initializations shown in (b).
	}
\end{figure*}

\setlength{\resLen}{1.5in}

\begin{figure*}[t]
	\centering
	\addtolength{\tabcolsep}{-7pt}
	\small
	\rev{\begin{tabular}{cccc}
		\multicolumn{1}{c}{(a) \textbf{Ground truth}} & (b) \textbf{Mesh-based} & (c1) \textbf{Ours (impl.)} & (c2) \textbf{Ours (full)}
		\\[-10pt]
		\adjincludegraphics[width=\resLen,trim={{.06\height} {.06\height} {.06\height} {.06\height}},clip]{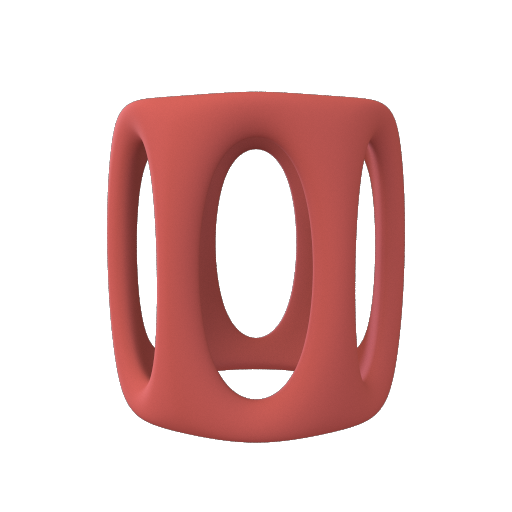} &
		\adjincludegraphics[width=\resLen,trim={{.06\height} {.06\height} {.06\height} {.06\height}},clip]{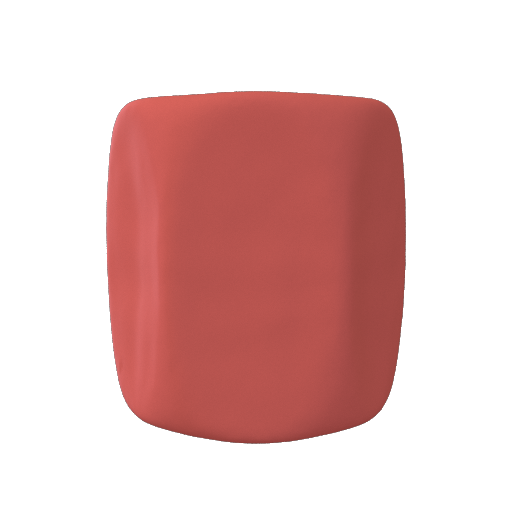} &
		\adjincludegraphics[width=\resLen,trim={{.06\height} {.06\height} {.06\height} {.06\height}},clip]{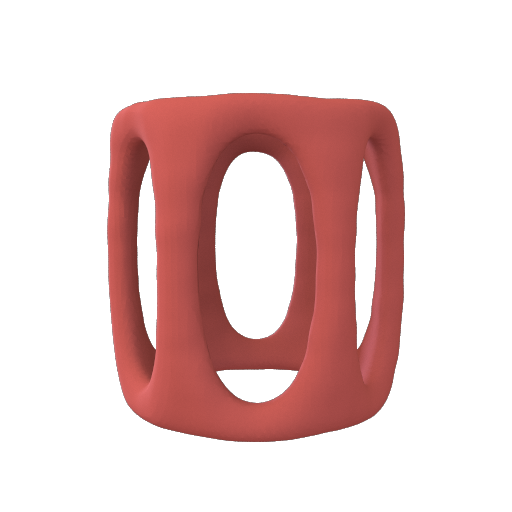} &
		\adjincludegraphics[width=\resLen,trim={{.06\height} {.06\height} {.06\height} {.06\height}},clip]{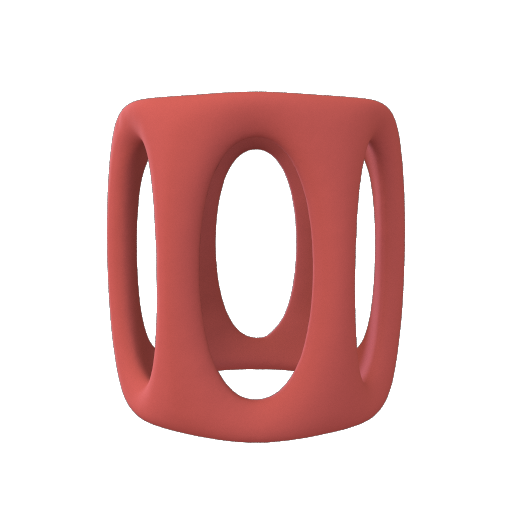} 
		\\
		\textbf{PSNR:} -- & 20.23 & 31.03 & 47.98 
		\\
		\adjincludegraphics[width=\resLen,trim={{.06\height} {.06\height} {.06\height} {.06\height}},clip]{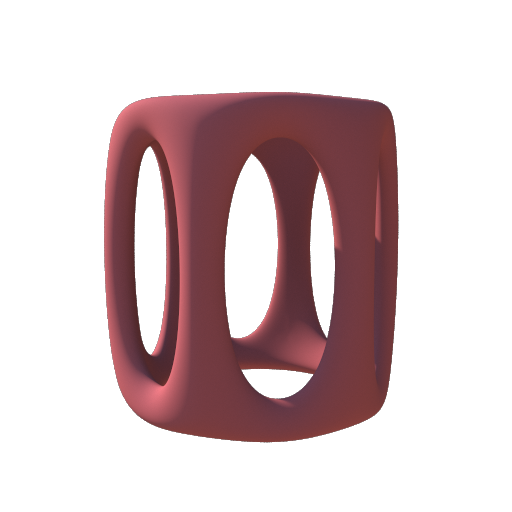} &
		\adjincludegraphics[width=\resLen,trim={{.06\height} {.06\height} {.06\height} {.06\height}},clip]{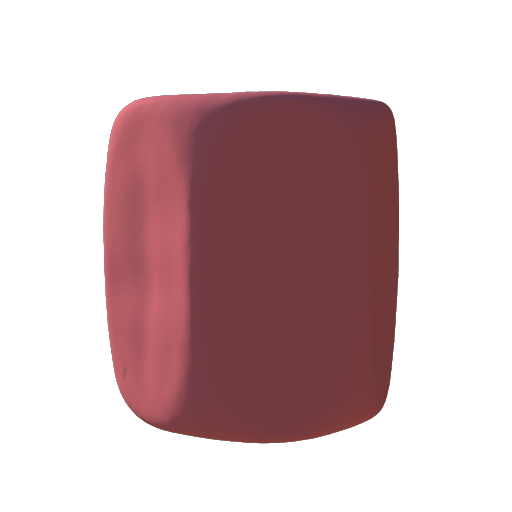} &
		\adjincludegraphics[width=\resLen,trim={{.06\height} {.06\height} {.06\height} {.06\height}},clip]{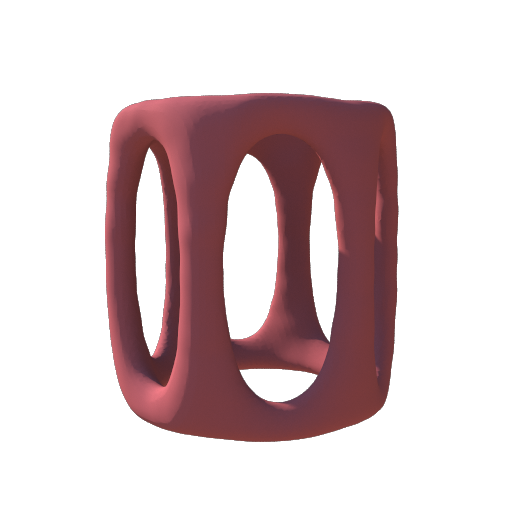} &
		\adjincludegraphics[width=\resLen,trim={{.06\height} {.06\height} {.06\height} {.06\height}},clip]{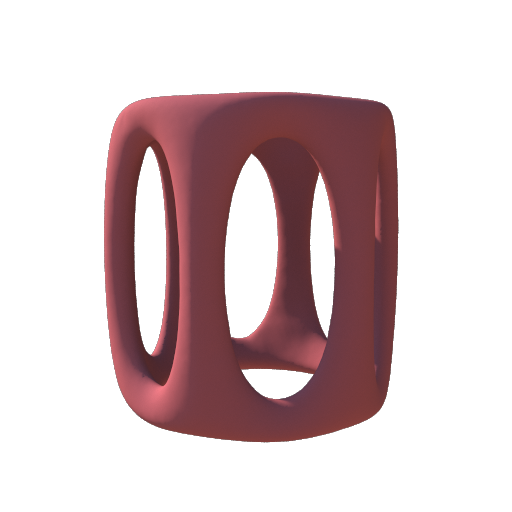} 
		\\
		\textbf{PSNR:} -- & 23.44 & 30.99 & 45.45 \\ \\
		\textbf{Chamfer dist.:} -- & 0.0661 & 0.0068 & 0.0008 \\
		\textbf{Hausdorff dist.:} -- & 0.4054 & 0.0222 & 0.0138 \\
		\textbf{Genus:} 7 & 0 & 7 & 7 \\
	\end{tabular}}
\caption{\label{fig:inv_comp3}
		\textbf{Inverse-rendering comparison (frame):}
		We show reconstruction results generated using mesh-based optimization in (b), our implicit stage in (c1), and our full pipeline in (c2). All methods share identical initializations similar to Figure~\protect\ref{fig:inv_comp2}-b.
	}
\end{figure*}

\setlength{\resLen}{1.7in}

\begin{figure*}[t]
	\centering
	\addtolength{\tabcolsep}{-7pt}
	\small
	\rev{\begin{tabular}{cccc}
		(a) \textbf{Ground truth} & (b) \textbf{Mesh-based} & (c1) \textbf{Ours (impl.)} & (c2) \textbf{Ours (full)} \\
		\adjincludegraphics[width=\resLen,trim={{.06\height} {.06\height} {.06\height} {.06\height}},clip]{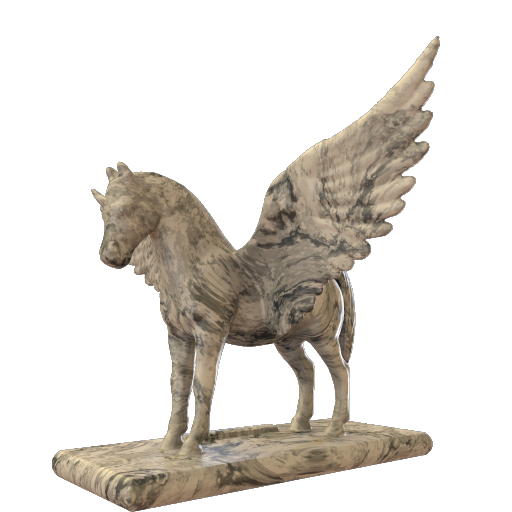} &
		\adjincludegraphics[width=\resLen,trim={{.06\height} {.06\height} {.06\height} {.06\height}},clip]{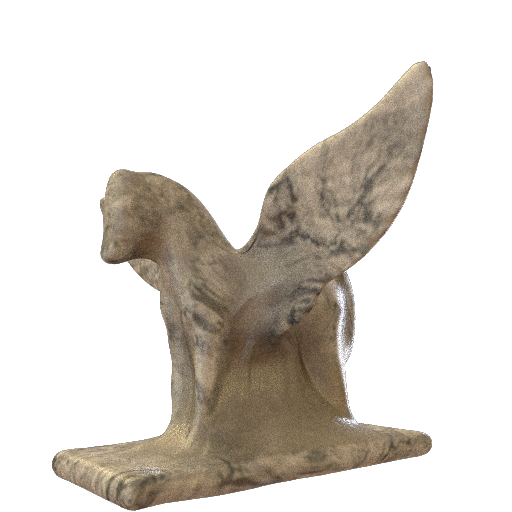} &
		\adjincludegraphics[width=\resLen,trim={{.06\height} {.06\height} {.06\height} {.06\height}},clip]{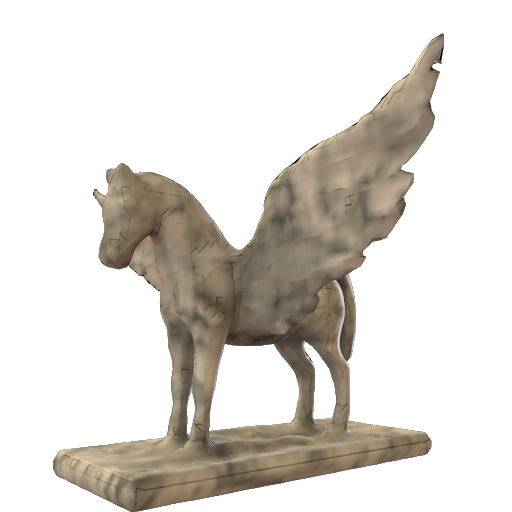} &
		\adjincludegraphics[width=\resLen,trim={{.06\height} {.06\height} {.06\height} {.06\height}},clip]{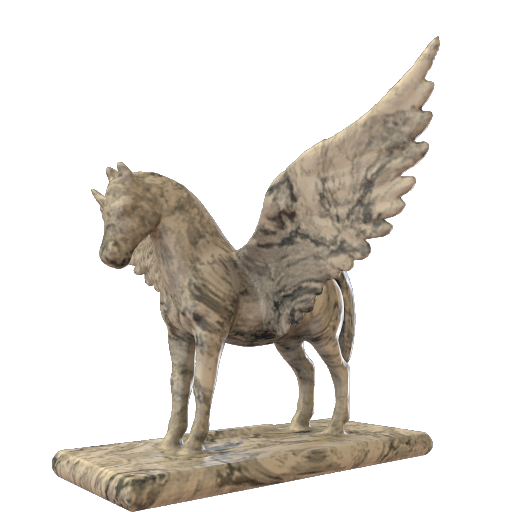} 
		\\
		\textbf{PSNR:} -- & 23.42 & 24.76 & 32.69 
		\\
		\adjincludegraphics[width=\resLen,trim={{.06\height} {.06\height} {.06\height} {.06\height}},clip]{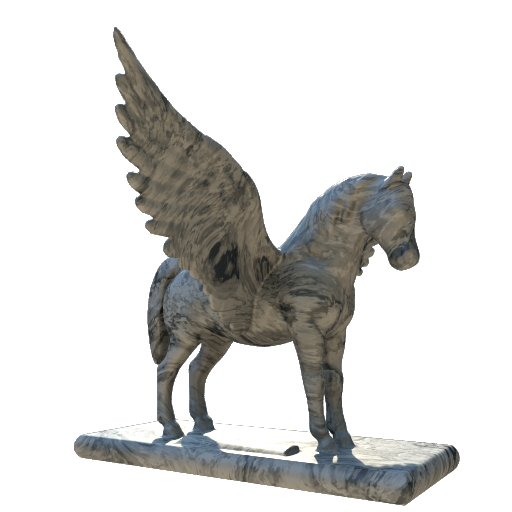} &
		\adjincludegraphics[width=\resLen,trim={{.06\height} {.06\height} {.06\height} {.06\height}},clip]{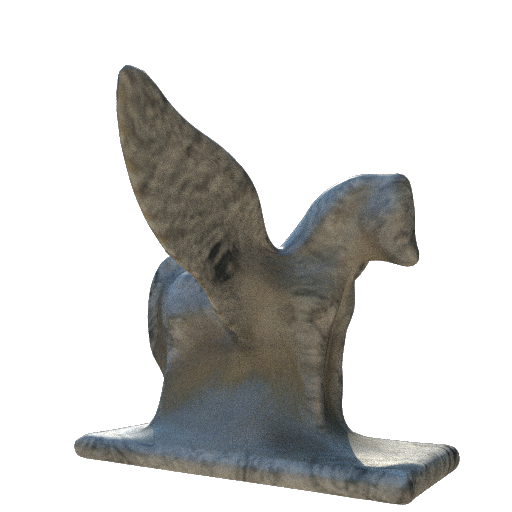} &
		\adjincludegraphics[width=\resLen,trim={{.06\height} {.06\height} {.06\height} {.06\height}},clip]{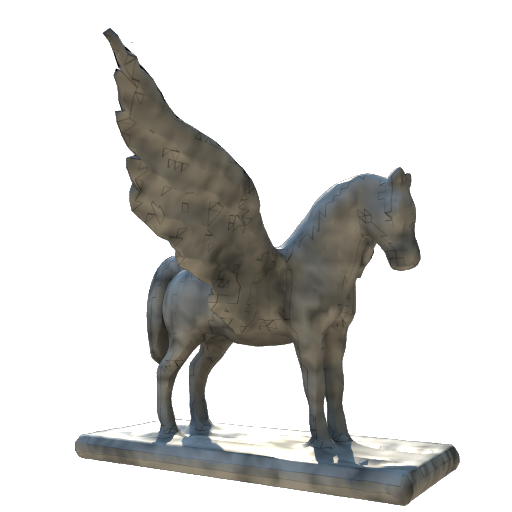} &
		\adjincludegraphics[width=\resLen,trim={{.06\height} {.06\height} {.06\height} {.06\height}},clip]{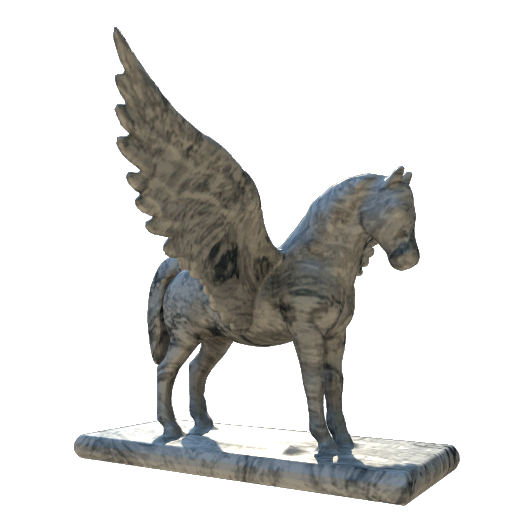} 
		\\
		\textbf{PSNR:} -- & 14.69 & 14.41 & 22.43 \\[5pt]
		\textbf{Chamfer dist.:} -- & 0.0153 & 0.0051 & 0.0009 \\
		\textbf{Hausdorff dist.:} -- & 0.1587 & 0.0572 & 0.0403 \\
		\textbf{Genus:} 3 & 0 & 3 & 3
		\\[5pt]
		\hline
		\\[-2pt]
		\raisebox{30pt}{\textbf{Our recon. maps:}} & \includegraphics[width=0.5\resLen]{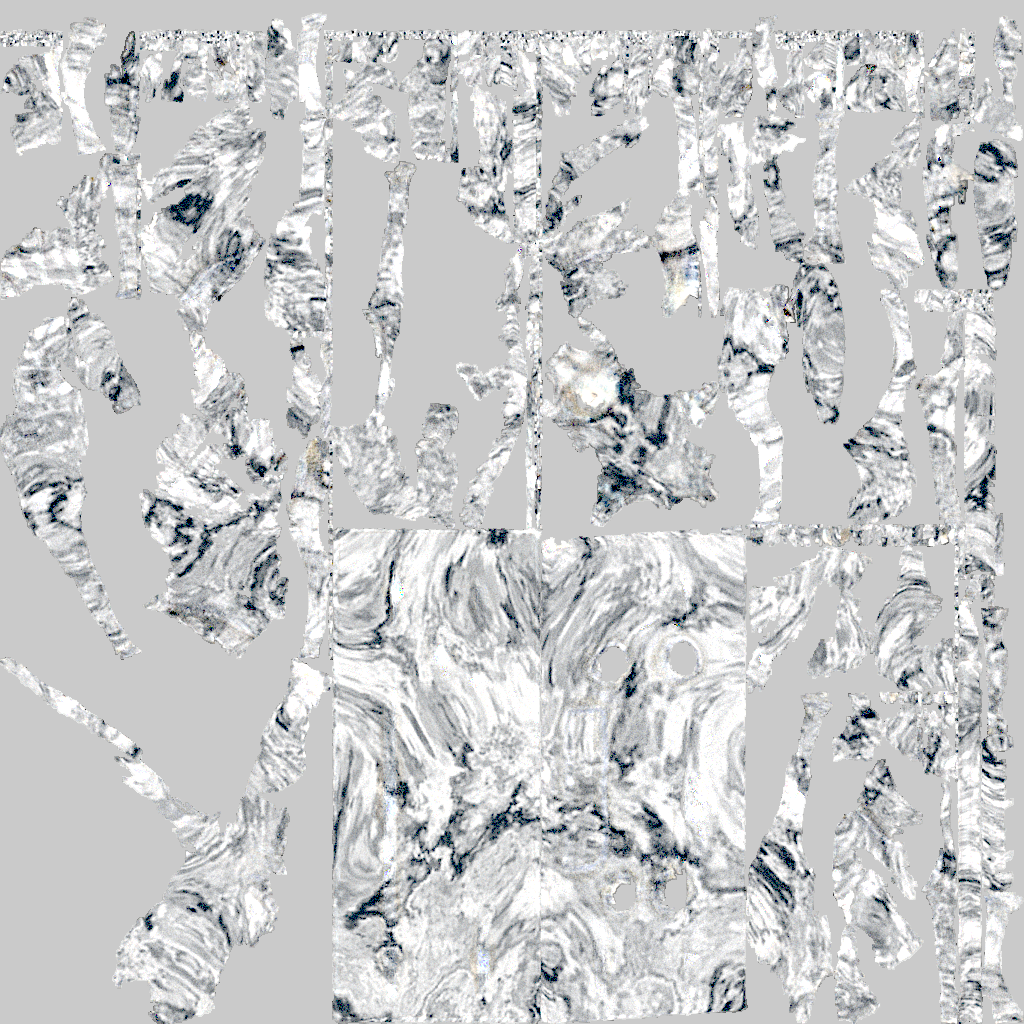} & \raisebox{30pt}{--} & \raisebox{30pt}{--}\\
		& diffuse & specular & roughness
	\end{tabular}}
	\caption{\label{fig:inv_comp4}
		\textbf{Inverse-rendering comparison (Pegasus):}
		We show reconstruction results generated using mesh-based optimization in (b), our implicit stage in (c1), and our full pipeline in (c2). All methods share identical initializations similar to Figure~\protect\ref{fig:inv_comp2}-b.
	}
\end{figure*}

\setlength{\resLen}{1.4in}

\begin{figure*}[t]
	\centering
	\addtolength{\tabcolsep}{-7pt}
	\small
	\rev{\begin{tabular}{cccc}
		(a) \textbf{Ground truth} & (b) \textbf{Mesh-based} & (c1) \textbf{Ours (impl.)} & (c2) \textbf{Ours (full)} \\
		\adjincludegraphics[width=\resLen,trim={{.1\width} {.08\height}  {.1\width} {.35\height}},clip]{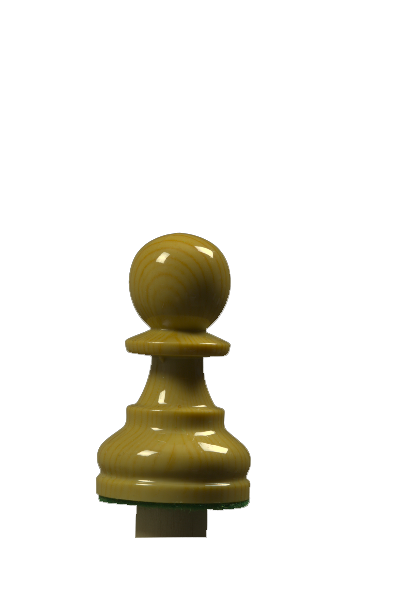} &
		\adjincludegraphics[width=\resLen,trim={{.1\width} {.08\height} {.1\width} {.35\height}},clip]{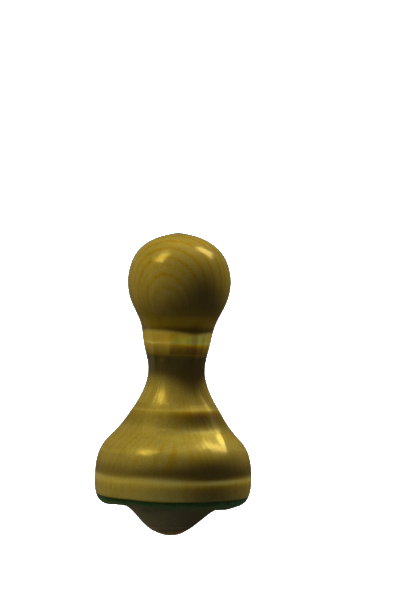} &
		\adjincludegraphics[width=\resLen,trim={{.1\width} {.08\height} {.1\width} {.35\height}},clip]{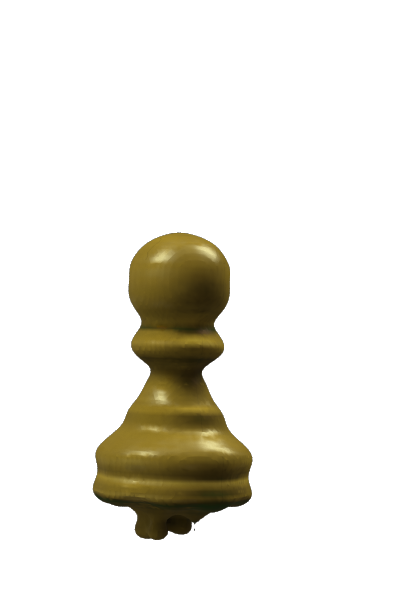} &
		\adjincludegraphics[width=\resLen,trim={{.1\width} {.08\height} {.1\width} {.35\height}},clip]{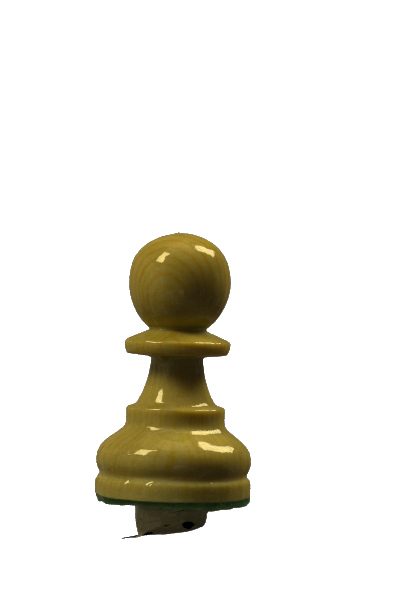} 
		\\
		\textbf{PSNR}: -- & 26.32 & 26.61 & 27.76
		\\[5pt]
		\adjincludegraphics[width=\resLen,trim={{.05\width} {0.15\height} {.05\width} {.25\height}},clip]{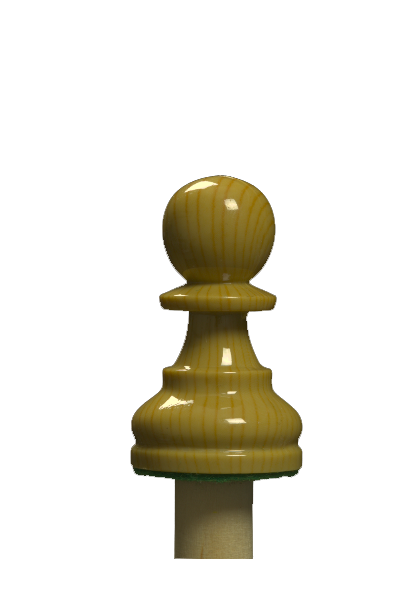} &
		\adjincludegraphics[width=\resLen,trim={{.05\width} {0.15\height} {.05\width} {.25\height}},clip]{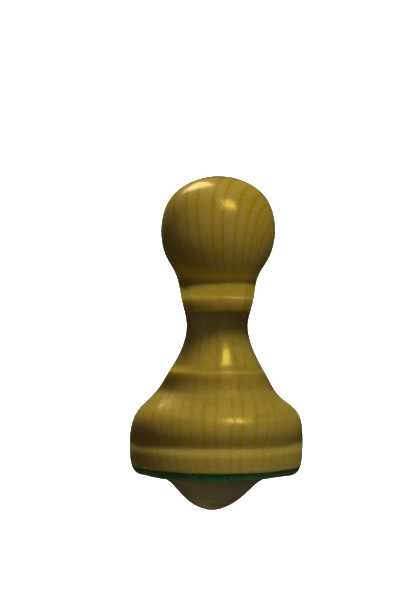} &
		\adjincludegraphics[width=\resLen,trim={{.05\width} {0.15\height} {.05\width} {.25\height}},clip]{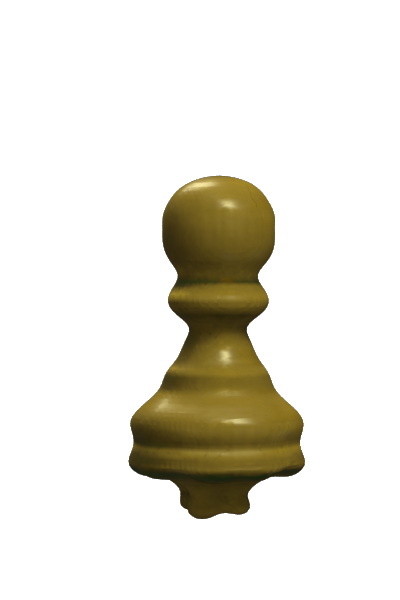} &
		\adjincludegraphics[width=\resLen,trim={{.05\width} {0.15\height} {.05\width} {.25\height}},clip]{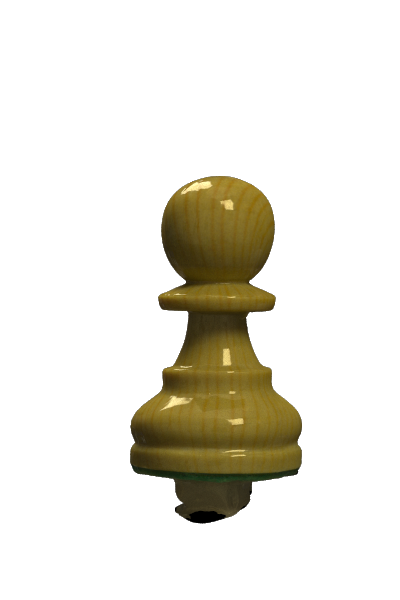} 
		\\
		\textbf{PSNR}: -- & 27.12 & 28.86 & 29.35 \\ \\ 
		\textbf{Genus}: -- & 0 & 0 & 0 \\
		\\[5pt]
		\hline
		\\[-2pt]
		\raisebox{30pt}{\textbf{Our recon. maps:}} & 
		\includegraphics[width=0.5\resLen]{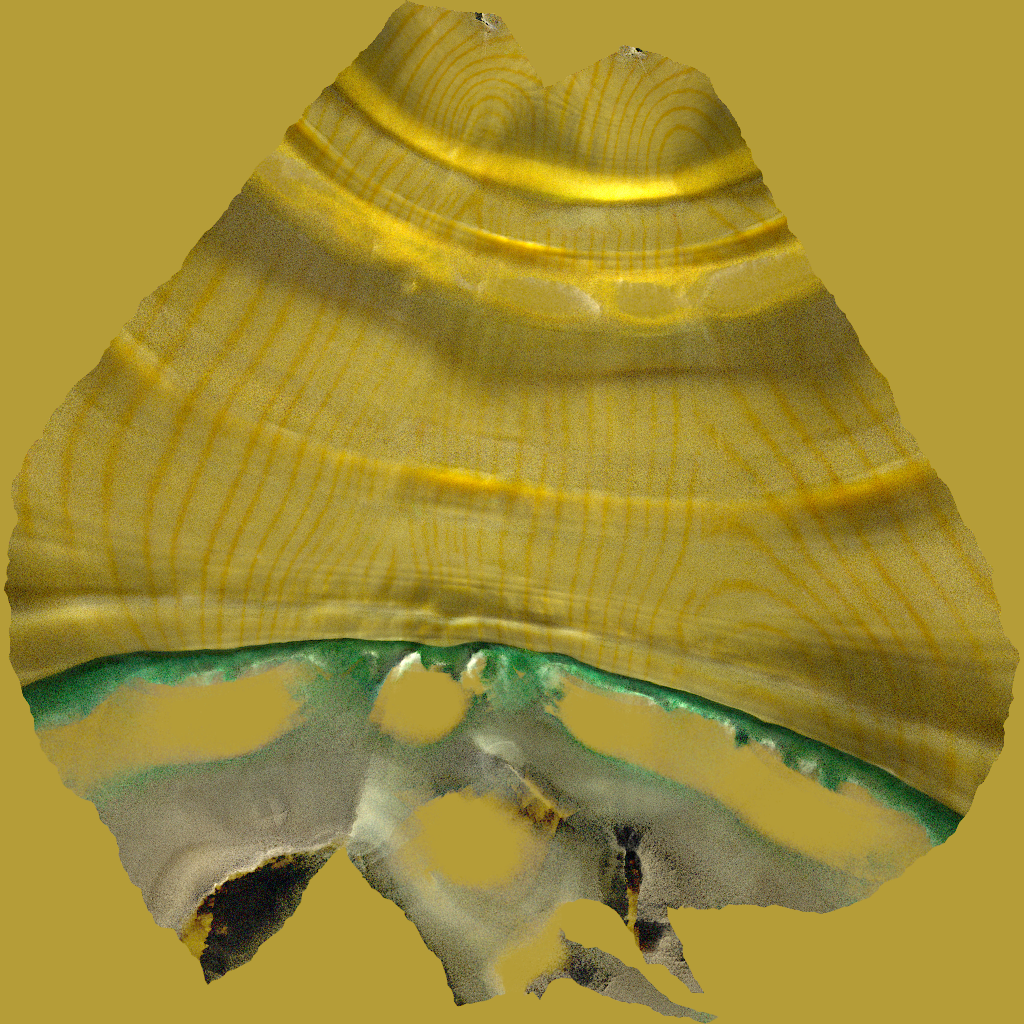} & 
		\includegraphics[width=0.5\resLen]{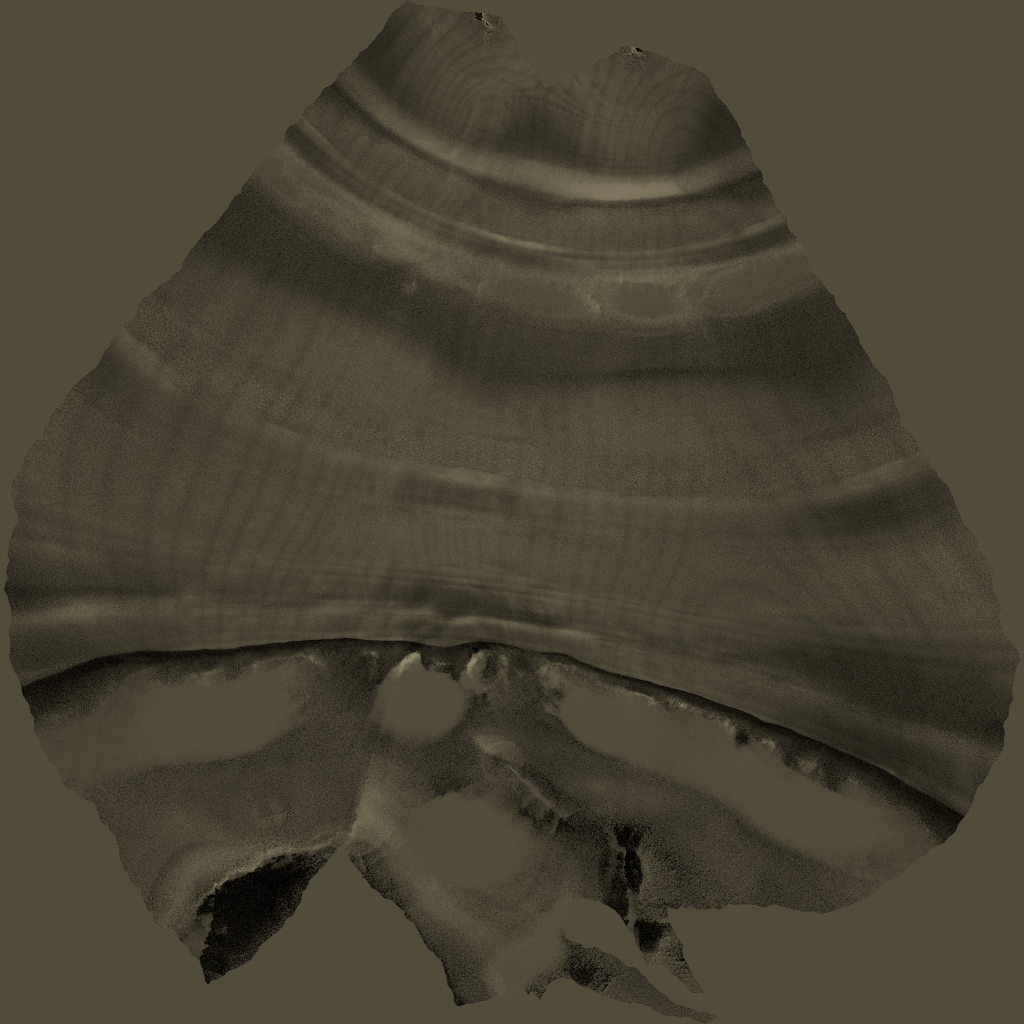} & 
		\includegraphics[width=0.5\resLen]{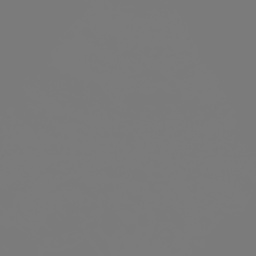} \\
		& diffuse & specular & roughness
	\end{tabular}}
	\caption{\label{fig:inv_comp5}
		\textbf{Inverse-rendering comparison (chess):}
		We show reconstruction results generated using mesh-based optimization in (b), our implicit stage in (c1), and our full pipeline in (c2). All methods share identical initializations similar to Figure~\protect\ref{fig:inv_comp2}-b.
	}
\end{figure*}

\setlength{\resLen}{1.7in}

\begin{figure*}[t]
	\centering
	\addtolength{\tabcolsep}{-7pt}
	\small
	\rev{\begin{tabular}{cccc}
		(a) \textbf{Ground truth} & (b) \textbf{Mesh-based} & (c1) \textbf{Ours (impl.)} & (c2) \textbf{Ours (full)} \\
		\adjincludegraphics[width=\resLen,trim={{0.1\width} {.01\height} {.1\width} {.2\height}},clip]{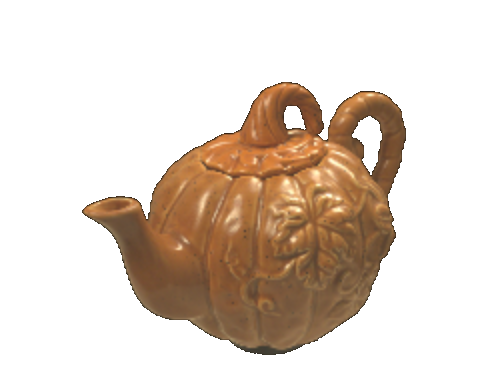} &
		\adjincludegraphics[width=\resLen,trim={{0.1\width} {.01\height} {.1\width} {.2\height}},clip]{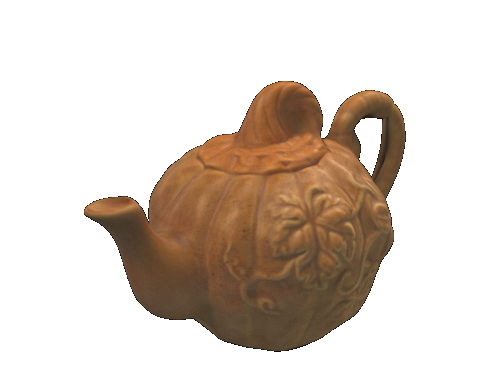} &
		\adjincludegraphics[width=\resLen,trim={{0.1\width} {.01\height} {.1\width} {.2\height}},clip]{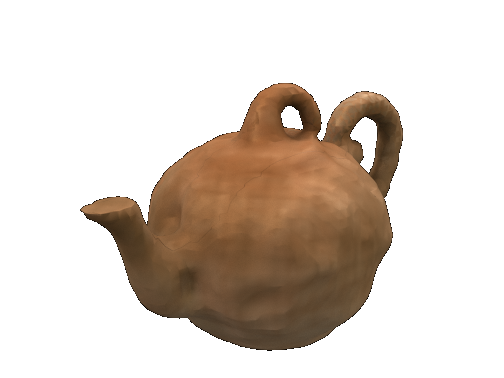} &
		\adjincludegraphics[width=\resLen,trim={{0.1\width} {.01\height} {.1\width} {.2\height}},clip]{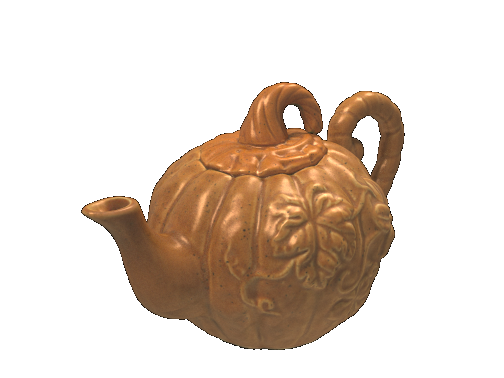} 
		\\[-5pt]
		\textbf{PSNR}: - & 25.56 & 24.53 & 29.45
		\\[5pt]
		\adjincludegraphics[width=\resLen,trim={0 0 0 {.2\height}},clip]{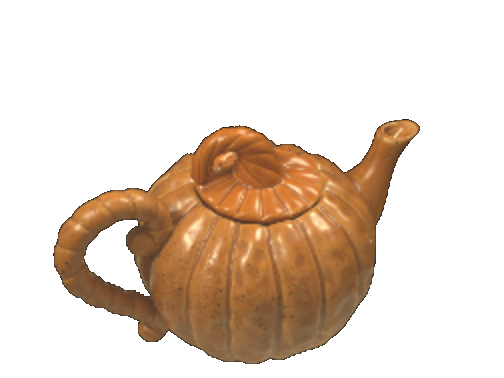} &
		\adjincludegraphics[width=\resLen,trim={0 0 0 {.2\height}},clip]{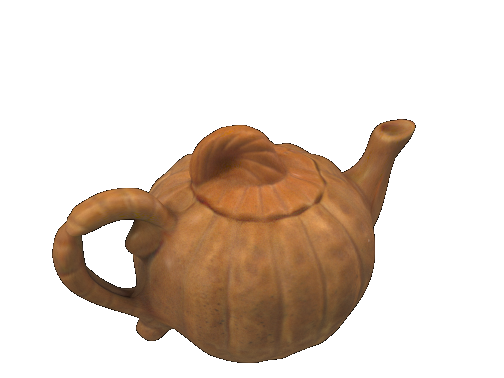} &
		\adjincludegraphics[width=\resLen,trim={0 0 0 {.2\height}},clip]{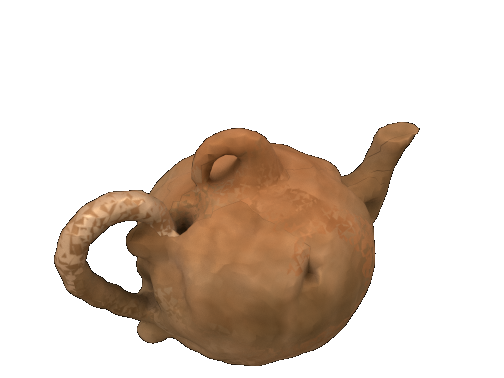} &
		\adjincludegraphics[width=\resLen,trim={0 0 0 {.2\height}},clip]{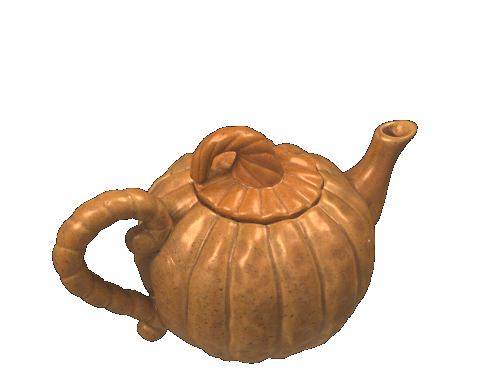} 
		\\
		\textbf{PSNR}: -- & 24.66 & 23.74 & 29.15 \\ \\ 
		\textbf{Genus}: -- & 0 & 2 & 2 \\
		\\[5pt]
		\hline
		\\[-2pt]
		\raisebox{30pt}{\textbf{Our recon. maps:}} & 
		\includegraphics[width=0.5\resLen]{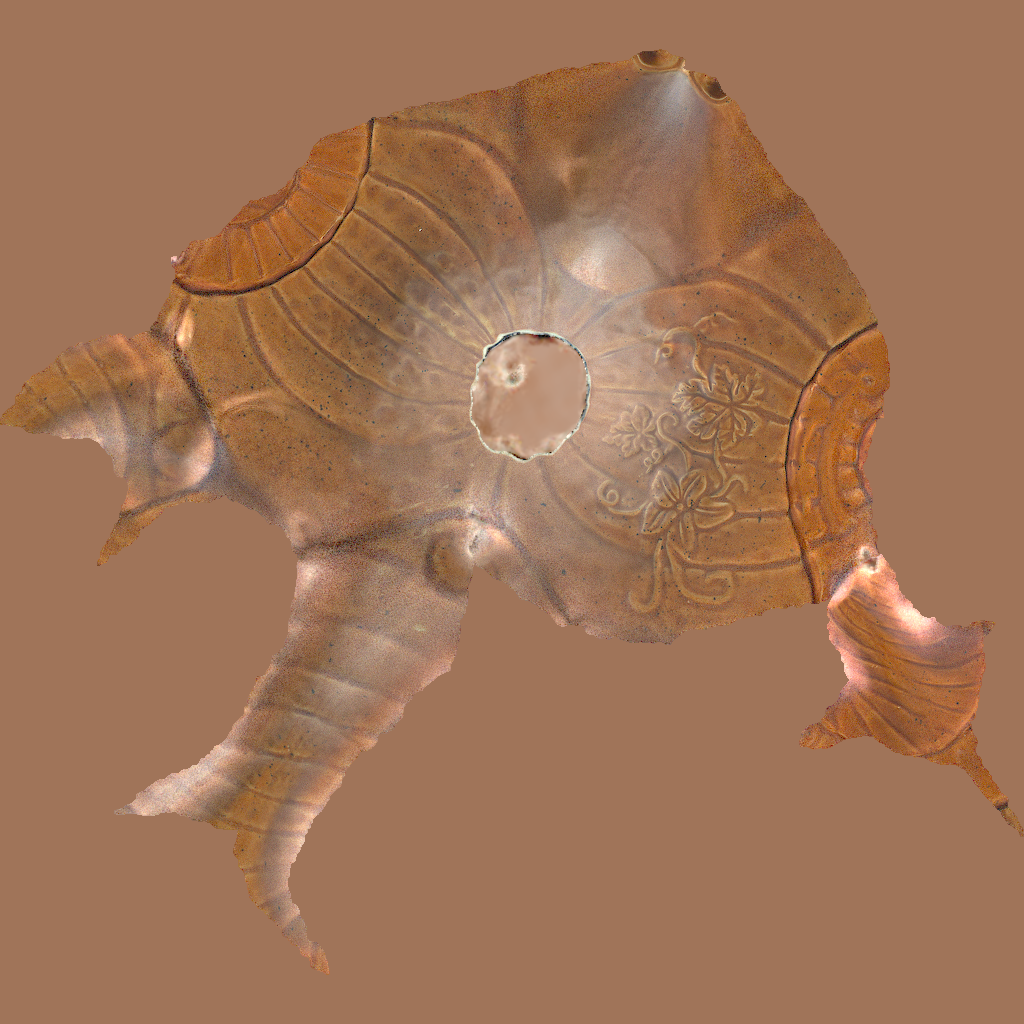} & 
		\includegraphics[width=0.5\resLen]{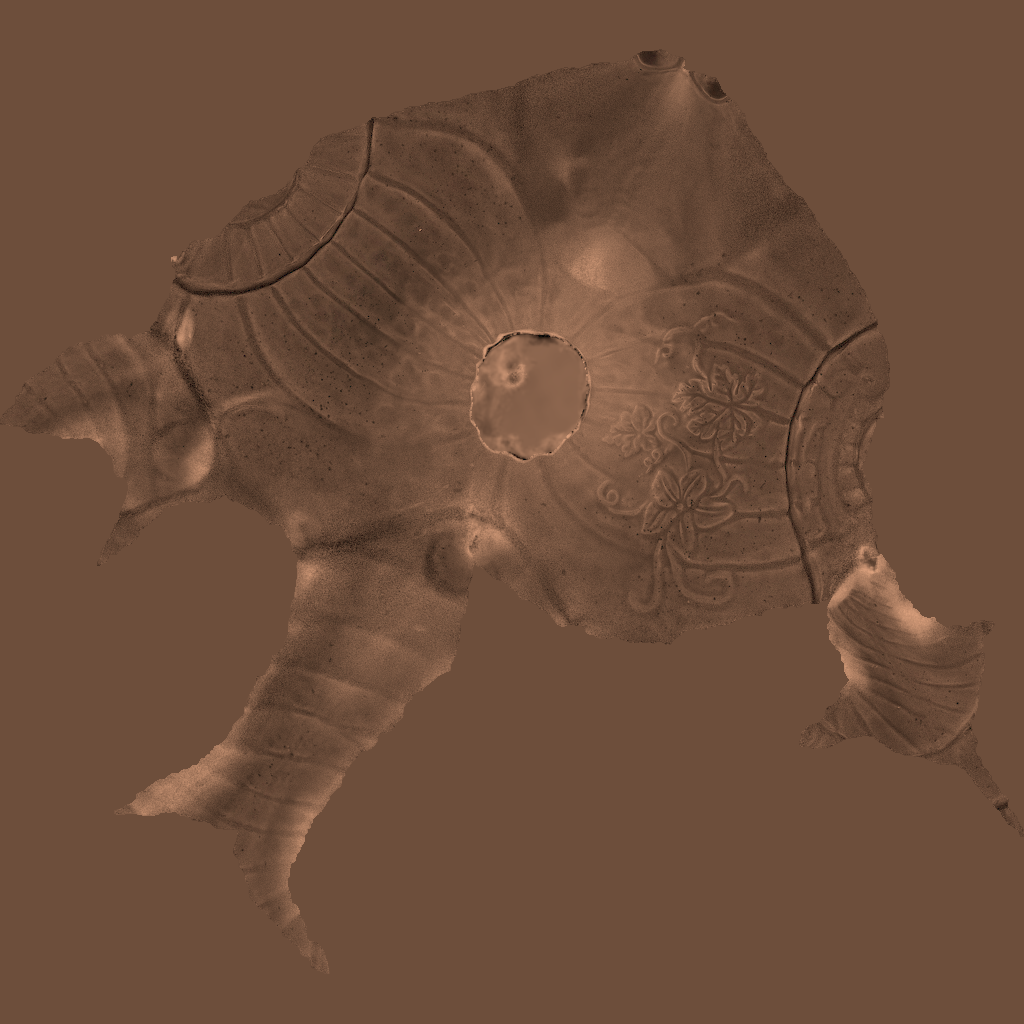} & 
		\includegraphics[width=0.5\resLen]{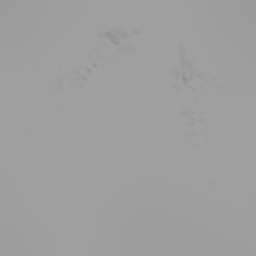} \\
		& diffuse & specular & roughness
	\end{tabular}}
	\caption{\label{fig:inv_comp6}
		\textbf{Inverse-rendering comparison (teapot):}
		We show reconstruction results generated using mesh-based optimization in (b), our implicit stage in (c1), and our full pipeline in (c2). All methods share identical initializations similar to Figure~\protect\ref{fig:inv_comp2}-b.
	}
\end{figure*}

\setlength{\resLen}{1.7in}

\begin{figure*}[t]
	\centering
	\addtolength{\tabcolsep}{-7pt}
	\small
	\rev{\begin{tabular}{cccc}
		(a) \textbf{Ground truth} & (b) \textbf{Mesh-based} & (c1) \textbf{Ours (impl.)} & (c2) \textbf{Ours (full)} \\
		\adjincludegraphics[width=\resLen,trim={{.06\height} {.06\height} {.06\height} {.06\height}},clip]{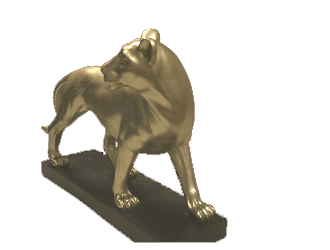} &
		\adjincludegraphics[width=\resLen,trim={{.06\height} {.06\height} {.06\height} {.06\height}},clip]{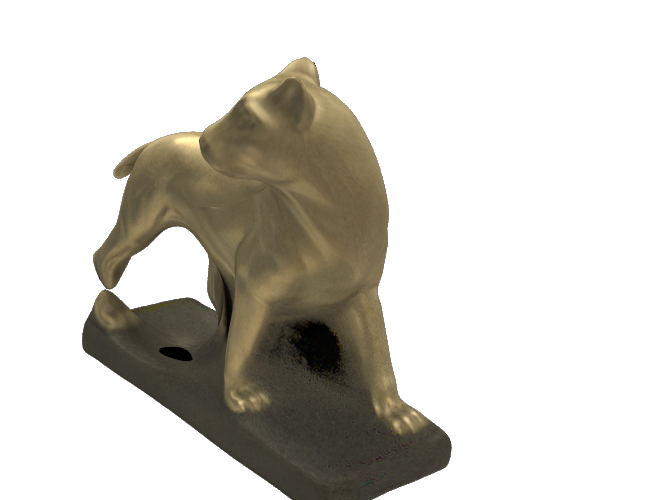} &
		\adjincludegraphics[width=\resLen,trim={{.06\height} {.06\height} {.06\height} {.06\height}},clip]{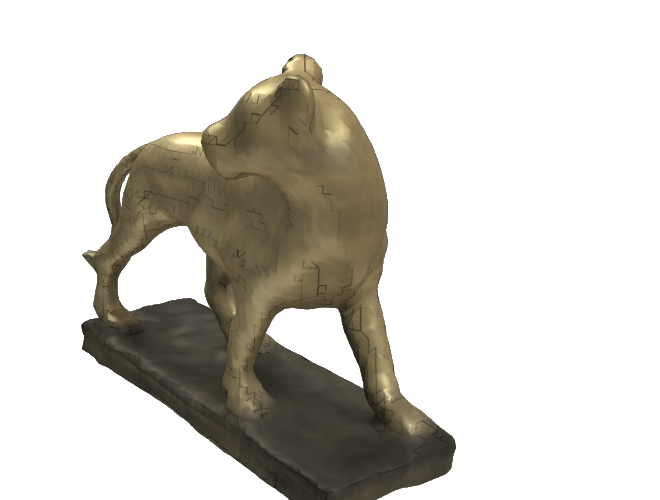} &
		\adjincludegraphics[width=\resLen,trim={{.06\height} {.06\height} {.06\height} {.06\height}},clip]{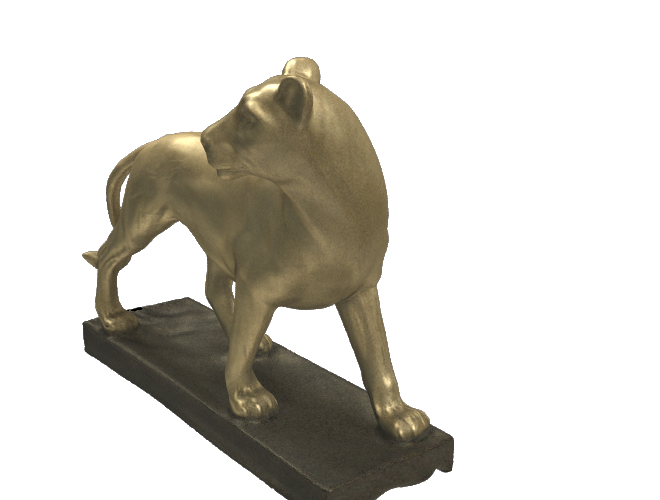} 
		\\
		\textbf{PSNR}: -- & 27.48 & 26.81 & 29.88
		\\
		\adjincludegraphics[width=\resLen,trim={{.06\height} {.06\height} {.06\height} {.06\height}},clip]{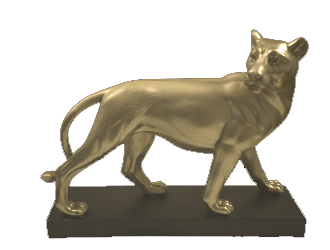} &
		\adjincludegraphics[width=\resLen,trim={{.06\height} {.06\height} {.06\height} {.06\height}},clip]{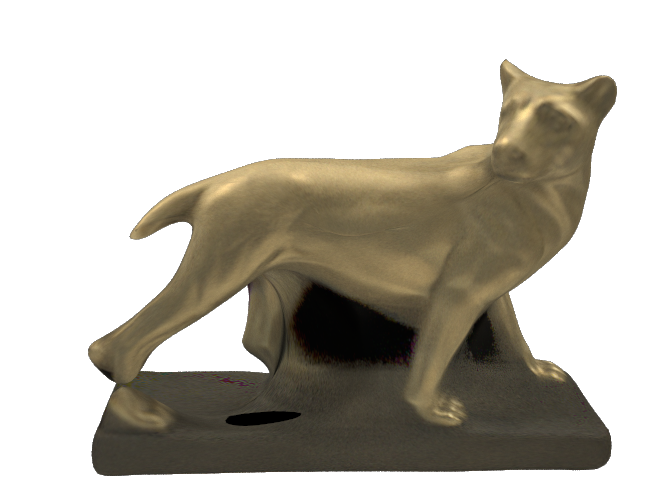} &
		\adjincludegraphics[width=\resLen,trim={{.06\height} {.06\height} {.06\height} {.06\height}},clip]{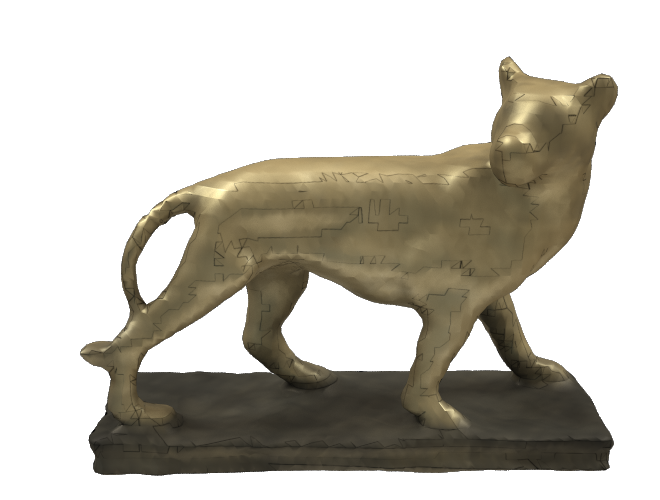} &
		\adjincludegraphics[width=\resLen,trim={{.06\height} {.06\height} {.06\height} {.06\height}},clip]{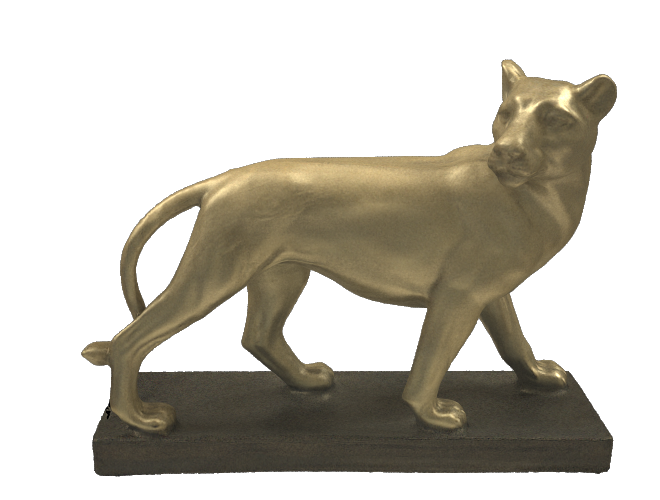} 
		\\
		\textbf{PSNR}: -- & 29.32 & 27.90 & 31.84 \\ \\
		\textbf{Genus}: -- & 0 & 5 & 4 \\
		\\[5pt]
		\hline
		\\[-2pt]
		\raisebox{30pt}{\textbf{Our recon. maps:}} & 
		\includegraphics[width=0.5\resLen]{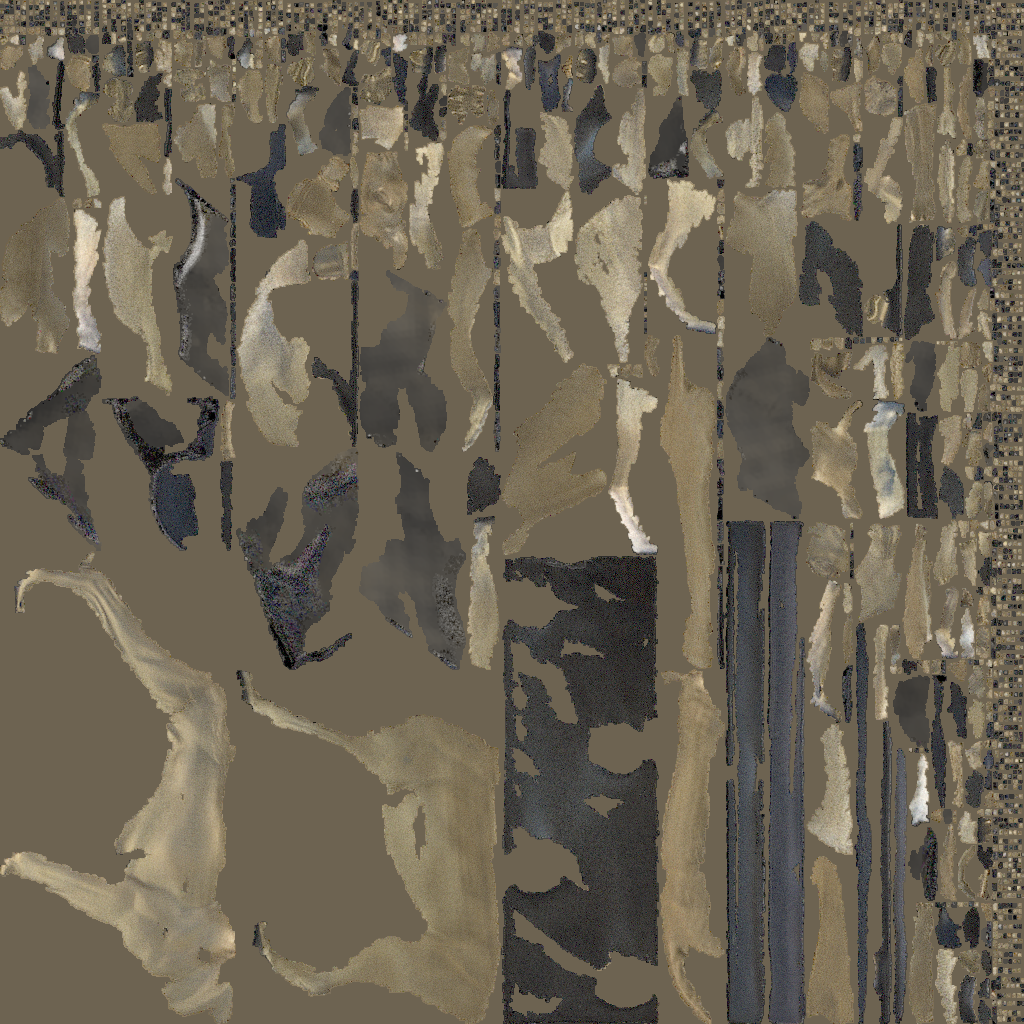} & 
		\includegraphics[width=0.5\resLen]{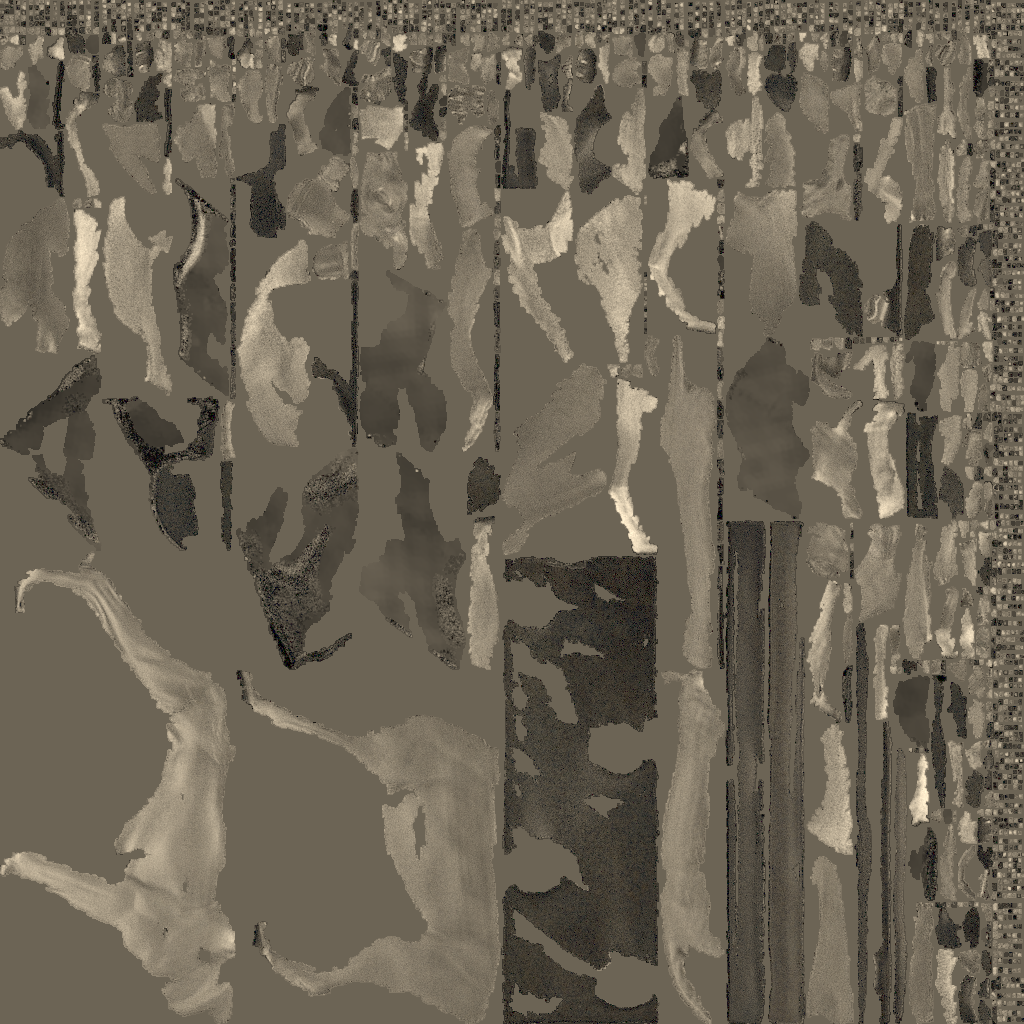} & 
		\includegraphics[width=0.5\resLen]{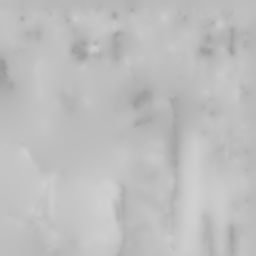} \\
		& diffuse & specular & roughness
	\end{tabular}}
	\caption{\label{fig:inv_comp7}
		\textbf{Inverse-rendering comparison (leopard):}
		We show reconstruction results generated using mesh-based optimization in (b), our implicit stage in (c1), and our full pipeline in (c2). All methods share identical initializations similar to Figure~\protect\ref{fig:inv_comp2}-b.
	}
\end{figure*}

\setlength{\resLen}{1.6in}

\begin{figure*}[t]
	\centering
	\addtolength{\tabcolsep}{-7pt}
	\small
	\rev{\begin{tabular}{cccc}
		(a) \textbf{Ground truth} & (b) \textbf{Mesh-based} & (c1) \textbf{Ours (impl.)} & (c2) \textbf{Ours (full)}
		\\[-15pt]
		\adjincludegraphics[width=\resLen,trim={{0.1\width} 0 {.1\width} {.2\height}},clip]{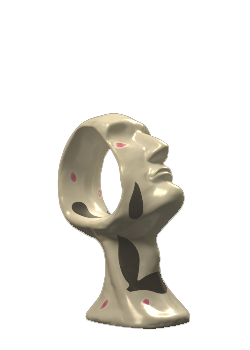} &
		\adjincludegraphics[width=\resLen,trim={{0.1\width} 0 {.1\width} {.2\height}},clip]{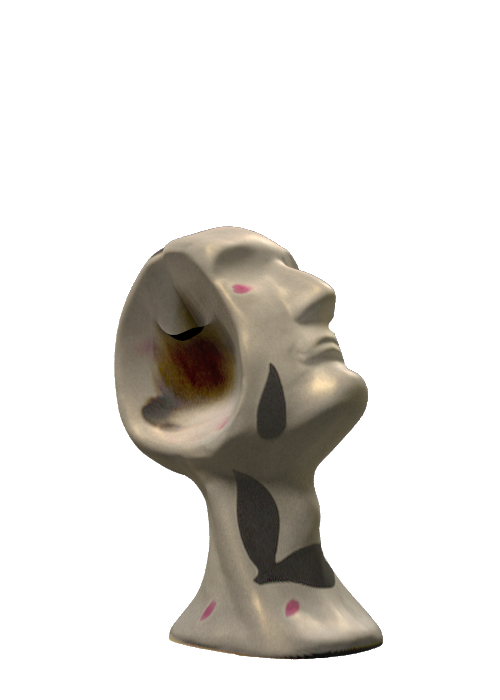} &
		\adjincludegraphics[width=\resLen,trim={{0.1\width} 0 {.1\width} {.2\height}},clip]{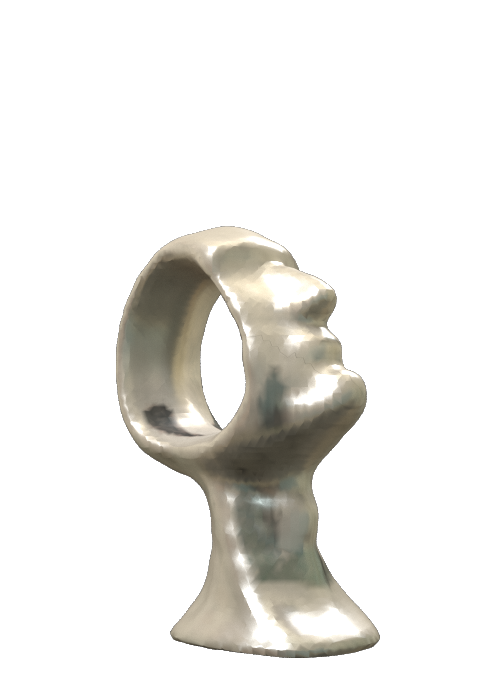} &
		\adjincludegraphics[width=\resLen,trim={{0.1\width} 0 {.1\width} {.2\height}},clip]{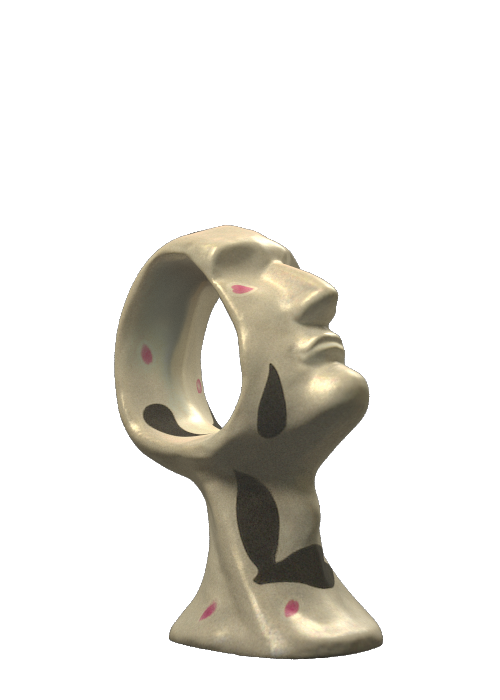} 
		\\[-5pt]
		\textbf{PSNR}: -- & 22.10 & 13.04 & 26.38
		\\[-10pt]
		\adjincludegraphics[width=\resLen,trim={{.06\height} {.06\height} {.06\height} {.06\height}},clip]{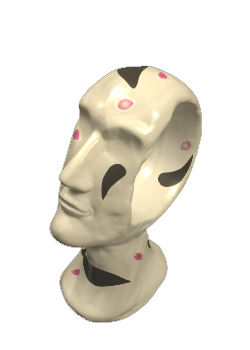} &
		\adjincludegraphics[width=\resLen,trim={{.06\height} {.06\height} {.06\height} {.06\height}},clip]{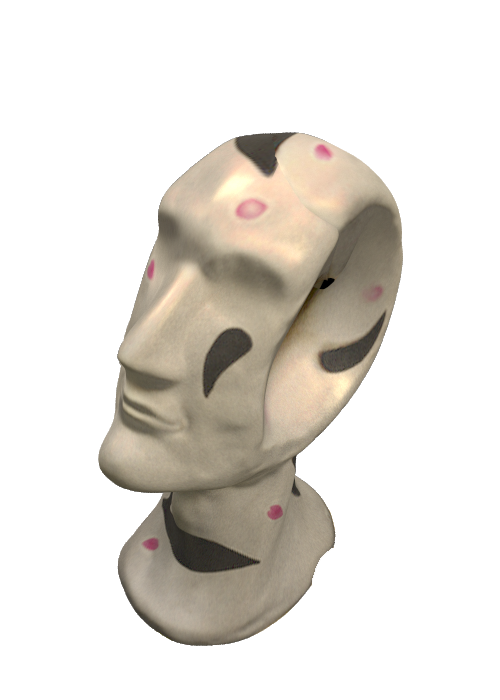} &
		\adjincludegraphics[width=\resLen,trim={{.06\height} {.06\height} {.06\height} {.06\height}},clip]{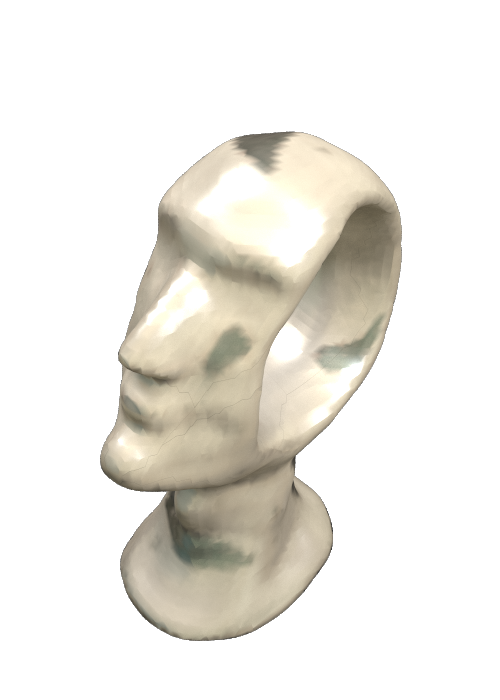} &
		\adjincludegraphics[width=\resLen,trim={{.06\height} {.06\height} {.06\height} {.06\height}},clip]{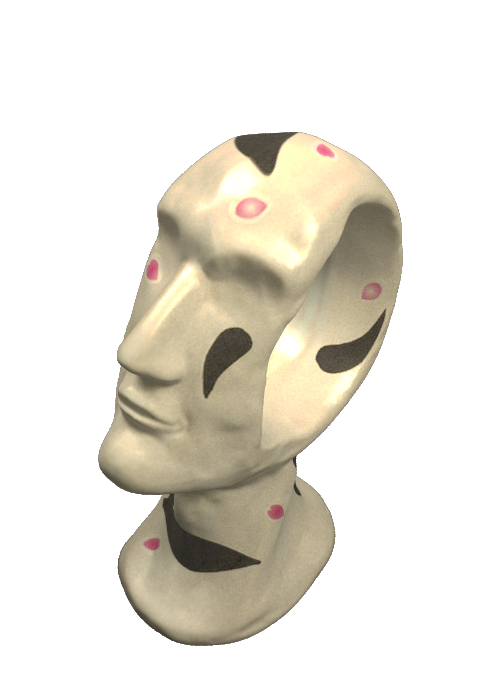} 
		\\
		\textbf{PSNR}: -- & 21.56 & 14.50 & 24.16 \\ \\
		\textbf{Genus}: -- & 0 & 1 & 1 \\
		\\[5pt]
		\hline
		\\[-2pt]
		\raisebox{30pt}{\textbf{Our recon. maps:}} & 
		\includegraphics[width=0.5\resLen]{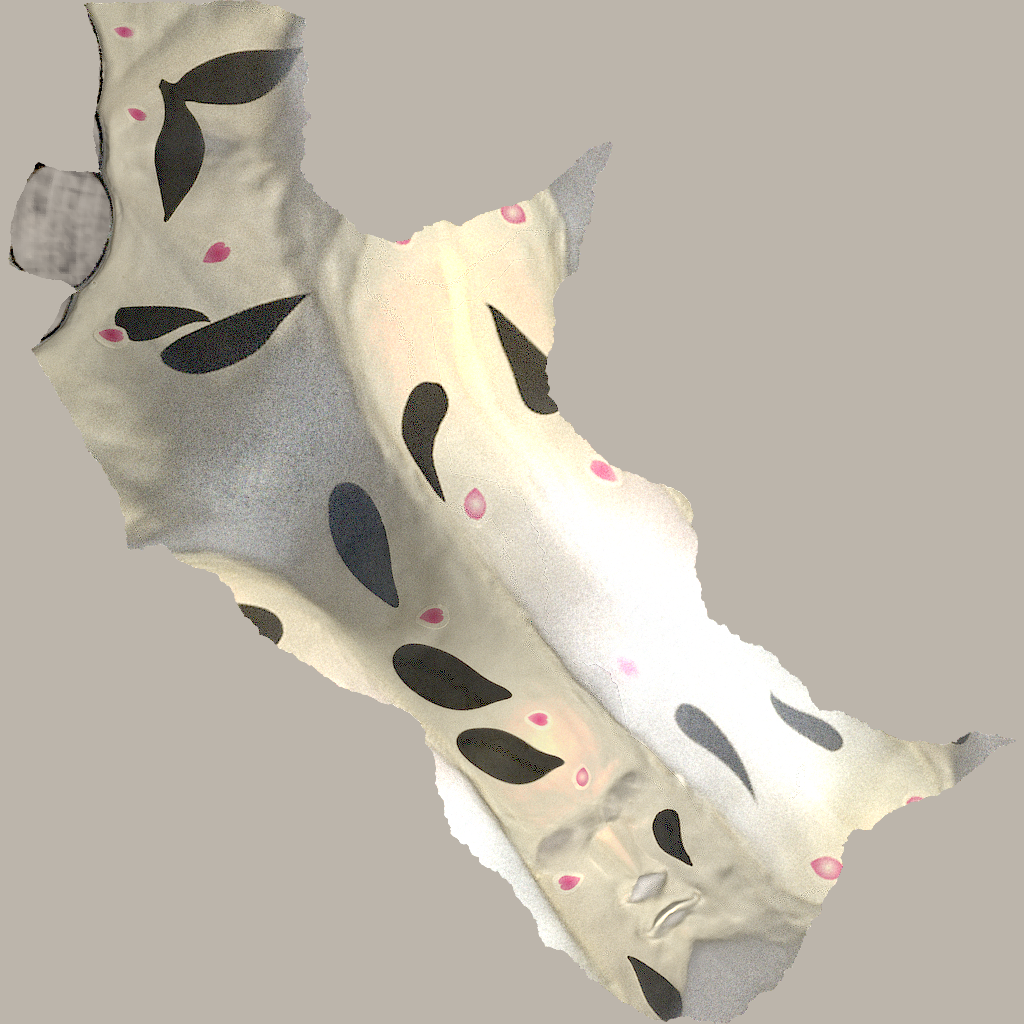} & 
		N/A & 
		\includegraphics[width=0.5\resLen]{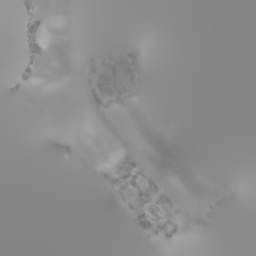} \\
		& diffuse & specular & roughness
	\end{tabular}}
	\caption{\label{fig:inv_comp8}
		\textbf{Inverse-rendering comparison (head):}
		We show reconstruction results generated using mesh-based optimization in (b), our implicit stage in (c1), and our full pipeline in (c2). All methods share identical initializations similar to Figure~\protect\ref{fig:inv_comp2}-b.
	}
\end{figure*}


\end{document}